\documentclass[12pt]{article}
\usepackage{graphicx}

\begin{document}

\title{Electromagnetic and weak form factors of nucleon and charged quasielastic scatterings of neutrino (antineutrino) and nucleon}
%\footnote{\bf \large Published in Physica Sinica, Vol.24, No.2, 124, March 1975
%(Chinese, translated by Jun Gao)}}

\author{Bing An Li \\
Department of Physics and Astronomy, University of Kentucky\\ Lexington, Kentucky,
40506, USA}
\maketitle

\newpage
\tableofcontents
\newpage

\pagebreak
\begin{abstract}
The study of electromagnetic and weak form factors of nucleon (charged quasielastic scatterings of neutrino (antineutrino) and nucleon) done 
in $70^\prime s$ and published in Chinese journals
is reviewed. In the approach of the study antiquark components are introduced to the wave functions of nucleon and the study shows that
the antiquark components of nucleon play an essential role in the EM and weak form factors of nucleon. 
The SU(6) symmetric wave functions of baryons in the rest frame ( s-wave in the rest frame)
have been constructed.    
In these wave functions there are both quark and antiquark 
components. Using Lorentz transformations these wave functions are boosted to moving
frame. In terms of effective Lagrangian these wave functions are used to study
the EM and weak form factors of nucleon and $p \rightarrow \Delta$. 
The ratio $\mu_p G^p_E/G^p_M$, $G^n_E$, $G^n_M$, 
$G^*_M$, $E1+$ and $S1+$ of $p \rightarrow \Delta$ are predicted. The axial-vector form factors of nucleon
is predicted to be $G_A(q^2)/G_A(0) = F^p_1(q^2)$, where the $F^p_1$ is the first Dirac form factor of proton.
This prediction agrees with data very well. 
The pseudoscalar form factor of nucleon is predicted. 
The model predicts there are three 
axial-form factors for $p\rightarrow\Delta$ and two of them play dominant roles. 
The cross sections of $\nu_\mu + n \rightarrow p + \mu^-\;\;\bar{\nu}_\mu + p \rightarrow n + \mu^+$,
$\Delta S = 1$ quasielastic neutrino scatterings, and $\nu_\mu + p \rightarrow \Delta^{++} + \mu^-$ are predicted.
Theoretical results are in agreement with data. The study shows that antiquark components of baryons play an essential role in understanding 
nucleon structure. 
\end{abstract}

\pagebreak

\section{Introduction}
The structure and properties of hadrons are always important topics of strong interaction, QCD.
Photon and neutrino are always used to explore the structure of hadrons.
The measurements and theoretical studies of the electromagnetic (EM) and weak form factors of nucleon last
for very long time and are still very active. Recently, Jlab Hall A Collaboration
has reported the measurements of the ratio of the electric and magnetic form factors of the
proton $\mu_p G^p_E/G^p_M$ by using the recoil polarization technique[1]. 
The measurement of this ratio by Jlab Hall C collaboration has been extended to 
$q^2 = 8.5\; \textrm{GeV}^2$ [2]. The
ratio $\mu_p G^p_E/G^p_M$ measured by the recoil polarization technique [1,2] shows a systematic decrease as $q^2$ increases.
These experimental results lead to many theoretical study on the shape of nucleon.
On the other hand, the experimental measurements [3] in the range of $q^2 < 1\; \textrm{GeV}^2$ show that this ratio is pretty flat [3] around one.
The understanding of this ratio from lower $q^2$ to high $q^2$ is a challenge. 
Besides this ratio the experimental data of the charge and magnetic form factors of neutron and the transit form factors of $p \rightarrow \Delta$
are very significant for understanding the structure of nucleon. Especially, 
the measurements of the electric quadruple $E1+$ and $S1+$ of $p \rightarrow \Delta$ are strongly related to the structure of nucleon.  
On the other hand the axial-vector form factor $G_A(q^2)$ of nucleon and the transit axial-vector form factors of $p \rightarrow \Delta$ 
play important role in understanding the structure of nucleon. A reasonable model of nucleon is needed to understand all these physical quantities.                 

In 70's we have done systematic study on electromagnetic structure of nucleon and $p \rightarrow \Delta$
and axial-vector form factors of $\nu + N \rightarrow N' + \mu$, $\nu + N \rightarrow Hyperon + \mu$, and $\nu + N \rightarrow \Delta + \mu$
scatterings in a relativistic quark model. These investigations are published in a Chinese journal [4,5,6,7]. 
In the approach of Refs. [4,5,6,7] the wave functions of baryons are assumed to be SU(6) symmetric in the rest frame of the baryon and
are boosted to moving frame by Lorentz transformation. In the rest frame the nucleon of $\underline{56}$-plet is in s-wave and spherical.
 
The approach used in Refs. [4,5,6,7] is interesting and the predictions of these form factors made by these study many years ago are still
compatible with current more accurate data in the range of $q^2 < 5\; \textrm{GeV}^2$. 
A review of these study with new fit are presented in this paper.
It is not the intention of this paper to review all the development of these topics. 

The behavior of $\mu_p G^p_E/G^p_M$ found by experiments [1,2,3] has been studied and
predicted by a relativistic quark model in 1975 [4,5]. The charge form factor and the magnetic form factor of the neutron, the transit magnetic  
form factor, and the the electric quadrupole
$E1+$ and Coulomb multiple $S1+$ of $p\rightarrow\Delta$ have been predicted by the same
model [4,5]. In 1977 this model has been applied to study semileptonic decays of baryons, $\nu +  n\rightarrow\mu^- +  p$, $\bar{\nu} +  p\rightarrow\mu^+ +  n$,
and $\nu + p\rightarrow\mu^- + \Delta^{++}$ [6,7]. The axial-vector form factor $G_A(q^2)$
of nucleon, the pseudoscalar form factor of nucleon, and the three axial-form factors of $p \rightarrow \Delta$
have been predicted. Theoretical predictions of the $G_A(q^2)$, 
the decay rates of the semileptonic decays of baryons, the scatterings of $\nu +  n\rightarrow\mu^- +  p$, $\bar{\nu} +  p\rightarrow\mu^+ +  n$,
and $\nu + p\rightarrow\mu^- + \Delta^{++}$ agree with data pretty well.
The calculation of the $\sigma$ term of nucleon is new. All these study [4,5,6,7] on the EM and weak form factors of nucleon 
show that the antiquark components of nucleon play an essential role in the structure of nucleon.
In the last two sections both the $\sigma$ term of proton and the contribution of the antiquark spinors to the density of antiquark of proton  
are calculated.

After the contents shown above there is Appendix in which the wave functions of low-lying excited baryons [4] are presented. The references
are presented too.
%This paper is organized as 1)Introduction; 
%2)Relativistic quark model; 3)wave functions of nucleon and $\Delta$;
%4)Trasit matrix elements and effective currents; 5)EM form factors of baryons; 
%6)Relationship between $f_1(x_1, x_2, x_3)$ and  $f_2(x_1, x_2, x_3)$;
%7)EM form factors of proton and neuttron;
%8)EM transition of $p\rightarrow\Delta$;
%9)Weak interactions and axial-vector form factors of baryon;
%$\nu_\mu + n \rightarrow \mu^- + p$ and $\bar{\nu}_\mu + p \rightarrow \mu^+ + n$; 
%7)semileptonic weak decays of baryons; 8) $\Delta S = 1$ quasielastic neutrino scattering;   
%9) $\nu_\mu + p \rightarrow \mu^- + \Delta^{++}$ scattering; 10)summary; 11)appendix.
 
\section{Relativistic quark model}
In 1964 Gell-Mann and Zweig [8] proposed the quark model of hadrons. In 1964 Greenberg [9] proposed the quantum number of color and the paraquark model.
In 1965 Han and Nambu applied the new quantum number, color, to quarks of integer charges to solve the problem of statistics of quarks [10].
In 1966 H. Y. Zhu at al. published a paper [11] of a relativistic quark model (straton model named in their papers).
In 1966 Y. Y. Liu [12] proposed that the quarks are a triplet of colors to solve the problem of statistics and fractional charges of quarks.
%In 1971 Feynman, Kislinger, and Ravndal [13] proposed a relativistic simple harmonic quark model.

Following Ref. [11], the EM and weak form factors of baryons are studied in Refs. [4,5,6,7] in 70's. 
The wave function of a baryon is composed of: color part, flavor part, the spin part, and the part of space and time.
A baryon is color singlet and the remaining parts of the wave functions are totally symmetric. 
The flavor part is determined by SU(3) symmetry. In Ref. [11]
the spin part of the wave functions of baryons are constructed by the spinors of quarks only.
They are Bargmann-Wigner wave functions.
%respectively. They are Bargmann-Wigner wave functions. For example, in the rest frame the spin part of the $0^{-+}$ meson is expressed as 
%\[\sum_{r_1,r_2} C^{00}_{{1\over2} r_1 {1\over2} r_2} \epsilon_{r1 r_2} u_{r_1} \bar{v}_{r_2} = {1\over 2\sqrt{2}}(1+\gamma_0)\gamma_5, \]
%where $\epsilon_{r_1 r_2} $ is an antisymmetric tensor and $\epsilon_{{1\over2} -{1\over2}} = 1 $, $u_{lambda}$ quark spinor of 4-components and
%$v_{\lambda}$ is the antiquark spinor of 4-components. Similarly, the spin part of baryon is constructed by $u_{\lambda_1}u_{\lambda_2}u_{\lambda_3}$
%and corresponding C-G coefficients.
Lorentz transformation is used to boost these wave functions to moving frame.
%For example, in the moving frame the spin part of the $0^{-+}$ meson is expressed as
%\[{1\over2\sqrt{2}}(1 + \frac{\gamma\cdot p}{m})\gamma_5 \]
%where m is the mass of the meson. 
In Ref. [4] a new set of wave functions of baryons are constructed by assuming the SU(6) symmetry in the 
rest frame of baryon. In these wave functions there are both quark and antiquark components. They are not Bargmann-Wigner wave functions.
Effective currents of electromagnetic and weak interactions are used to calculate transit matrix elements.
The picture of single quark transition has been applied to study the 
electromagnetic and weak processes of hadrons.
%Four quark interactions have been applied to study the processes of strong interactions of hadrons [14]. Besides the mass relations
%obtained by the symmetry breaking of SU(3) and SU(6) new mass relations are obtained
%\[\frac{m^2_K - m^2_\pi}{m^2_\rho - m^2_\pi} = {3\over 4}\frac{M_\Xi - M_\Sigma}{M_\Delta - M_N}.\]
%M_\Xi - M_\Sigma = M_{\Xi^* - M_\Sigma^*}.\]
These wave functions are different from the ones used in Ref. [11].

\section{Wave functions of nucleon and $\Delta$}
There are 
three parts in the calculations of the form factors of baryons: wave functions, effective 
Lagrangian, and the  matrix elements. A review of new set of wave functions of baryons constructed in Ref. [4] is presented in this section.
In Ref. [4] the general expressions of the wave functions of the ${1\over2}^+$ and the ${3\over2}^+$ baryons 
are constructed by requesting that the wave functions satisfy the SU(6) symmetry in the rest frame. 
 
Baryons are bound states of quarks in nonperturbative QCD. Instead solving nonperturbative QCD, symmetry is applied in this study.
It is well known that the SU(6) symmetry [13] works well for low lying hadrons. The ${1\over2}^+$ and ${3\over2}^+$ baryons are $\underline{56}$-plet of the SU(6) group.
The predictions of some of the properties of baryons by SU(6) symmetry are in good agreement with date. 
It is known that kinetic term violates SU(6) symmetry.  In this study the wave functions of the baryons are required to satisfy the SU(6) symmetry in the rest frame of
the baryons only and are boosted to moving frame by Lorentz transformation.
 
The transition matrix elements of EM and weak interactions between baryons are expressed as
\begin{equation}
< B' | j_\mu (0) | B > = \int dx'_1 dx'_2 dx'_3 dx_1 dx_2 dx_3 \bar{B}'(x'_3, x'_2, x'_1) G_\mu (x'_3, x'_2, x'_1, x_1, x_2, x_3) B(x_1, x_2, x_3),
\end{equation}
where $j_\mu$ is either the electromagnetic or weak current of quarks, which will be shown below, $G_\mu$ is the kernel of the matrix element and will be discussed.
In Eq. (1) the wave functions of baryon have indices: color, flavor, and spin and they are Bethe - Salpeter amplitudes.
For example, the wave function of a ${1\over2}^+$ baryon is defined as
\begin{eqnarray}
B^{i'j'k'}_{\alpha i, \beta j, \gamma k}(x_1,x_2,x_3)^m_{l,\lambda} = < 0 |T\{\psi^{i'}_{\alpha i}(x_1)\psi^{j'}_{\beta j}(x_2)
\psi^{k'}_{\gamma k}(x_3)\} | B(p)_\lambda >,\nonumber \\
\bar{B}^{i'j'k'}_{ijk}(x_1,x_2,x_3)^m_{l,\lambda} = < B(p)_\lambda |T\{\bar{\psi}^{i'}_{\alpha i}(x_1)\bar{\psi}^{j'}_{\beta j}(x_2)
\bar{\psi}^{k'}_{\gamma k}(x_3)\} | 0 >, 
\end{eqnarray}
where $\psi(x)$ is a quark field, $i' j' k'$ are color indices, i j k are flavor indices. For a ${1\over2}^+$ baryon the flavor state of the baryon is represented 
by $\left(\begin{array}{c}
           m\\
           l
    \end{array} \right) (see Appendix)$ .    

Baryon is color singlet and the wave function of color is
\[{1\over\sqrt{6}}\epsilon_{i'j'k'},\]
Therefore the remaining part of the wave function, 
the indices $x_1 \alpha i,\; x_2 \beta j,\; x_3 \gamma k$ are total symmetric.
The baryons of SU(6) $\underline{56}$-plet have positive parity. These wave functions must be 
Lorentz covariant. 
According to $SU(3)$ symmetry, ${1\over2}^+$ baryons are octet and the flavor part 
of the wave function  
takes following forms
\begin{equation}
\epsilon_{ijm}\delta_{kl},\;\;\epsilon_{jkm}\delta_{il},\;\;
\epsilon_{kim}\delta_{jl}.
\end{equation}
They satisfies the identity. 
\begin{equation}
\epsilon_{ijm}\delta_{kl}+\epsilon_{jkm}\delta_{il}+\epsilon_{kim}\delta_{il}=0.
\end{equation}
Besides the color part the rest part of the wave 
function of a baryon must be total symmetric.
The wave function of ${1\over2}^+$ baryon is expressed as
\begin{eqnarray}
\lefteqn{B_{\alpha\beta\gamma,ijk}(x,y)^m_{l,\lambda}={1\over 6\sqrt{2}}\epsilon_{i'j'k'}
\{(\epsilon_{ijm}\delta_{kl}+\epsilon_{ikm}\delta_{jl})
\Gamma^{{1\over2}}_{\alpha\beta,\gamma}(x,y)_\lambda}\nonumber \\
&&+(\epsilon_{jkm}\delta_{il}+\epsilon_{ikm}\delta_{jl})
\Gamma^{{1\over2}}_{\beta\gamma,\alpha}(x,y)_\lambda\},
\end{eqnarray}
where x and y are relative coordinates, $x = x_1 - x_2$ and $y = {1\over 2}(x_1 + x_2) - x_3 $, 
and $\Gamma^{{1\over2}}_{\alpha\beta,\gamma}(x,y)_\lambda$
is antisymmetric in  $(x_1,\alpha);\; (x_2,\beta)$ and $\Gamma^{{1\over2}}_{\beta\gamma,\alpha}(x,y)_\lambda$
is antisymmetric in $(x_2,\beta);\; (x_3,\gamma)$. There is $\Gamma^{{1\over2}}_{\gamma\alpha,\beta}(x,y)_\lambda$
which is antisymmetric in  $(x_3,\gamma);\; (x_1,\alpha)$.
These three $\Gamma$ functions must satisfy the identity
\begin{equation}
\Gamma^{{1\over2}}_{\alpha\beta,\gamma}(x,y)_\lambda  
+\Gamma^{{1\over2}}_{\beta\gamma,\alpha}(x,y)_\lambda  
+\Gamma^{{1\over2}}_{\gamma\alpha,\beta}(x,y)_\lambda=0.
\end{equation}  
These three $\Gamma$ functions have the $O_2$ mixing symmetry ,
where $O_2$ is the projection operator of permutation group of three indices (see Appendix). 
The wave function of ${1\over2}^+$ baryon (5) is totally antisymmetric.

${3\over2}^+$ baryons are SU(3) decuplet whose flavor wave functions are
\begin{equation}
d^{lmn}_{ijk}
\end{equation}
where lmn are the flavor wave function of the ${3\over2}$ baryon and ijk are the flavor indices of the three quarks and they are totally symmetric respectively, the values of $d^{lmn}_{ijk}$ 
are presented in the Appendix. The general expression of the wave functions of  
${3\over2}^+$ 
baryons are written as
\begin{equation}
B^{lmn}_{\alpha\beta\gamma, ijk}(x,y)_\lambda={1\over 6\sqrt{2}}\epsilon_{i'j'k'}
d^{lmn}_{ijk}\Gamma^{{3\over2}}_{\alpha\beta\gamma}(x,y)_\lambda,
\end{equation}
where the $\Gamma^{{3\over2}}_{\alpha\beta\gamma}(x,y)_\lambda$ function are totally
symmetric in 
$(x_1\alpha)$, $(x_2\beta)$, $(x_3\gamma)$.

The $\Gamma$ functions of Eqs. (5,8) can be determined. Besides the symmetric properties mentioned above, 
these functions must satisfy following properties:
\begin{enumerate}
\item they are Lorentz covariant,
\item $\Gamma^{{1\over2}}_{\alpha\beta,\gamma}(x,y)_\lambda$,
      $\Gamma^{{1\over2}}_{\beta\gamma,\alpha}(x,y)_\lambda$,
      and $\Gamma^{{1\over2}}_{\gamma\alpha,\beta}(x,y)_\lambda$ are spin-${1\over2}$ and their parity are positive,
\item $\Gamma^{{3\over2}}_{\alpha\beta\gamma}(x,y)_\lambda$ is spin - ${3\over2}$ and positive parity,
\item according to the SU(6), the ${1\over2}^+$ and ${3\over2}^+$ baryons are $\underline{56}$-plet and they are s-wave states
      in the rest frame.
\end{enumerate}
Using these properties, the general expressions of these $\Gamma$ functions in the rest frame can be constructed and 
they are boosted to moving frame by Lorentz transformation.
In the rest frame of the baryon the $\Gamma$ function of Eqs.(5,8) can be written as
\begin{equation}
\Gamma_{\alpha\beta\gamma}(x,y)_\lambda=\sum_c C^{J\lambda}_{{1\over2}c,l
\lambda-c}\{A_{\alpha\beta}(x,y)D_{\gamma\gamma'}(x,y)\}_{l\lambda-c} 
u_{c,\gamma'},
\end{equation}
where $C^{J\lambda}_{{1\over2}c,l \lambda-c}$ is the Clebsch-Gordan coefficient (C - G coefficient),
$\{A_{\alpha\beta}(x,y)D_{\gamma\gamma'}(x,y)\}_{l\lambda-c}$ has angular momentum l 
and the third component is $\lambda-c$. The Lorentz transformations of matrices A and D are
\begin{eqnarray}
\lefteqn{(AC)'=\Lambda(AC)\Lambda^{-1},}\nonumber \\
&&D'=\Lambda D\Lambda^{-1},
\end{eqnarray}
where $\Lambda$ is the Lorentz transformation, C is the operator of charge conjugate,
AC and D are linear combinations of $\gamma$ matrices, $u_{c,\gamma'}$ is the spinor in the rest frame, 
the indices c is the helicity and the $\gamma'$ is the Lorentz index.

In the rest frame we can use following $\gamma$ matrices
\[Scalar:\;\;I,\;\gamma_0,\]
\[Pseudoscalar:\;\;\gamma_5,\;\gamma_0\gamma_5,\]
\[Vector:\;\;\gamma_j,\;\;\gamma_0 \gamma_j,\]
\[Axial-vector:\;\;\gamma_j\gamma_5,\;\;\gamma_0\gamma_j\gamma_5\]
and $x_j$ and $y_j$ ( $j = 1,2,3$)
to construct the general $\Gamma$ functions (5,8). Because
\[\gamma_0 u_{c,\gamma'} = u_{c,\gamma'}\]
in the $D_{\gamma\gamma'}$ of Eq. (9) there is no $\gamma_0$ matrix.
The $\Gamma$ function can be constructed as 
\begin{eqnarray}
\lefteqn{\Gamma_{\alpha\beta\gamma}(x,y)_\lambda=\{(f_1+f_2\gamma_0+f_3 \vec{x}\cdot
\vec{\gamma}+f_4\vec{y}\cdot\gamma+f_5 x_i\sigma_{i4}+f_6 y_i\sigma_{i4}}\nonumber \\
&&+f_7 x_i y_j\sigma_{ij}+f_8x_i y_j\gamma_k\epsilon_{ijk}\gamma_5)\gamma_5C\}_{\alpha
\beta}\nonumber \\
&&\{1+f_9\vec{x}\cdot\gamma+f_{10}\vec{y}\cdot\gamma+f_{11}x_i y_j\sigma_{ij}\}
_{\gamma\gamma'}u_{\lambda,\gamma'}\nonumber \\
&&+ \{(g_1+g_2\gamma_0+g_3 \vec{x}\cdot
\vec{\gamma}+g_4\vec{y}\cdot\gamma+g_5 x_i\sigma_{i4}+g_6 y_i\sigma_{i4}\}\nonumber \\
&&+g_7 x_i y_j\sigma_{ij}+g_8x_i y_j\gamma_k\epsilon_{ijk}\gamma_5)C\}_{\alpha
\beta}\nonumber \\
&&\{(1+g_9\vec{x}\cdot\gamma+g_{10}\vec{y}\cdot\gamma+g_{11}x_i y_j\sigma_{ij})\gamma_5\}
_{\gamma\gamma'}u_{\lambda,\gamma'}\nonumber \\
&&+ \{(h_1\gamma_k+h_2\gamma_0\gamma_k+h_3 x_i\sigma_{ik}+h_4 y_i\sigma_{ik}
+h_5 x_i\gamma_j\epsilon_{ijk}\gamma_5\nonumber \\
&&+h_6 y_i\gamma_j\epsilon_{ijk}\gamma_5
+h_7 x_i y_j\epsilon_{ijk}\gamma_5+h_8 x_i y_j\epsilon_{ijk}\gamma_0\gamma_5)C\}
_{\alpha\beta}\nonumber \\
&&\{(\gamma_k+h_9 x_i\sigma_{ik}+h_{10} y_i\sigma_{ik})\gamma_5\}_{\gamma\gamma'}
u_{\lambda,\gamma'}\nonumber \\
&&+ \{(k_1\gamma_k+k_2\gamma_0\gamma_k+k_3 x_i\sigma_{ik}+k_4 y_i\sigma_{ik}
+k_5 x_i\gamma_j\epsilon_{ijk}\gamma_5\nonumber \\
&&+k_6 y_i\gamma_j\epsilon_{ijk}\gamma_5
+k_7 x_i y_j\epsilon_{ijk}\gamma_5+k_8 x_i y_j\epsilon_{ijk}\gamma_0\gamma_5)\gamma_5 C\}
_{\alpha\beta}\nonumber \\
&&\{\gamma_k+k_9 x_i\sigma_{ik}+k_{10} y_i\sigma_{ik}\}_{\gamma\gamma'}
u_{\lambda,\gamma'},
\end{eqnarray}
where $f_1 ...f_{11}, g_1 ...g_{11}, h_1...h_{10}, k_1...k_{10}$ are Lorentz invariant
functions of x, y, p.
Applying the projection operators $Y_s,\;O_2$ (Appendix) to the indices $(\alpha x_1), (\beta x_2), (\gamma x_3)$ respectively 
we can obtain
\begin{eqnarray} 
\Gamma^{{1\over2}}_{\alpha\beta, \gamma}(x,y)_{\lambda} = O_2 \Gamma_{\alpha\beta\gamma}(x,y)_\lambda,\nonumber \\
\Gamma^{{3\over2}}_{\alpha\beta\gamma}(x,y)_{\lambda} = Y_s \Gamma_{\alpha\beta\gamma}(x,y)_\lambda,
\end{eqnarray}
where $Y_s$ is the total symmetric projector. The same way $\Gamma^{{1\over2}}_{\beta\gamma, \alpha}(x,y)_{\lambda}$ of Eq. (5)
is obtained too. 

In the rest frame the wave functions of the ${1\over2}^+$ and the ${3\over2}^+$ baryons are the bases of the $\underline{56}$-plet of SU(6) symmetry. 
The generators of the SU(6)
group, $\lambda_a \sigma_l$, $\lambda_a$, and $\sigma_l$, can transform one base to another,
where $\lambda_a$ are the generators of the SU(3) group (flavor) and $\sigma_l$ are the SU(2)
generators (spin). 
Using these operators, the wave functions of the ${1\over2}^+$ and the ${3\over2}^+$ 
baryons can be transformed from one to another and the relationships between
these two wave functions can be found.   
Under an infinitesimal SU(6) transformation, for example $\lambda_a \sigma_l$ there are
\begin{eqnarray}
\lefteqn{B'_{\alpha\beta\gamma,ijk}(x,y)_{\lambda U}=B_{\alpha\beta\gamma,ijk}
(x,y)_{\lambda U}}\nonumber \\
&&+i\epsilon^a_l (\lambda_a \sigma_l)^{i'}_{i,\alpha\alpha'}B_{\alpha'\beta\gamma,i'jk}
(x,y)_{\lambda U}\nonumber \\
&&+i\epsilon^a_l(\lambda_a\sigma_l)^{j'}_{j,\beta\beta'}B_{\alpha\beta'\gamma,ij'k}
(x,y)_{\lambda U}\nonumber \\
&&+i\epsilon^a_l(\lambda_a\sigma_l)^{k'}_{k,\gamma\gamma'}B_{\alpha\beta\gamma',ijk'}
(x,y)_{\lambda U},
\end{eqnarray}
where U stands for the flavor state (SU(3)), $B'_{\alpha\beta\gamma,ijk}(x,y)_{\lambda U}$ is 
the wave function of the baryon after the infinitesimal SU(6) transformation, 
$\epsilon^\alpha_l$ is the infinitesimal parameter of the transformation, the color part of the wave function 
has been omitted. Similar transformations under $\lambda_a$ or $\sigma_l$ can be found respectively.
The wave functions
of ${1\over2}^+$ and ${3\over2}^+$ (5,8) already satisfy the SU(3) symmetry. 
Because of the requirement of SU(6) symmetry under an infinitesimal 
transformation of SU(2) (spin)
no baryon states with angular momentum higher than ${3\over2}$ can
be generated.
Without losing generality we take proton as an example,\\
 \(i=1,\; j = 1,\; k = 2,\; a = 2,\; U = (^3_1)\) and \\
\begin{center}
\( \lambda_2 = \left (\begin{array}{ccc}
          0\;\;\;1\;\;\;0\\
          0\;\;\;0\;\;\;0\\
          0\;\;\;0\;\;\;0
          \end{array}\right )\).
\end{center} 
Using Eq. (13), 
under an infinitesimal SU(2) (spin) transformation we obtain
\begin{eqnarray}
B'_{\alpha\beta\gamma,112}(x,y)_{\lambda U}&=&B_{\alpha\beta\gamma,112}(x,y)
+i\epsilon^2_l\sigma_{l,\alpha\alpha'}\{\Gamma^{{1\over2}}_{\alpha'\beta,\gamma}
(x,y)_\lambda+\Gamma^{{1\over2}}_{\beta\gamma,\alpha'}(x,y)_\lambda\}\nonumber \\
&&+i\epsilon^2_l\sigma_{l,\beta\beta'}\{\Gamma^{{1\over2}}_{\alpha \beta', \gamma}(x,y)_\lambda - \Gamma^{{1\over2}}_{\gamma\alpha,\beta'}
(x,y)_\lambda\}.
\end{eqnarray}
For ${3\over2}^+$ baryon we take $\Delta^0$, \(U = (122)\), as an example
\begin{eqnarray}
B'_{\alpha\beta\gamma,122}(x,y)_{\lambda U} = B_{\alpha\beta\gamma,122}
(x,y)_{\lambda U}\nonumber \\
+ \frac{i}{\sqrt{3}}\epsilon^2_l\sigma_{l,\alpha\alpha'}
\Gamma^{{3\over2}}_{\alpha'\beta\gamma}(x,y)_\lambda
+\frac{i}{\sqrt{3}}\epsilon^2_l\sigma_{l,\beta\beta'}
\Gamma^{3\over2}_{\alpha\beta'\gamma}(x,y)_\lambda.
\end{eqnarray}
To satisfy SU(6) symmetry only ${1\over2}^+$ and ${3\over2}^+$ states are allowed on the 
right-hand sides of Eqs. (14,15) and no states with angular momenta higher than ${3\over2}$.

The spin operator $\sigma_l$ makes transformations between the $\Gamma^{{3\over2}}_{\alpha \beta \gamma}(x,y)_\lambda$ of the decuplet baryons
and the $\Gamma^{{1\over2}}_{\alpha \beta,\gamma}(x,y)_\lambda, \Gamma^{{1\over2}}_{\beta\gamma,\alpha}(x,y)_\lambda,
\Gamma^{{1\over2}}_{\gamma \alpha, \beta}(x,y)_\lambda$ of the octet baryons. 

The transformation of the space-spin wave function of the ${1\over2}^+$ baryons (11) is studied.
In the expression of $\Gamma_{\alpha\beta\gamma}(x,y)_\lambda$ (11) there are scalar terms like
\[\vec{x}\cdot\gamma,\;\;\vec{y}\cdot\gamma,\;\;x_i y_j \sigma_{ij},\; etc.....\]
Applying the spin operator $\sigma_i$ to these terms, p-waves will be generated. For example,
\begin{equation}
\sigma_i \vec{x}\cdot\gamma = \{ x_i + \sigma_{ij} x_j\} \gamma_0 \gamma_5.
\end{equation}
The term $x_i$ is a vector and p-wave.
The couplings of these p-waves with other states will produce states whose angular momentum are higher than ${3\over2}$
which do not belong to $\underline{56}$-plet and SU(6) symmetry is broken. 
Therefore, to keep SU(6) symmetry in the rest frame the terms related to p-waves in Eq. (11) must been erased. 
This is consistent with that because  of the SU(6) symmetry the internal kinetic motions of the quarks of the baryons of the $\underline{56}$-plet 
are ignored and they are in s-waves only. 

The general $\Gamma$ function of ${1\over2}^+$ baryon in s-wave
is obtained from Eq. (11)
\begin{eqnarray}
\lefteqn{\Gamma^{{1\over2}}_{\alpha\beta,\gamma}(x,y)_\lambda=
\{(f_1+f_2\gamma_0)
\gamma_5 C\}_{\alpha\beta}u_{\lambda,\gamma}  
+\{(g_1+g_2\gamma_0)
C\}_{\alpha\beta}\{\gamma_5 u_\lambda\}_\gamma}\nonumber \\  
&&+ \{(h_1+h_2 \gamma_0)
\gamma_k C\}_{\alpha\beta}\{\gamma_k\gamma_5 u_\lambda\}_\gamma  
+\{(k_1+k_2\gamma_0)
\gamma_k\gamma_5 C\}_{\alpha\beta}\{\gamma_k u_\lambda\}_\gamma,
\end{eqnarray}
where $f_1, f_2, g_1, g_2, h_1, h_2, k_1, k_2$ are totally symmetric Lorentz scalar 
functions of $x_1,x_2,x_3$. 
Except the color wave function, the remaining wave functions (5) of ${1\over2}^+$ baryons are totally 
symmetric and as mentioned above that the $\Gamma^{{1\over2}}_{\alpha\beta,\gamma}$ satisfies
\begin{equation}
O_2\Gamma^{{1\over2}}_{\alpha\beta,\gamma}(x,y)= 
\Gamma^{{1\over2}}_{\alpha\beta,\gamma}(x,y).
\end{equation}
From Eq. (18) we obtain
\begin{eqnarray}
\lefteqn{g_2=h_1=h_2=k_2=0,}\nonumber \\ 
&&f_1 - f_2 - g_1 - 3k_1=0.
\end{eqnarray}

Similarly, the $\Gamma$ function of ${3\over2}^+$ baryon in s-wave is constructed as
\begin{equation}
\Gamma_{\alpha\beta\gamma}(x,y)_\lambda=\{(f'_1+f'_2\gamma_0)
\gamma_k C\}_{\alpha\beta}\psi^\lambda_{k,\gamma} 
+\{(g'_1+g'_2\gamma_0)\gamma_k\gamma_5 C\}_{\alpha\beta}
\{\gamma_5\psi^\lambda_k\}_\gamma,
\end{equation}
where $f'_1, f'_2, g'_1, g'_2$ are totally symmetric Lorentz scalar functions of 
$x_1,x_2,x_3$, $\psi^\lambda_k$ is the Rarita-Schwinger spinor. $\Gamma_{\alpha
\beta\gamma}(x,y)$ must be totally symmetric and satisfies
\begin{equation}
O_1\Gamma_{\alpha\beta\gamma}(x,y)=0.
\end{equation}
Eq.(21) leads to 
\begin{equation}
g'_1 = 0,\;\;\;g'_2 = f'_2 - f'_1.
\end{equation}
The operators $\sigma_l$ of SU(6) can do the
transformation between the $\Gamma$ functions of spin-${1\over2}$ and spin-${3\over2}$ baryons.
For ${1\over2}^+$ baryon it is obtained
\begin{eqnarray}
(\sigma_l)_{\alpha'\alpha}\Gamma^{{1\over2}}_{\alpha'\beta,\gamma}(x,y)
_\lambda&=&-i\{(f_1+f_2\gamma_0)\gamma_l C\}_{\alpha\beta}u_{\lambda,\gamma}
+ig_1\{\gamma_l\gamma_0\gamma_5 C\}_{\alpha\beta}(\gamma_5 u_\lambda)_\gamma\nonumber \\
+ik_1\{\delta_{kl}\gamma_0 C+i\gamma_0\sigma_{lk}C\}_{\alpha\beta}(\gamma_k u_
\lambda)_\gamma.
\end{eqnarray}
The RHS of eq.(23) can be divided into spin ${1\over2}$ and ${3\over2}$ two parts.
The spin ${3\over2}$ part is written as
\begin{equation}
\{(f_2+f_1\gamma_0)\gamma_k C\}_{\alpha\beta}\psi^\lambda_{k,\gamma}
+g_1(\gamma_0\gamma_k C)_{\alpha\beta}(\gamma_5\psi^\lambda_k)
_\gamma-ik_1(\gamma_0\sigma_{kl}C)_{\alpha\beta}(\gamma_l\psi^\lambda_k)_\gamma.
\end{equation} 
The indices $\alpha\beta\gamma$ of Eq.(24) must be totally symmetric. 
$(\gamma_0\sigma_{kl}C)_{\alpha\beta}$ is antisymmetric in ${\alpha\beta}$. Therefore,
\begin{equation}
k_1 = 0.
\end{equation}
The totally symmetric requirement leads to 
\begin{equation}
g_1 = f_1 - f_2.
\end{equation}
Substituting Eqs.(25,26) into Eq.(24) and comparing with Eq.(20), we obtain
\begin{equation}
f'_1 = f_2,\;\;f'_2 = f_1,\;\;g'_2 = f_1-f_2.
\end{equation}
Using all Eqs.(25,26,27), the spin ${1\over2}$ part of Eq.(20) can be written as
\begin{equation}
\sum_{\lambda,r_3}C^{{1\over2}c}_{1\lambda{1\over2}r_3}\{[(f_2+f_1\gamma_0)
\gamma_l C]_{\alpha\beta}e^\lambda_l u_{r_3}+(f_1-f_2)(\gamma_0\gamma_l\gamma_5 C)
_{\alpha\beta}e^\lambda_l(\gamma_5 u_{r_3})_r\},
\end{equation}
where $e^\lambda_l$ is the polarization vector.
It can be  proved that Eq.(28) is the same as the expression(17) which satisfies Eqs.(18,19).
Taking \(c={1\over2}\) and using the four spinors, the expression(28) can be 
rewritten as
\begin{eqnarray}
\lefteqn{2\sqrt{2}\sum_{r_1 r_2 r_3}C^{{1\over2}{1\over2}}_{1\lambda{1\over2}r_3}
C^{1\lambda}_{{1\over2}r_1{1\over2}r_2}\{f_+ u_{r_1,\alpha}u_{r_2,\beta}
u_{r_3,\gamma}-f_- \epsilon_{r_1 r'_1}\epsilon_{r_2 r'_2}v_{r'_1,\alpha}v_{r'_2,\beta}
u_{r_3,\gamma}\}}\nonumber \\
&&-f_-(\epsilon_{r_1 r'_1}v_{r'_1,\alpha}u_{r_2,\beta}
+\epsilon_{r_2 r'_2}u_{r_1,\alpha}v_{r'_2,\beta})\epsilon_{r_3 r'_3}v_{r'_3,\gamma}\}\nonumber \\
&&=-\frac{i}{\sqrt{3}}\{[(f_1+f_2\gamma_0)\gamma_5 C]_{\beta\gamma}u_{{1\over2}
,\alpha}\nonumber \\
&&+[(f_1+f_2\gamma_0)\gamma_5C]_{\alpha\gamma}u_{{1\over2},\beta}+(f_1-f_2)
C_{\alpha\gamma}(\gamma_5u_{{1\over2}})_\beta\},
\end{eqnarray}
where
\[f_+={1\over2}(f_1+f_2),\;\;f_-={1\over2}(f_1-f_2),\]
\[\epsilon_{{1\over2},-{1\over2}}=1,\;\;\epsilon_{-{1\over2},{1\over2}}=-1,
\;\;\epsilon_{r_1,r_2}=-\epsilon_{r_2,r_1},\]
\begin{eqnarray}
u_{{1\over2}}=\left(\begin{array}{c}
                      1\\0\\0\\0
                      \end{array}
                      \right),\;\;
u_{-{1\over2}}=\left(\begin{array}{c}
                      0\\1\\0\\0
                      \end{array}
                      \right),\;\;
v_{{1\over2}}=\left(\begin{array}{c}
                      0\\0\\0\\-i
                      \end{array}
                      \right),\;\;
v_{-{1\over2}}=\left(\begin{array}{c}
                      0\\0\\i\\0
                      \end{array}
                      \right).
\end{eqnarray}
Multiplying Eq.(29) by corresponding SU(3) wave function
\[\epsilon_{ikm}\delta_{jl}+\epsilon_{jkm}\delta_{il}\]
and symmetrizing the indices ($\alpha$ i), ($\beta$ j), ($\gamma$ k), 
the wave function of ${1\over2}^+$ baryon (5) is indeed obtained.
It is important to notice that all the four spinors (30) are in s-wave. Because of the 
requirement of SU(6) symmetry in the rest frame of the nucleon the kinetic terms of these spinors 
or internal motions of quarks are ignored. After ignoring the kinetic terms, the $u_{\pm}$ are the spinors of quarks
and the $v_{\pm}$ are the spinors of antiquarks.

It is worth to point out that if a spin operator $\sigma_l$ acts on the index 
$\gamma$ of 
$\Gamma^{{1\over2}}_{\alpha\beta,\gamma}$ it doesn't 
change the the spin of the baryon and only changes the third component. 

Using C-G coefficients, following two expressions are obtained after a $\sigma_l$ acts on the 
$\Gamma^{{3\over2}}_{\alpha\beta\gamma}(x,y)_\lambda$ of ${3\over2}$ baryon
\begin{enumerate}
\item
\begin{equation}
\sum_{\lambda_1,\lambda_2}C^{{1\over2}\lambda}_{{3\over2}\lambda_1,1\lambda_2}
e^{\lambda_2}_l (\sigma_l)_{\gamma\gamma'}\Gamma^{{3\over2}}_{\alpha\beta\gamma'}
(x,y)_\lambda=\frac{2\sqrt{2}}{3}i\{\Gamma^{{1\over2}}_{\alpha\gamma,\beta}(x,y)
_\lambda
+\Gamma^{{1\over2}}_{\beta\gamma,\alpha}(x,y)_\lambda\}.
\end{equation}
$\alpha$ and $\beta$ are symmetric. Multiplying Eq.(31) by $\epsilon_{ikm}\delta_{jl}
+\epsilon_{jkl}\delta_{il}$ and symmetrizing the indices $\alpha i, \beta j, 
\gamma k$, the wave function of ${1\over2}^+$ baryon is obtained.
\item 
\begin{equation}
\sum_{\lambda_1,\lambda_2}C^{{3\over2}\lambda}_{{3\over2}\lambda_1,1\lambda_2}
e^{\lambda_2}_l(\sigma_l)_{\gamma\gamma'}\Gamma^{{3\over2}}_{\alpha\beta\gamma'}(x,y)
_{\lambda_1}=-\sqrt{{5\over3}}\Gamma^{{3\over2}}_{\alpha\beta\gamma}(x,y)_\lambda,
\end{equation}
where $\Gamma^{{3\over2}}_{\alpha\beta\gamma}(x,y)_\lambda$ is the wave function of ${3\over2}$ baryon.
\end{enumerate}

Eqs.(31,32) show that the spin operators make the transformations between the ${1\over2}^+$ and ${3\over2}^+$ wave functions of baryons only.  
Therefore, the ${1\over2}^+$ and ${3\over2}^+$ wave functions of baryons satisfy
the SU(6) symmetry.

Finally, the wave functions of ${1\over2}^+$ and ${3\over2}^+$ baryons, which
satisfy SU(6) symmetry in the rest frame are written as
\begin{eqnarray}
\lefteqn{B_{\alpha\beta\gamma,ijk}(x,y)^m_{l,\lambda}=\frac{1}{6\sqrt{2}}
\epsilon_{i'j'k'}\{(\epsilon_{ijm}\delta_{kl}+\epsilon_{ikm}\delta_{jl})
\Gamma^{{1\over2}}_{\alpha\beta,\gamma}(x,y)_\lambda}\nonumber \\ 
&&+(\epsilon_{jkm}\delta_{il}+\epsilon_{ikm}\delta_{jl})
\Gamma^{{1\over2}}_{\beta\gamma,\alpha}(x,y)_\lambda\},\nonumber \\
&&B^{lmn}_{\alpha\beta\gamma,ijk}(x,y)_{\lambda}=\frac{1}{6\sqrt{2}}
\epsilon_{i'j'k'}d^{lmn}_{ijk}
\Gamma^{{3\over2}}_{\alpha\beta\gamma}(x,y)_\lambda,
\end{eqnarray} 
where
\begin{eqnarray}
\lefteqn{\Gamma^{1\over2}_{\alpha\beta,\gamma}(x,y)_\lambda=\{(f_1+f_2\gamma_0)
\gamma_5 C\}_{\alpha\beta}u_{\lambda,\gamma}+(f_1-f_2)C_{\alpha\beta}
\{\gamma_5 u_\lambda\}_\gamma,}\nonumber \\
&&\Gamma^{3\over2}_{\alpha\beta\gamma}(x,y)_\lambda=\{(f_2+f_1\gamma_0)
\gamma_k C\}_{\alpha\beta}\psi^\lambda_{k,\gamma}+(f_1-f_2)(\gamma_0\gamma_k
\gamma_5C)_{\alpha\beta}
\{\gamma_5 \psi^\lambda_k\}_\gamma.
\end{eqnarray}
Eqs. (33,34) show that the SU(6) symmetry leads to that the general expressions of the wave functions 
of ${1\over2}^+$ and ${3\over2}^+$ baryons contain $f_1(x,y)$ and $f_2(x,y)$ two functions only, which are Lorentz invariant and
symmetric in $x_1,\;x_2,\;x_3$. In these two functions there are only s-waves in the
rest frame. Therefore, according to SU(6) symmetry in the rest frame, the ${1\over2}^+$ and the ${3\over2}^+$ baryons are 
spherical. 

Using Lorentz transformation, the $\Gamma$ functions (33,34) are boosted to moving frame 
and expressed as
\begin{eqnarray}
\Gamma^{{1\over2}}_{\alpha\beta,\gamma}(x,y,p)_\lambda=\{(f_1(x,y)+f_2(x,y)
{-i\over m}\gamma\cdot p)
\gamma_5 C\}_{\alpha\beta}u(p)_{\lambda,\gamma}\nonumber \\
+\{f_1(x,y)-f_2(x,y)\}C_{\alpha\beta}
\{\gamma_5 u(p)_\lambda\}_\gamma,\nonumber \\
\Gamma^{{3\over2}}_{\alpha\beta\gamma}(x,y,p)_\lambda=\{(f_2(x,y)+f_1(x,y)
{-i\over m}\gamma\cdot p)
\gamma_\mu C\}_{\alpha\beta}\psi^\lambda_\mu(p)_\gamma\nonumber \\
+\{f_1(x,y)-f_2(x,y)\}
{-i\over m}(\gamma\cdot p\gamma_\mu
\gamma_5C)_{\alpha\beta}
\{\gamma_5 \psi^\lambda_\mu(p)\}_\gamma.
\end{eqnarray}
If taking 
\[f_1(x,y) = f_2(x,y) =f(x,y),\]
Eq. (34) becomes 
\begin{eqnarray}
\Gamma^{{1\over2}}_{\alpha\beta,\gamma}(x,y)_\lambda&=&f(x,y)\{(1+\gamma_0)\gamma_5C\}
_{\alpha\beta} u_{\lambda,\gamma}\nonumber \\
\Gamma^{{3\over2}}_{\alpha\beta\gamma}(x,y)_\lambda&=&f(x,y)\{(1+\gamma_0)\gamma_kC\}
_{\alpha\beta} \psi^\lambda_{k,\gamma}.
\end{eqnarray}
The terms $C_{\alpha\beta} \{\gamma_5 u(p)_\lambda\}_\gamma$ and ${1\over m}(\gamma_0\gamma_\mu
\gamma_5C)_{\alpha\beta} \{\gamma_5 \psi^\lambda_\mu(p)\}_\gamma$ of Eq. (34) disappear.

These wave functions (36) can be constructed by the two spinors of quarks, $u_{1\over2},\;u_{-{1\over2}}$ (30) only and there are no contribution 
from the spinors of antiquarks $v_{1\over2},\;v_{-{1\over2}}$.
Therefore, the effects of antiquark spinors make \(f_1\neq f_2\).  
In order to explore the cause of the difference between $f_1$ and $f_2$ the spectral representation of 
the wave function of baryon is used
\begin{eqnarray}
\lefteqn{B^{i'j'k'}_{\alpha\beta\gamma,ijk}(x,y)_{\lambda U}=6Y_a\theta(x_0)\theta(-
{x_0\over2},-y_0){1\over(2\pi)^6}\int d^4 p_n d^4 p_l d M^2_n dM^2_l}\nonumber \\ 
&&\delta(p^2_n+m^2_n)\delta(p^2_l+M^2_l)\theta(p_{n0})\theta(p_{l0})exp[i(p_n-{1\over2}p_l)x
-i(p_l-{1\over3}p)y]\nonumber \\
&&f^{i'j'k'}_{\alpha\beta\gamma,ijk}(p_n,p_l,p,M^2_n,M^2_l)_{\lambda U},
\end{eqnarray}
where
\begin{eqnarray}
f^{i'j'k'}_{\alpha\beta\gamma,ijk}(p_n,p_l,p,M^2_n,M^2_l)_{\lambda U} =
\sum_{nl}'<0|\psi^{i'}_{\alpha i}(0)|n><n|\psi^{j'}_{\beta j}(0)|l>
<l|\psi^{k'}_{\gamma k}(0)|B_{\lambda U}(p)>,
\end{eqnarray}
where $\sum'_{nl}$ is the $\sum_{nl}$ after the factors ${1\over(2\pi)^6}
\frac{d^3 p_n}{2E_n}\frac{d^3 p_l}{2E_l}d M^2_n dM^2_l$ being taken away, $Y_a$ is the 
antisymmetric operator(see Appendix) to antisymmetrizing $(x_1\alpha ii'), (x_2\beta jj'),
(x_3\gamma kk')$.
 
Using the wave function (37), the expression of $f_1$ is obtained as
\begin{eqnarray}
{1\over\sqrt{2}}f_1(x,y)\epsilon_{i'j'k'}\epsilon_{ijm}\delta_{kl'}u_{\lambda,
\gamma}=\frac{1}{(2\pi)^6}\int d^4 p_n d^4 p_l d M^2_n d M^2_l
\delta(p^2_n+M^2_n)\delta(p^2_l+M^2_l)\theta(p_{n0})\theta(p_{l0})\nonumber \\
\sum'_{n,l}\{\theta(x_0)\theta(-{x_0\over2}+y_0)e^{i(p_n-{1\over2}p_l)x+i(p_l
-{2\over3}p)y]}
<0|(\gamma_5 C\psi^{i'}_i)_\beta|n><n|\psi^{j'}_{\beta,j}|l><l|\psi^{k'}_{\gamma,k}
|B^{m'}_{\lambda,l'}>\nonumber \\
-\theta({x_0\over2}+y_0)\theta({x_0\over2}-y_0)e^{{i\over2}(p_n+p_l-p)x+i(p_n-p_l
-{1\over3}p)y}
<0|(\gamma_5 C\psi^{i'}_i)_\beta|n><n|\psi^{k'}_{\gamma,k}|l><l|\psi^{j'}_{\beta,j}
|B^{m'}_{\lambda,l'}>\nonumber \\
-\theta(-{x_0\over2}+y_0)\theta(-{x_0\over2}-y_0)e^{-{i\over2}(p_n+p_l-p)x+i(p_n-p_l
-{1\over3}p)y}
<0|(\gamma_5 C\psi^{j'}_j)_\beta|n><n|\psi^{k'}_{\gamma,k}|l><l|\psi^{i'}_{\beta,i}
|B^{m'}_{\lambda,l'}>\nonumber \\
+\theta(-x_0)\theta({x_0\over2}+y_0)e^{-i(p_n-{1\over2}p)x+i(p_l
-{2\over3}p)y}
<0|(\gamma_5 C\psi^{j'}_j)_\beta|n><n|\psi^{i'}_{\beta,i}|l><l|\psi^{k'}_{\gamma,k}
|B^{m'}_{\lambda,l'}>\nonumber \\
+\theta(-{x_0\over2}-y_0)\theta(x_0)e^[-{i\over2}(p_n+p-2p_l)x-i(p_n
-{1\over3}p)y]
<0|\psi^{k'}_{\gamma,k}|n><n|(\gamma_5 C\psi^{i'}_i)_{\beta}|l><l|\psi^{j'}_{\beta,j}
|B^{m'}_{\lambda,l'}>\nonumber \\
-\theta({x_0\over2}-y_0)\theta(-x_0)e^{{i\over2}(p_n+p-2p_l)x-i(p_n
-{1\over3}p)y}
<0|\psi^{k'}_{\gamma,k}|n><n|(\gamma_5 C\psi^{j'}_j)_{\beta}|l><l|\psi^{i'}_{\beta,i}
|B^{m'}_{\lambda,l'}>.
\end{eqnarray}
Replacing 
\begin{equation}
\gamma_5 C\psi\;\; by \;\;\gamma_5 C\gamma_0\psi
\end{equation}
in Eq. (39),
\begin{equation}
{1\over\sqrt{2}} f_2 (x,y) \epsilon_{i'j'k'} \epsilon_{ijm'} \delta _{k l'} u_{\lambda, \gamma}
\end{equation}
is obtained. In general the quark field $\psi$ has four components. 
If only quark spinors ($u_{\pm{1\over2}}$ (30)) are taken into account then \(\gamma_0 \psi = \psi\)
and \(f_1 = f_2\) as mentioned above. For antiquark spinors ($v_{\pm{1\over2}}$ (30)) \(\gamma_0\psi = -\psi\) 
and \(f_1 \neq f_2\). 

Theoretically, the relationship between the antiquark spinors $v_{\pm}$ (30) and possible 
antiquark density of the nucleon is a dynamical question. The dynamical nature of this question is not explored in this paper.
However, some arguments are made. 
in Ref. [14] antiquark density of a nucleon are
defined as
\begin{equation}
\bar{q}_i = {1\over2}<p|\bar{\psi}_i\psi_i-\bar{\psi}_i \gamma_0 \psi_i|p>.
\end{equation}
Using the arguments above, this model predicts that
\[\bar{q}_i \neq 0.\]
Detailed calculation of the quantity $\bar{q}_i\;i=u,\;d,\;s$ will be presented in the section of antiquark components.

In this paper when "the contribution of 0antiquark components" is mentioned it really means that
the contributions of the antiquark spinors $v_{\pm}$.

The wave function of baryon, $\bar{B}_{\alpha\beta\gamma,ijk}(x_1,x_2,x_3)_{\lambda U}$ is required in the calculation of the matrix elements and is defined as
(color indices are dropped)
\begin{equation}
\bar{B}_{\alpha\beta\gamma,ijk}(x_1,x_2,x_3)_{\lambda U}=<B_{\lambda U}(p)|
T\{\bar{\psi}_{\alpha,i}(x_1)\bar{\psi}_{\beta,j}(x_2)\bar{\psi}_{\gamma,k}(x_3)\}|0>.
\end{equation} 
Using time reversal, it is found
\begin{equation}
\bar{B}^{i'j'k'}_{\alpha\beta\gamma,ijk}(x_1,x_2,x_3)_{\lambda U}=\eta(-1)^{J+1}(\gamma_5 C)
_{\alpha\alpha'}(\gamma_5 C)_{\beta\beta'}(\gamma_5 C)_{\gamma\gamma'}B^{k'j'i'}
_{\gamma'\beta'\alpha',kji}(-x_3,-x_2,-x_1)_{-\lambda U},
\end{equation}
where $\eta$ is a phase factor and it is determined as
\begin{eqnarray}
J = {1\over2},\;\;\eta = -i,\nonumber \\
J = {3\over2},\;\;\eta = i.
\end{eqnarray}
Using Eqs. (44,35,33), another sets of wave functions in the moving frame are obtained
\begin{eqnarray}
\lefteqn{\bar{B}_{\alpha\beta\gamma,ijk}(x,y)^m_{l,\lambda}=-
\frac{1}{6\sqrt{2}}
\epsilon_{i'j'k'}\{(\epsilon_{ijm}\delta_{kl}+\epsilon_{ikm}\delta_{jl})
\bar{\Gamma}^{{1\over2}}_{\alpha\beta,\gamma}(x,y)_\lambda}\nonumber \\ 
&&+(\epsilon_{jkm}\delta_{il}+\epsilon_{ikm}\delta_{jl})
\bar{\Gamma}^{{1\over2}}_{\beta\gamma,\alpha}(x,y)_\lambda,\nonumber \\
&&\bar{B}^{lmn}_{\alpha\beta\gamma,ijk}(x,y)_{\lambda}=\frac{1}{2\sqrt{2}}
\epsilon_{i'j'k'}d^{lmn}_{ijk}
\bar{\Gamma}^{{3\over2}}_{\alpha\beta\gamma}(x,y)_\lambda,
\end{eqnarray} 
\begin{eqnarray}
\bar{\Gamma}^{1\over2}_{\alpha\beta,\gamma}(x,y,p)_\lambda=
\{C[f_1(-x,-y)+f_2(-x,-y)
{i\over m}\gamma\cdot p]
\gamma_5 \}_{\alpha\beta}\bar{u}(p)_{\lambda,\gamma}\nonumber \\
+[f_1(-x,-y)-f_2(-x,-y)]
C_{\alpha\beta}
\{\bar{u}(p)_\lambda \gamma_5\}_\gamma,\nonumber \\
\bar{\Gamma}^{3\over2}_{\alpha\beta\gamma}(x,y,p)_\lambda=
\{C[f_2(-x,-y)+f_1(-x,-y)
{i\over m}\gamma\cdot p]
\gamma_\mu\}_{\alpha\beta}\bar{\psi}^\lambda_\mu(p)_\gamma\nonumber \\
+[f_1(-x,-y)-f_2(-x,-y)]
{i\over m}(C\gamma_mu\gamma\cdot p
\gamma_5)_{\alpha\beta}
\{\bar{\psi}^\lambda_\mu(p)\gamma_5\}_\gamma.
\end{eqnarray}

In Ref. [4] in the rest frame of baryon the $O(3)\times SU(6)$ symmetry has been
applied to construct the wave functions of one p-wave and one s-wave(1s1p);
one p - wave and one - p wave (1p1p) and one s-wave and 1 d wave (1s1d). 
For convenience these wave functions are presented in the Appendix. 

\section{Transit matrix elements and the effective currents}
In the relativistic quark model [11] the effective Lagrangian 
are used to study
the electromagnetic and weak properties of baryon. 
The electromagnetic (EM) effective Lagrangian of quarks is defined as 
\begin{equation}
{\cal L}=-ie\bar{\psi}Q\{\hat{A}(x)-\frac{i\kappa}{4m_N}\sigma_{\mu\nu}F^{\mu\nu}\}
\psi,
\end{equation}
where \[Q = \left( \begin{array}{ccc}
             {2\over3}\;\;\;    0 \;\;\;      0\\
              0 \;\;\;      -{1\over3}\;\;\;  0\\
              0 \;\;\;          0 \;\;\;      -{1\over3}
             \end{array}  \right ),\]
$\kappa$ is the anomalous magnetic moment of quark and it is a parameter in this model.

The general expression of the transit matrix element between baryons is shown in Eq. (1).
The kernel $G_\mu$ of Eq. (1) is the effect of strong interactions (nonperturbative QCD).  
In the quark model [11] the mechanism of single quark transition is claimed. 
On the other hand, in order to satisfy the requirement of SU(6) symmetry the kinetic terms in the kernel $G_\mu$ must be ignored. 
Under this mechanism the kernel $G_\mu$ is reduced to 
\[M(x'_1,x'_2,x_1,x_2) J_\mu (0),\]
where $M(x'_1,x'_2,x_1,x_2)$ is a scalar function and the $J_\mu(0)$ is the EM or weak current of quarks. The EM current is shown in Eq. (48)
and the weak current will be shown in section 9. 
The transit matrix element (1) is rewritten as
\begin{eqnarray}
<B_\lambda(p')_{U^{\prime}}|J_\mu(0)|B_\lambda(p)_U> = P_{k_1 k'_1}\Gamma_{\mu,
\gamma\gamma^{\prime} }  \nonumber \\
\int dx'_1 dx'_2 dx_1 dx_2 M(x'_1,x'_2,x_1,x_2)\bar{B}^{\lambda^{\prime}}
_{\alpha\beta\gamma,ijk_1}(x'_1,x'_2,0)_{U^{\prime}} B^\lambda_{k'_1ji,\gamma'\beta\alpha}(0,
x_2,x_1)_U,
\end{eqnarray}
where the function $M(x'_1,x'_2,x_1,x_2)$ is unknown and for EM interactions
\begin{eqnarray}
P = Q,\;\;\Gamma_\mu=\gamma_\mu+\frac{\kappa}{2m_N}\sigma_{\mu\nu}q_\nu,
\end{eqnarray}
where $q_\mu$ is the transfer momentum. For weak interactions P and $\Gamma\mu$ have different expressions (see section 9). 
%As mentioned above that SU(6) symmetry
%in the rest frame of baryon is assumed in the study [4,5,6,7]. If the function M contains $\gamma$
%matrix, like $\gamma\cdot x$, SU(6) symmetry is broken. Therefore, in a simplest way 
%the function $M(x'_1,x'_2,x_1,x_2)$ is assumed to be a scalar function to satisfy SU(6) symmetry . 
In Ref. [11] the $M(x'_1,x'_2,x_1,x_2)$ is taken as a constant.

\section{EM form factors of baryons }
In Ref. [5] the wave functions (33,46), the matrix element (49), and the effective EM current (50) are applied to study 
the EM form factors of nucleon. The results are presented in this section.

The EM form factors of a baryon are defined as
\begin{eqnarray}
< B(p')_{\lambda'} | J_\mu (0) | B(p)_\lambda > = \bar{u}(p')_{\lambda'}\{ F_1(q^2) \gamma_\mu + F_2 (q^2) \frac{\kappa\sigma_{\mu\nu} q_\nu}{2 m_p}\} u(p)_\lambda,\\
G_E(q^2) = F_1 - \tau \kappa F_2,\;\;\;G_M(q^2) = F_1 + \kappa F_2,\\
\tau = {q^2\over 4 m^2_p}.
\end{eqnarray}
The electric charge of proton is one.
Therefore, there is normalization condition 
\begin{equation}
G^p_E(0) = 1,\;\;F^p_1(0) = 1.
\end{equation}

By using the matrix element (49), the effective EM current (50), and the wave functions of baryons (33,46), these matrix elements are calculated 
and the EM form factors $G_E$ and $G_M$ of baryons are determined in this model. 

The matrix elements of electric currents of $\frac 12^{+}$ baryon are obtained
\begin{eqnarray}
&<&B_{\lambda ^{^{\prime }}}^{\frac 12}(p_f)_{l_1}^{l_1^{^{\prime }}}\mid
J_\mu (0)\mid B_\lambda ^{\frac 12}(p_i)_{l_2}^{l_2^{^{\prime }}}>
 = -\frac{ie}{24}
\{A_1I_1-A_2I_2\},
\end{eqnarray}
where m, $m'$ and E, $E'$ are the
initial and final mass and energy of the baryon respectively,
\begin{equation}
A_1 = Tr\overline{B} Q B,\;\;A_2 = Tr\overline{B} B Q.
\end{equation}
B, $\overline{B}$ are the $SU_3$ flavor matrices of the initial and
final baryon, which are presented in the Appendix.
\begin{eqnarray}
I_1 &=&-20\{D_2(q^2)(1-\frac{m_{+}}{5m})+D_2^{^{\prime }}(q^2)(1-\frac{m_{+}%
}{5m^{^{\prime }}})  \nonumber \\
&&+\frac 1{2mm^{^{\prime }}}(m_{-}^2+q^2+\frac{\kappa m_{+}}{5m_p}%
q^2)D_3(q^2)\}\overline{u}_{\lambda ^{^{\prime }}}(p_f)\gamma _\mu u_\lambda
(p_i)  \nonumber \\
&&-20\{2D_1(q^2)-(1-\frac{2m_p}{5\kappa m})D_2(q^2)-(1-\frac{2m_p}{5\kappa
m^{^{\prime }}})D_2^{^{\prime }}(q^2)  \nonumber \\
&&+\frac 1{2mm^{^{\prime }}}(m_{+}^2+\frac 35q^2)D_3(q^2)\}\frac \kappa
{2m_p}\overline{u}_{\lambda ^{^{\prime }}}(p_f)q_\nu \sigma _{\mu \nu
}u_\lambda (p_i)  \nonumber \\
&&-4i\{\frac 1mD_2(q^2)-\frac 1{m^{^{\prime }}}D_2^{^{\prime }}(q^2)+\frac
\kappa {2mm^{^{\prime }}m_p}(m^{^{\prime }2}-m^2)D_3(q^2)\}  \nonumber \\
&&q_\mu \overline{u}_{\lambda ^{^{\prime }}}(p_f)u_\lambda (p_i),
\nonumber \\
I_2 &=&4\{(1-2\frac{m_{+}}m)D_2(q^2)+(1-2\frac{m_{+}}{m^{^{\prime }}}%
)D_2^{^{\prime }}(q^2)  \nonumber \\
&&+\frac 1{2mm^{^{\prime }}}(m_{-}^2+q^2+2\frac{\kappa m_{+}}{m_p}%
q^2)D_3(q^2)\}\overline{u}_{\lambda ^{^{\prime }}}(p_f)\gamma _\mu u_\lambda
(p_i)  \nonumber \\
&&-4\{2D_1(q^2)-(1-\frac{4m_p}{\kappa m})D_2(q^2)-(1-\frac{4m_p}{\kappa
m^{^{\prime }}})D_2^{^{\prime }}(q^2)  \nonumber \\
&&+\frac 1{2mm^{^{\prime }}}(m_{+}^2-3q^2)D_3(q^2)\}\frac \kappa {2m_p}%
\overline{u}_{\lambda ^{^{\prime }}}(p_f)q_\nu \sigma _{\nu \mu }u_\lambda
(p_i)  \nonumber \\
&&+8i\{\frac 1mD_2(q^2)-\frac 1{m^{^{\prime }}}D_2^{^{\prime }}(q^2)+\frac{%
\kappa (m^{^{\prime }2}-m^2)}{2mm^{^{\prime }}m_p}D_3(q^2)\}  \nonumber \\
&& q_\mu \overline{u}_{\lambda ^{^{\prime }}}(p_f)u_\lambda (p_i),
\end{eqnarray}
where
\begin{equation}
q_\mu =p_\mu-p'_{f\mu },\;\;m_{+}=m+m^{^{\prime }},\;\;
m_{-}=m^{^{\prime }}-m,
\end{equation}
\begin{eqnarray}
D_1(q^2) &=&-\int f_1^{^{\prime
}}(-x'_1,-x'_2,0)M(x'_1,x'_2,x_1,x_2)f_1(0,x_2,x_1)d^4x'_1d^4x'_2d^4x_1d^4x_2,  
\nonumber \\
D_2(q^2) &=&-\int f_1^{^{\prime
}}(-x'_1,-x'_2,0)M(x'_1,x'_2,x_1,x_2)f_2(0,x_2,x_1)dx'_1dx'_2dx_1dx_2,  \nonumber \\
D_2^{^{\prime }}(q^2) &=&-\int f_2^{^{\prime
}}(-x'_1,-x'_2,0)M(x'_1,x'_2,x_1,x_2)f_1(0,x_2,x_1)d^4x'_1d^4x'_2d^4x_1d^4x_2,  \nonumber \\
D_3(q^2) &=&-\int f_2^{^{\prime
}}(-x'_1,-x'_2,0)M(x'_1,x'_2,x_1,x_2)f_2(0,x_2,x_1)d^4x'_1d^4x'_2d^4x_1d^4x_2.
\end{eqnarray}
$m, m^{^{\prime }}$ are the mass of the initial and final
baryon respectively. 
Eq.(59) shows that when $p'\longleftrightarrow p$ is taken, we have
\begin{equation}
D_2(q^2)\longleftrightarrow D_2^{^{\prime }}(q^2),
\end{equation}
therefore, when \(m = m'\)
\begin{equation}
D_2(q^2) = D_2^{^{\prime }}(q^2).
\end{equation}
In Eq. (57) there are three unknown functions, $D_1(q^2),\;D_2(q^2),\; D_3(q^2)$. 

%Similarly, the matrix elements of $%
%\frac 12^{+}$ baryon$ ->\frac 23^{+}$ baryon are obtained
%\begin{eqnarray}
%<B_{\lambda ^{^{\prime }}}^{\frac 32}(p_f)^{lmn}\mid J_\mu (0)\mid
%B_\lambda ^{\frac 12}(p_i)_{l_1}^{l_1^{^{\prime }}}>
%= ied_{l_1jk}^{lmn}%
%\varepsilon _{jk^{^{\prime }}l_1^{^{\prime }}}Q_{kk^{^{\prime
%}}}\{2D_2(q^2)+\kappa [\frac{m_{+}}{m_p}D_3(q^2)+2\frac m{m_p}D_1(q^2)\nonumber \\
%-\frac m{m_p}D_2(q^2)-\frac m{m_p}D_2^{^{\prime }}(q^2)]\}
%\frac
%1{mm^{^{\prime }}}p_\rho q_\sigma \varepsilon _{\rho \sigma \nu \mu }
%\overline{\psi }_\nu ^{\lambda ^{^{\prime }}}(p')u_\lambda (p)\nonumber \\  
%+ie
%d_{l_1jk}^{lmn}\varepsilon _{jk^{^{\prime }}l_1^{^{\prime
%}}}Q_{kk^{^{\prime }}}\{D_3(q^2)-D_2(q^2)+\frac{\kappa m}{2m_p}[D_2(q^2)
%+D_2^{^{\prime }}(q^2)-2D_1(q^2)]\}\nonumber \\
%\frac 1{mm^{^{\prime }}}(p'_{\mu }q_\nu
%-p'\cdot q\delta _{\mu \nu })\overline{\psi }_\nu ^{\lambda ^{^{\prime
%}}}(p')\gamma _5u_\lambda (p)  \nonumber \\
%+ie
%d_{l_1jk}^{lmn}\varepsilon _{jk^{^{\prime }}l_1^{^{\prime
%}}}Q_{kk^{^{\prime }}}\{D_2^{^{\prime }}(q^2)-\frac{m^{^{\prime }}}mD_2(q^2)+%
%\frac{m_{-}}mD_3(q^2)\}  
%\times \overline{\psi }_\mu ^{\lambda ^{^{\prime }}}(p')\gamma
%_5u_\lambda (p).
%\end{eqnarray}
%m, m' are the rest mass of $\frac 12^{+}$ baryon and $\frac
%23^{+}$ baryon respectively,
%\[P_\mu =p_{\mu }+p'_{\mu }.\]

In Eq.(57), when \(m = m'\) is taken, the terms in $I_1$ and $I_2$, which are
proportional to $q_\mu $ vanish. Thus, when \(m = m'\), the current
matrix element of $\frac 12^{+}$ baryon automatically satisfies the
current conservation. In general cases in order to satisfy
the current conservation, the following
condition must be satisfied
\begin{equation}
D_2^{^{\prime }}(q^2)-\frac{m^{^{\prime }}}mD_2(q^2)+\frac{m_{-}}mD_3(q^2)=0.
\end{equation}
For ${1\over2}^+$ baryons the only matrix element with $m'\neq m$ is $\Sigma
^0->\Lambda $. For this process, we have
\begin{equation}
A_1 = A_2= \frac 1{2\sqrt{3}}.
\end{equation}
The condition(62) guarantees current
conservation for the EM process $\Sigma ^0->\Lambda + \gamma$.

{\bf Ward identity in this model}\\
It is well known that current conservation is satisfied in the electromagnetic interactions.
Why the condition (62) is required in this model ? This question must be answered. As mentioned 
that in this study a relativistic quark model is exploited. In this model SU(6) symmetry for the wave functions in the rest frame,
effective current, and the kernel of the transit matrix elements are assumed. In order to satisfy the current conservation 
the Ward identity must be satisfied after these assumptions. The Ward identity is the constraint on these assumptions.
In Ref. [15] the conditions for EM transit matrix elements of baryons to satisfy the Ward identity 
are revealed. As a matter of factor, Eq. (62) is the condition for satisfying the Ward identity.

The condition (62) not only guarantees the current conservation of $B \rightarrow B'$ 
it will be shown in this paper that the same condition (62) guarantees the current conservation for $p \rightarrow \Delta$
and the same condition (62)
prohibits the appearance of the second class current in weak transition of $B \rightarrow B'$ (see section 9). 
This condition (62) makes the vector form factors of $\Sigma^+ \rightarrow \Lambda + e^+ + \nu$ and $\Sigma^- \rightarrow \Lambda + e^- + \bar{\nu}$
to be proportional to $q^2$ (Tab. 3) and they are in agreement with the data (see subsection 9.4).
   
The electromagnetic form factors of $\frac 12^{+}$ baryons are
obtained from the current matrix elements (57)
\begin{eqnarray}
G_E(q^2) &=&-\frac 23(A_1+2A_2)(1+\frac{q^2}{4m^2})\{D_2(q^2)-\frac{\kappa
q^2}{4mm_p}D_3(q^2)\}  \nonumber \\
&&+\frac 13(A_2+5A_1)\{D_2(q^2)+\frac{q^2}{4m^2}[D_3(q^2)+\frac{\kappa m}{m_p%
}D_2(q^2)  \nonumber \\
&&-\frac{\kappa m}{m_p}D_1(q^2)-\kappa \frac m{m_p}(1+\frac{q^2}{4m^2}%
)D_3(q^2)]\}. \\
G_M(q^2) &=&\frac 13(A_2+5A_1)\{{m_p\over m}[ D_2(q^2)+\frac{q^2}{4m^2}D_3(q^2)] + \kappa
[D_1(q^2)  \nonumber \\
&&-D_2(q^2)+(1+\frac{q^2}{4m^2})D_3(q^2)]\},
\end{eqnarray}
where m is the mass of the baryon and ${e\over 2 m_p}$ is the unit of the $G_M(q^2)$. 
The expression of the magnetic moment of $\frac 12^{+}$ baryon is
obtained from Eq.(65)
\begin{eqnarray}
\mu  = G_M(0) = \frac 13(A_2+5A_1)\{\frac{m_p}m+\kappa
[D_1(0)+D_3(0)-D_2(0)]\}.
\end{eqnarray}
Eq. (66) shows that there are two parts in the magnetic moment of baryon: the term ${m_p\over m}$ 
is resulted in the recoil effect of the baryon and the second term is the contribution
of the anomalous magnetic moment of quark.
Eqs. (64,65) show that the two invariant functions $f_1,\;f_2$ and the function 
$M(x'_1,x'_2,x_1,x_2)$ all appear in the three unknown functions of $D_{1,2,3}(q^2)$.

%The matrix element of weak current $n\rightarrow p$ is obtained from Eq.()
%\begin{eqnarray}
%<p|j_\mu(0)|n> = {1\over3}\sqrt{{m\over E}}\bar{u}_{\lambda'}(p')\{5G^p_M(q^2)
%\gamma_\mu+5\lambda G_A(q^2)\gamma_\mu\gamma_5
%+{i\over m}P_\mu J(q^2)\}u_\lambda(p),
%\end{eqnarray}
%where
%\begin{eqnarray}
%P = p+p',\nonumber\\
%G_A(q^2) = F^p_1(q^2)&=&D_2(q^2)+\tau D_3(q^2),\\
%J(q^2)=D_2(q^2)+{5\over2}\kappa\{D_1(q^2)-D_2(q^2)+(1+{3\over5}\tau)D_3(q^2)\},
%\end{eqnarray}
%\[\tau=\frac{q^2}{4m^2}\].
%Eq.() is the prediction of this model. 

%Using the expressions of $G^p_E$ and $G^p_M$
%\begin{eqnarray}
%G^p_E(q^2)=F_1(q^2)-\kappa\tau F_2(q^2),\\
%G^p_M(q^2)=F_1(q^2)+\kappa F_2(q^2),
%\end{eqnarray}
%it is obtain
%\begin{equation}
%G_A(q^2)=\frac{1}{1+\tau}\{G^p_E(q^2)+\tau G^p_M(q^2)\}.
%\end{equation}
%On the other hand, the ratio $\mu_p G^p_E G^p_M$ is expressed as[]
%\begin{equation}
%\frac{\mu_p G^p_E(q^2)}{G^p_M(q^2)}=(1+\tau)\frac{\mu_p G_A(q^2)}{G^p_M(q^2)}-
%\mu_p\tau.
%\end{equation}
%We take the dipole form factor for $G^p_M(q^2)$
%\begin{equation}
%{1\over\mu_p}G^p_M(q^2)=\frac{1}{(1+{q^2\over0.71})^2}.
%\end{equation}

\section{Relationship Between $f_1(x_1,x_2,x_3)$ and $f_2(x_1,x_2,x_3)$}
The SU(6) symmetry in the rest frame leads to that in the wave functions of baryons there are two
functions $f_1(x_1,x_2,x_3)$ and $f_2(x_1,x_2,x_3)$ and 
$f_1(x_1,x_2,x_3)\neq f_2(x_1,x_2,x_3)$ is resulted in the effects of antiquarks.
In order to explore the effects of antiquarks in the form factors of baryons 
the possible relationship between the two invariant
functions $f_1(x_1,x_2,x_3)$ and $f_2(x_1,x_2,x_3)$
in the frame of center-of-mass is studied [5].

In the rest frame the $\Gamma^{{1\over2}}_{\alpha \beta ,\gamma }(p)_\lambda $ (34)
can be rewritten as
\begin{eqnarray}
\Gamma _{\alpha \beta ,\gamma }(x_1,x_2,x_3)_\lambda
&=&g_1(x_1,x_2,x_3)\{(1+\gamma _0)\gamma _5C\}_{\alpha \beta
}u_{\lambda ,\gamma }  \nonumber \\
&&+g_2(x_1,x_2,x_3)\{[(1-\gamma _0)\gamma _5C]_{\alpha \beta
}u_{\lambda ,\gamma }+2C_{\alpha \beta }(\gamma _5u_\lambda
)_\gamma \},  \nonumber \\ g_1(x_1,x_2,x_3) &=&\frac
12\{f_1(x_1,x_2,x_3)+f_2(x_1,x_2,x_3)\},  \nonumber \\
g_2(x_1,x_2,x_3) &=&\frac 12\{f_1(x_1,x_2,x_3)-f_2(x_1,x_2,x_3)\},
\end{eqnarray}
The wave functions of baryons (2) are the Bethe - Salpeter (BS) amplitudes and they satisfy  
corresponding BS equations. 
In order to make the wave functions to satisfy the SU(6) symmetry in the rest frame 
the BS equation must satisfy the SU(6) symmetry in the rest frame too.

The BS equation of a ${1\over2}^+$ baryon is written as
\begin{eqnarray}
&&(i\widehat{p}_1+M)_{\alpha \alpha ^{^{\prime }}}(i\widehat{p}_2+M)_{\beta
\beta ^{^{\prime }}}(i\widehat{p}_3+M)_{\gamma \gamma ^{^{\prime
}}}B_{\alpha ^{^{\prime }}\beta ^{^{\prime }}\gamma ^{^{\prime
}},ijk}^{\frac 12\lambda }(p_1,p_2,p_3)_l^m  \nonumber \\
&=&-i(i\widehat{p}_3+M)_{\gamma \gamma ^{^{\prime }}}\int U(q)B_{\alpha
\beta \gamma ^{^{\prime }},ijk}^{\frac 12\lambda }(p_1-q,p_2+q,p_3)_l^md^4q
\nonumber \\
&&-i(i\widehat{p}_1+M)_{\alpha \alpha ^{^{\prime }}}\int U(q)B_{\alpha
^{^{\prime }}\beta \gamma ,ijk}^{\frac 12\lambda }(p_1,p_2-q,p_3+q)_l^md^4q
\nonumber \\
&&-i(i\widehat{p}_2+M)_{\beta \beta ^{^{\prime }}}\int U(q)B_{\alpha
\beta ^{^{\prime }}\gamma ,ijk}^{\frac 12\lambda }(p_1+q,p_2,p_3-q)_l^md^4q
\nonumber \\
&&-\int V(q_1,q_2,q_3)\delta ^4(q_1+q_2+q_3)B_{\alpha \beta \gamma
,ijk}^{\frac 12\lambda }(p_1+q_1,p_2+q_2,p_3+q_3)  \nonumber \\
&&\times d^4q_1d^4q_2d^4q_3.
\end{eqnarray}
It is assumed that $U(q)$ of two bodies interactions
and $V(q_1,q_2,q_3)$ of three bodies interactions are independent of the momentum of
the baryon and they are scalars to keep possible SU(6) symmetry.
The $p_1, p_2, p_3$ of Eq. (68) satisfy
\begin{equation}
p_1 + p_2 + p_3=p,
\end{equation}
where p is the momentum of $\frac 12^{+}$ baryon.

It is not the intention to solve this equation and the Eq. (68)  is used to study 
the relationship between $f_{1,2}$ functions.
For this purpose in order to keep SU(6) symmetry the spatial part of the matrix $\hat{p}_i$ of Eq. (68)
must be ignored.
Substituting the wave function
of $\frac 12^{+}$ (33,34) into
Eq.(68), we obtain
\begin{eqnarray}
&&(M-\gamma _0p_{10})_{\alpha \alpha ^{^{\prime }}}(M-\gamma
_0p_{20})_{\beta \beta ^{^{\prime }}}(M-\gamma _0p_{30})_{\gamma \gamma
^{^{\prime }}}\Gamma^{{1\over2}} _{\alpha ^{^{\prime }}\beta ^{^{\prime }}\gamma
^{^{\prime }}}(p_1,p_2,p_3)_\lambda   \nonumber \\
&=&-i(M-\gamma _0p_{30})_{\gamma \gamma ^{^{\prime }}}\int U(q)\Gamma^{{1\over2}}
_{\alpha \beta ,\gamma ^{^{\prime }}}(p_1-q,p_2+q,p_3)_\lambda d^4q
\nonumber \\
&&-i(M-\gamma _0p_{10})_{\alpha \alpha ^{^{\prime }}}\int U(q)\Gamma^{{1\over2}}
_{\alpha ^{^{\prime }}\beta ,\gamma }(p_1,p_2-q,p_3+q)_\lambda d^4q
\nonumber \\
&&-i(M-\gamma _0p_{20})_{\beta \beta ^{^{\prime }}}\int U(q)\Gamma^{{1\over2}}
_{\alpha \beta ^{^{\prime }},\gamma }(p_1+q,p_2,p_3-q)_\lambda d^4q
\nonumber \\
&&-\int V(q_1,q_2,q_3)\delta ^4(q_1+q_2+q_3)\Gamma^{{1\over2}} _{\alpha \beta
,\gamma }(p_1+q_1,p_2+q_2,p_3+q_3)  \nonumber \\
&&\times d^4q_1d^4q_2d^4q_3.
\end{eqnarray}
where $\Gamma^{{1\over2}} _{\alpha \beta, \gamma }(p_1,p_2,p_3)_\lambda $ is
the expression of $\Gamma^{{1\over2}} _{\alpha \beta \gamma
}(x_1,x_2,x_3)_\lambda $ (34) in the momentum
representation. Calculations lead to
\begin{eqnarray}
&&(M-p_{10})(M-p_{20})(M-p_{30})g_1(p_1,p_2,p_3)  \nonumber \\
&=&-i\int U(q)\{(M-p_{30})g_1(p_1-q,p_2+q,p_3)  \nonumber \\
&&+(M-p_{10})g_1(p_1,p_2-q,p_3+q)  \nonumber \\
&&+(M-p_{20})g_1(p_1+q,p_2,p_3-q)\}d^4q  \nonumber \\
&&-\int V(q_1,q_2,q_3)\delta ^4(q_1+q_2+q_3)  \nonumber \\
&&\times g_1(p_1+q_1,p_2+q_2,p_3+q_3)d^4q_1d^4q_2d^4q_3,
\end{eqnarray}
\begin{eqnarray}
&&(M+p_{10})(M+p_{20})(M-p_{30})g_2(p_1,p_2,p_3)  \nonumber \\
&=&-i\int U(q)\{(M-p_{30})g_2(p_1-q,p_2+q,p_3)  \nonumber \\
&&+(M+p_{10})g_2(p_1,p_2-q,p_3+q)  \nonumber \\
&&+(M+p_{20})g_2(p_1+q,p_2,p_3-q)\}d^4q  \nonumber \\
&&-\int V(q_1,q_2,q_3)\delta ^4(q_1+q_2+q_3)  \nonumber \\
&&\times g_2(p_1+q_1,p_2+q_2,p_3+q_3)d^4q_1d^4q_2d^4q_3,
\end{eqnarray}
\begin{eqnarray}
&&(M+p_{10})(M-p_{20})(M+p_{30})g_2(p_1,p_2,p_3)  \nonumber \\
&=&-i\int U(q)\{(M+p_{30})g_2(p_1-q,p_2+q,p_3)  \nonumber \\
&&+(M+p_{10})g_2(p_1,p_2-q,p_3+q)  \nonumber \\
&&+(M-p_{20})g_2(p_1+q,p_2,p_3-q)\}d^4q  \nonumber \\
&&-\int V(q_1,q_2,q_3)\delta ^4(q_1+q_2+q_3)  \nonumber \\
&&\times g_2(p_1+q_1,p_2+q_2,p_3+q_3)d^4q_1d^4q_2d^4q_3,
\end{eqnarray}
\begin{eqnarray}
&&(M-p_{10})(M+p_{20})(M+p_{30})g_2(p_1,p_2,p_3)  \nonumber \\
&=&-i\int U(q)\{(M+p_{30})g_2(p_1-q,p_2+q,p_3)  \nonumber \\
&&+(M-p_{10})g_2(p_1,p_2-q,p_3+q)  \nonumber \\
&&+(M+p_{20})g_2(p_1+q,p_2,p_3-q)\}d^4q  \nonumber \\
&&-\int V(q_1,q_2,q_3)\delta ^4(q_1+q_2+q_3)  \nonumber \\
&&\times g_2(p_1+q_1,p_2+q_2,p_3+q_3)d^4q_1d^4q_2d^4q_3.
\end{eqnarray}
Since $V(q_1,q_2,q_3)$ are totally symmetric functions of $q_1,
q_2, q_3$ and $g_1(p_1,p_2,p_3)$ are
totally symmetric functions of $p_1, p_2, p_3$ too.
From Eqs.(73,74), we see that
$g_2(p_1,p_2,p_3)$ have following symmetries: (1) totally
symmetric in $p_1, p_2, p_3$. (2) since $U(q)$ and
$V(q_1,q_2,q_3)$ are independent of the momentum p
, the equation is invariant under the transformations
$p_{20}\rightarrow -p_{20}$, $p_{30}\rightarrow -p_{30}$;
$p_{10}\rightarrow -p_{10}$, $p_{30}\rightarrow -p_{30}$;
$p_{10}\rightarrow -p_{10}$, $p_{20}\rightarrow -p_{20}$. By
using the second symmetry of $g_2(p_1,p_2,p_3)$, Eq.(74)
becomes Eq.(73) under the transformation $p_{10}\rightarrow
-p_{10}$, $p_{20}\rightarrow -p_{20}$, thus $g_1(p_1,p_2,p_3)$ and
$g_2(p_1,p_2,p_3)$ satisfy the same equation. $g_1(p_1,p_2,p_3)$
is related to $g_2(p_1,p_2,p_3)$ by
\begin{equation}
g_1(p_1,p_2,p_3)=c g_2(p_1,p_2,p_3),
\end{equation}
where c is a constant. Eq. (75) leads to
\begin{equation}
f_2(x_1,x_2,x_3) = a f_1(x_1,x_2,x_3),
\end{equation}
where a is another constant.
Using Eq. (76), the three functions $D_{1,2,3}(q^2)$ of Eqs. (64,65) are reduced to one unknown function and a parameter a
\begin{equation}
D_1(q^2) = {1\over a} D_2(q^2),\;\;D_3(q^2) = a D_2(q^2).
\end{equation}
Obviously, the discussion above is not a proof of Eq. (76) and
it is an argument that the relationship (76) is possible. 

Substituting Eq. (76) into the condition of current conservation (62),
two solutions are obtained
\begin{equation}
a = 1
\end{equation}
or
\begin{equation}
a=\frac 1{1-\frac{m_0}m}
\end{equation}
$m_0$ is a parameter, m is the physical mass of the baryon.
As mentioned above $a = 1$ means $f_1 = f_2$ and there is no antiquark effects. In Refs. [5,6,7]
the effects of antiquarks are investigated. Therefore,
$a \neq 1$ (79) is taken in the studies [5,6,7]. It will be shown that the physical results of this model favor
the solution (79) or favor the existence of antiquark components in baryon.

\section{EM form factors of proton and neutron}
The EM form factors of proton and neutron are derived from Eqs.(64,65) [5]
\begin{eqnarray}
G_E^p(q^2) = D_2(q^2)+\frac{q^2}{4m_N^2}\{D_3(q^2)+\kappa
[D_2(q^2)-D_1(q^2)-(1+\frac{q^2}{4m_N^2})D_3(q^2)]\}, \\
G_M^p(q^2) = D_2(q^2)+\kappa [D_1(q^2)+D_3(q^2)-D_2(q^2)]+(1+\kappa )\frac{%
q^2}{4m_N^2}D_3(q^2),  \\
G_E^n(q^2) = -\frac
23\frac{q^2}{4m_N^2}\{D_3(q^2)-D_2(q^2)+\kappa
[D_2(q^2)-D_1(q^2)]\},  \\
G_M^n(q^2) =-\frac 23G _M^p(q^2).
\end{eqnarray}
The charge normalization of the proton 
\begin{equation}
G_E^p(0) = 1
\end{equation}
determines
\begin{equation}
D_2(0) = 1.
\end{equation} 
In Eqs. (80-83) the mass difference between proton and neutron is ignored. The isospin symmetry is reserved.

The relationship between the magnetic moments of the proton and neutron 
\begin{eqnarray}
\mu_p = G^p_M (0) = 1 + \kappa \{ D_1(0) + D_3(0) -1\}, \\
\mu_n = G^n_M (0) = -\frac23 \mu_p
\end{eqnarray}
is revealed from Eqs. (81,83).
Eq. (87) is the prediction of the SU(6) symmetry and it agrees with the data well. 
This result shows that the wave function (33,34) has, indeed, SU(6) symmetry.
Beside Eq. (87) this model predicts
the relation between the two magnetic form factors of proton and neutron (83).
Proton and neutron are doublet of isospin. The deviation of Eq. (87) from experimental value
is originated in the isospin symmetry breaking which is small. 

Eq. (82) shows $G^n_E(0) = 0$ and
the charge condition for neutral neutron is automatically satisfied.
Nonzero charge form factor of neutron (82) is predicted by this model.
\subsection{ Magnetic form factors of nucleon}
The radius of the magnetic form factors of proton and neutron can be defined
\begin{eqnarray}
G_M^p(q^2) = \mu_p \{1 - \frac16 (r^2)^p_M q^2\},\nonumber \\
G_M^n(q^2) = \mu_n \{1 - \frac16 (r^2)^n_M q^2\}.
\end{eqnarray}
Eq. (88) predicts that 
\begin{equation}
r^M_p = r^M_n.
\end{equation}
There are three reports of the value of the $r^M_p$: $0.777 \pm 0.013 \pm 0.010 fm$ [16], $0.876 \pm 0.010 \pm 0.016 fm$
[17], $0.854 \pm 0.005 fm$ [18], $0.867 \pm 0.009_{exp} \pm 0.018_{fit} fm$ [18]. One measurement of the $r^M_n$ has been reported as $0.80 \pm 0.10 fm$ [18].
The prediction (89) agrees with the data within the experimental errors.

From Eqs. (83,87) 
\begin{equation}
\frac{G^p_M}{\mu_p} = \frac{G^n_M}{\mu_n}
\end{equation}
is obtained. The measurements of ${1\over \mu_p} G^p_M/ G_D$ and ${1\over \mu_n}G^n_M \ G_D$ can be found in a review articles [19],
where 
\begin{equation}
G_D = \frac{1}{(1 + {q^2\over 0.71 \textrm{GeV}^2})^2}.
\end{equation}
The prediction (90) is not in contradiction with data within about few percent in the region of $q^2 < 10\; \textrm{GeV}^2$.
\subsection{Charge form factor of neutron and the ratio of the EM form factors of proton}
The electric form factor $G^n_E$ is not zero (82).
If $f_1 = f_2$ is taken, 
the Eqs. (59) become
\begin{eqnarray}
D_1(q^2) = D_3 (q^2) = D_2(q^2).
\end{eqnarray}
Substitute Eq. (92) into Eq. (82)
\begin{eqnarray}
G^n_E = 0
\end{eqnarray}
is obtained.
Therefore, in this model non-zero $G^n_E$ is resulted in the effects of the antiquark components in the wave function of the neutron.

On the other hand, if $f_1 = f_2$ is taken and using Eq. (92), the EM form factors of proton (80,81) become
\begin{eqnarray}
G_E^p(q^2) = D_2(q^2) (1 + \tau) (1- \kappa q^2),\\
G_M^p(q^2) = D_2(q^2) (1+ \kappa) ( 1+ \tau),
\end{eqnarray}
and
\begin{eqnarray}
\mu_p = 1 + \kappa,\\
R = \frac{\mu_p G_E^p(q^2)}{G_M^p(q^2)} = 1 - \kappa q^2 = 1-(\mu_p -1 ) q^2.
\end{eqnarray}
are obtained. 
The ratio  R (97) decreases very fast and
\begin{equation}
R \leq 0,\;\; when\;\;q^2 \geq \frac{4 m^2_p}{\mu_p -1} = 1.96\; \textrm{GeV}^2.
\end{equation}
This ratio (97) is in strong disagreement with current data [1,2,3].

Therefore, $f_1 = f_2$ is rejected by the ratio of the EM form factors of proton and 
the charge form factor of neutron.

Now we need to study the case of $f_1 \neq f_2$. Using Eqs.(77,85), it is found 
\begin{eqnarray}
\mu_p = 1 + \kappa \{1 + a -{1\over a}\},\\
\mu_n = -{2\over 3} \mu_p,\\
G_E^p(q^2) = D_2(q^2) \{1 +\tau ( a + 1 -\mu_p) - a \kappa \tau^2 \},\\ 
G_M^p(q^2) = \mu_p D_2(q^2) \{1 + {a\over \mu_p} (1 + \kappa) \tau \},\\
G_E^n(q^2) = \mu_n \tau D_2(q^2) {1\over \mu_p} (a-1) (1 + {\kappa\over a})\}.
\end{eqnarray}
The magnetic form factor of the neutron is expressed as Eq. (83). The geometric picture of the form factors of nucleon
in this model can be constructed as following. 
According to the SU(6) symmetry the hadrons of $\underline{56}$-plet
are in s-wave in the rest frame. In the rest frame the ${1\over2}^+$ and ${3\over2}^+$ 
hadrons have spherical shapes. The wave function of a moving baryon is obtained by boosting the wave function from the rest frame to moving frame
by Lorentz transformation. Because of Lorentz contraction the shapes of these baryons are 
changed to the shape of a football from a sphere. When the transfer momentum $q^2$ increases the length
of the ball along the direction of motion is shorten and the overlap of the two balls, 
one is at rest and the second is in motion, is decreasing. This effect makes the overlap function,
the form factor, decreases with $q^2$.

The $D_2(q^2)$, the $\kappa$, and the parameter a or $m_0$ are unknown and they are taken as three inputs in this model.
Inputting the $G_M^p(q^2)$,
the $D_2(q^2)$ can be determined
\begin{equation}
D_2(q^2) = {1\over \mu_p} G_M^p(q^2)\{1 + {a\over \mu_p} (1 + \kappa) \tau \}^{-1}.
\end{equation}
Inputting Eq. (104) into the charge form factor of the neutron, 
the $G_E^n(q^2)$ (103) has triple poles. Because of the factor $\frac{q^2}{4 m^2}$ the $G_E^n(q^2)$
increases with $q^2$ in small region of $q^2$ and because of the triple poles of the $D_2(q^2)$ it decreases 
with $q^2$ in the range of larger $q^2$.

The ration of \(R = \frac{\mu_p G_E^p(q^2)}{G_M^p(q^2)}$ is obtained
\begin{equation}
R = \frac{1 +\tau ( a + 1 -\mu_p) - a \kappa \tau^2}{1 + {a\over \mu_p} (1 + \kappa) \tau}.
\end{equation}
If taking $a = 1$ in Eq. (105), the ratio R will go back to Eq. (97). It is interesting to notice that besides the two factors in Eq. (105) which
decreases with $q^2$ if $(a + 1 - \mu_p) > 0 $ the factor $(a + 1 - \mu_p) \tau$ increases with $q^2$.  
In this article two versions of comparison with data are presented: \\
1) the original comparison presented
in Ref. [5];\\ 
2) comparison with new data.

{\bf Previous comparison with data [5]}\\
In Ref. [5] $\mu_p$ (99)
and $\mu_\Lambda$
\begin{equation}
\mu_\Lambda = -{1\over3}\{\frac{m_p}{m_\lambda} + \kappa (a_\Lambda + {1\over a_\Lambda} -1)\} = -0.64.
\end{equation}
are taken as two inputs and
the two parameters $\kappa$ and $ m_0$ (79) are
determined to be
\begin{equation}
\kappa = 0.481,\;\;m_0 = 0.778\; m_p,\;\;a = 4.51.
\end{equation} 
By using Eq. (107),
\begin{eqnarray}
G_E^p(q^2) = D_2(q^2)\{1+\tau (2.71-2.17\tau )\},   \\
G_M^p(q^2) = \mu _p D_2(q^2)\{1+2.39\tau \},   \\
G_E^n(q^2) = 1.39 \mu _n \tau D_2(q^2),\\
D_2(q^2) = {1\over \mu_p} G^p_M(q^2)(1+2.39\tau)^{-1} ,\\
D_2(q^2) = \frac 1{(1+\frac{q^2}{0.71})^2(1+2.39\tau )}
\end{eqnarray}
are obtained, where ${1\over \mu_p} G^p_M(q^2) = 1/(1+\frac{q^2}{0.71})^2$ is taken.
The ratio of the electric and
magnetic form factor of proton is obtained
\begin{equation}
R = \frac{\mu _p G_E^p(q^2)}{G_M^p(q^2)}=\frac{1+\tau (2.71-2.17\tau )}{%
1+2.39 \tau }.
\end{equation}
In the range of small $q^2$ $R = 1 + 0.091 q^2$ is revealed. Therefore,
the ratio (113) shows that in the range of small $q^2$ the ratio increases with $q^2$ slowly, then decreases with $q^2$.
This behavior is the prediction of this model.
Comparisons with data are shown in Fig.1 and 2.
\begin{figure}
\begin{center}
\includegraphics[width=7in, height=7in]{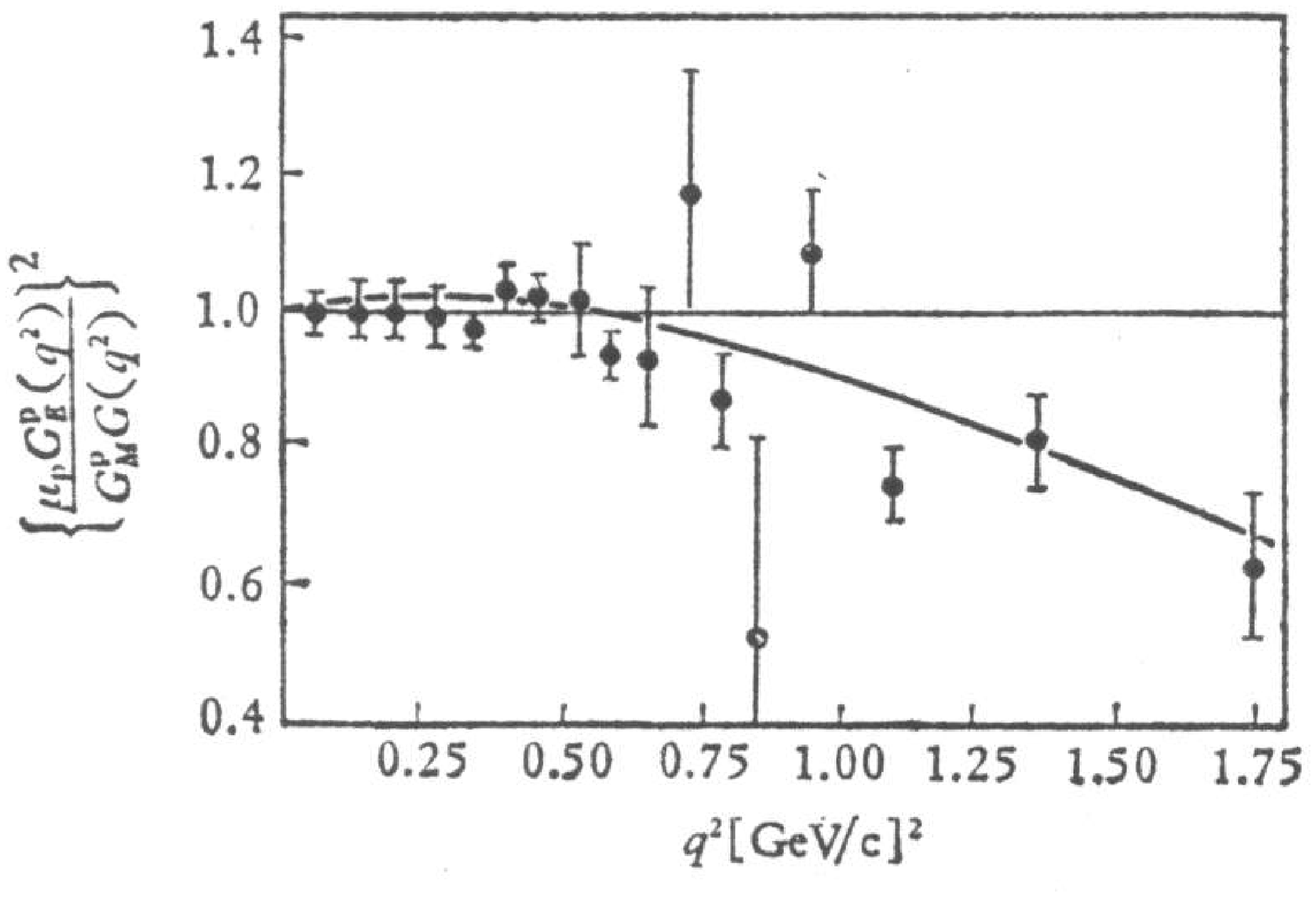}
FIG. 1
\end{center}
\end{figure}

\begin{figure}
\begin{center}
\includegraphics[width=7in, height=7in]{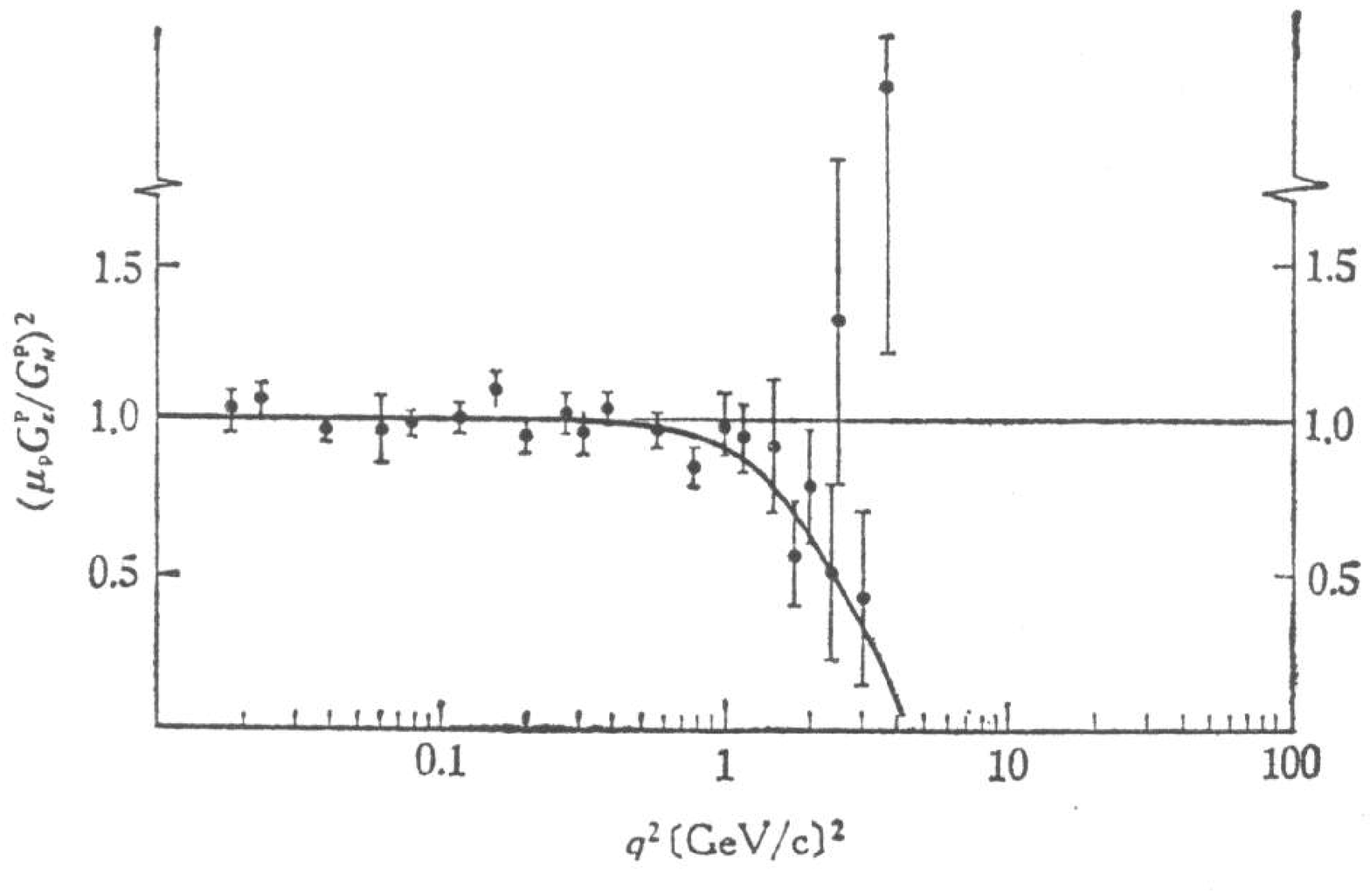}
FIG. 2
\end{center}
\end{figure}

The experimental data of Fig. 1 is from Ref.[20], and that for
Fig. 2 is from Ref.[21].
Fig. 1 and Fig.2 show that in the range of $0 < q^2 < 0.55\; \textrm{GeV}^2$ the ratio R is about one (a little bit 
greater than one). After $q^2 = 0.55\; \textrm{GeV}^2$ the ratio is decreasing with $q^2$. At  $q^2 = 5.45 \;\textrm{GeV}^2$ the ratio reaches zero and after this
value of $q^2$ the ratio is negative.

The expression of the electric form factor of
neutron is obtained
\begin{equation}
G_E^n(q^2)=1.39 \mu_n \tau (1+2.39\tau )^{-1}(1+\frac{q^2}{0.71})^{-2}.
\end{equation}
The expression of the $G^E_n(q^2)$ obtained in this model is just the Galster type [22]
\begin{equation}
G_E^n(q^2) = A \mu_n \tau G_D(q^2) (1 + B \tau)^{-1},
\end{equation}
where $G_D(q^2)$ is the expression of Eq. (91), the two parameters A and B are determined to be
\begin{eqnarray}
A = (a - 1)(1 + {\kappa\over a}) {1\over \mu_p},\\
B = 1 + ( 1 + \kappa){a\over \mu_p}.
\end{eqnarray}
Using $a = 4.51$ (107),
\begin{equation}
A = 1.39,\;\;\;B= 2.39 
\end{equation}
are determined.
Therefore, this model predicts a smaller negative charge form factor of neutron.
The slope of $G_E^n(q^2)$ at $q^2=0$ is
\begin{equation}
\frac{dG_E^n(q^2)}{dq^2}\mid _{q^2=0}=1.39\frac{\mu _n}{4m_N^2}%
=-0.73\;\textrm{GeV}^{-2}.
\end{equation}
The experimental data are
\begin{equation}
-0.579\pm 0.018^{[23]},\;-0.512\pm 0.049^{[24]},\;-0.495\pm 0.010^{[25]}.
\end{equation}
The data (120) are in the unit of $\textrm{GeV}^{-2}$.
Comparisons of the $G^n_E(q^2)$ (114) with the experimental data are shown in
Fig. 3 and Fig. 4.
\begin{figure}
\begin{center}
\includegraphics[width=7in, height=7in]{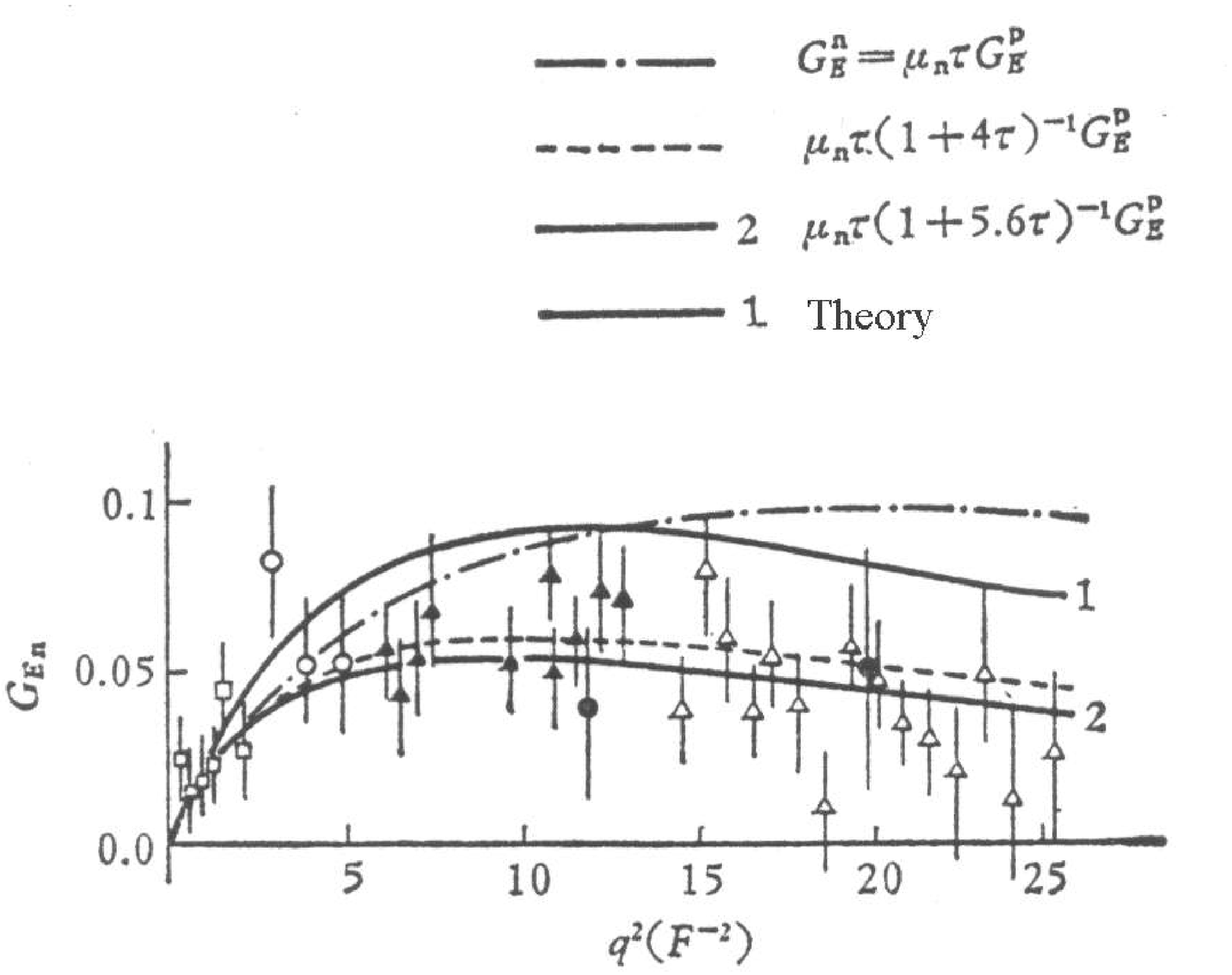}
FIG. 3
\end{center}
\end{figure}

\begin{figure}
\begin{center}
\includegraphics[width=7in, height=7in]{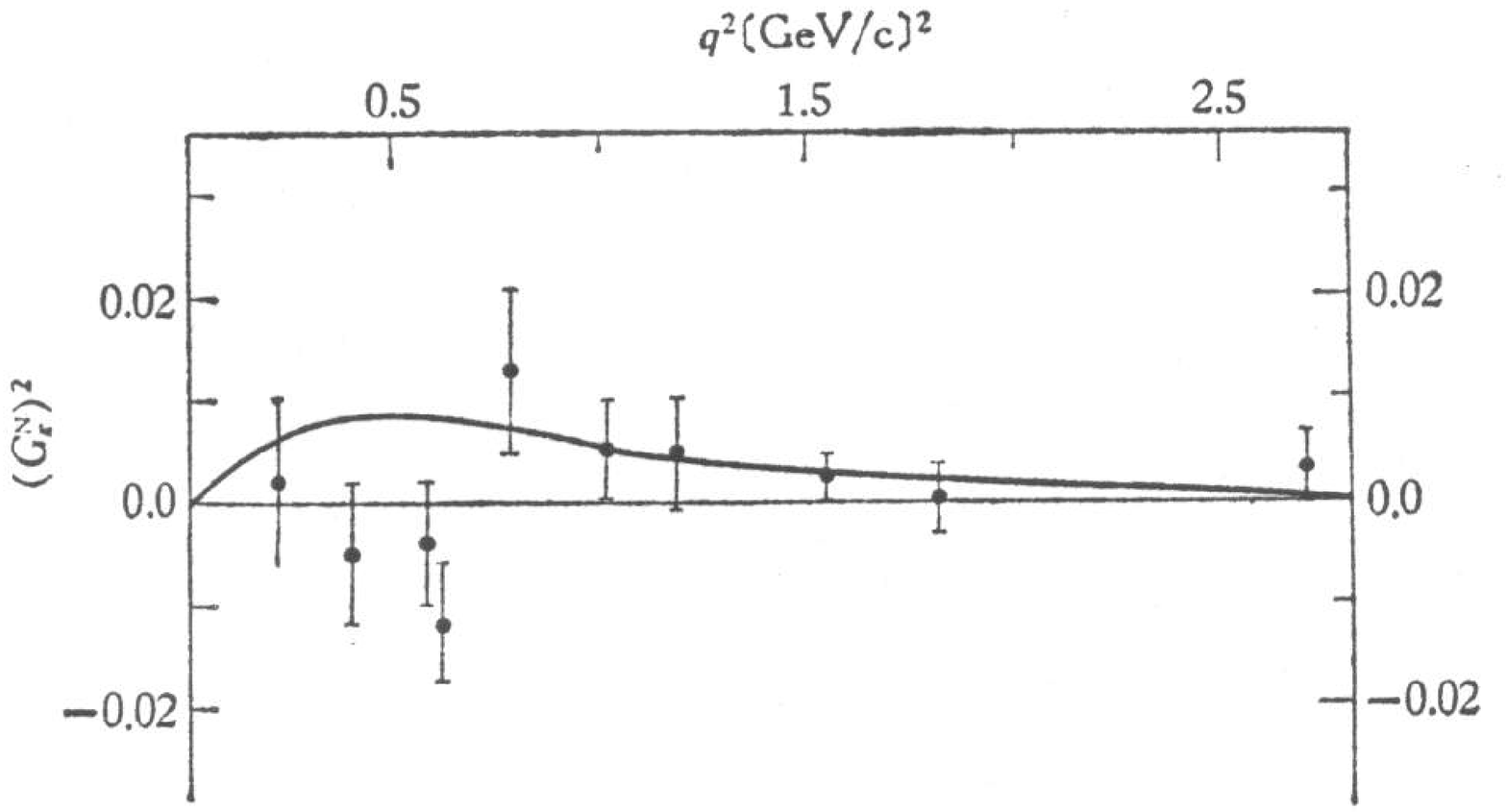}
FIG. 4
\end{center}
\end{figure}

The experimental data of Fig. 3 comes from Ref. [26] and that for
Fig. 4 comes from Ref. [21].
Comparing with the $G^p_E(q^2)$, this model [4,5] predicts a smaller charge form factor of neutron and $|G_E^n(q^2)| < 0.1$.
In the range of  $q^2 < 0.443\; \textrm{GeV}^2$  $G_E^n(q^2)$ increases with $q^2$ and after it decreases. 
However, the $G^n_E(q^2)$ predicted is greater than the experimental data.

{\bf New fit with recent date}\\

There are new data for $\frac{\mu_p G_E^p(q^2)}{G_M^p(q^2)}$ and $G_E^n(q^2)$. The comparison of the $\frac{\mu_p G_E^p(q^2)}{G_M^p(q^2)}$ (105)
with new data is shown in Fig. 5
\begin{figure}
\begin{center}
\includegraphics[width=7in, height=7in]{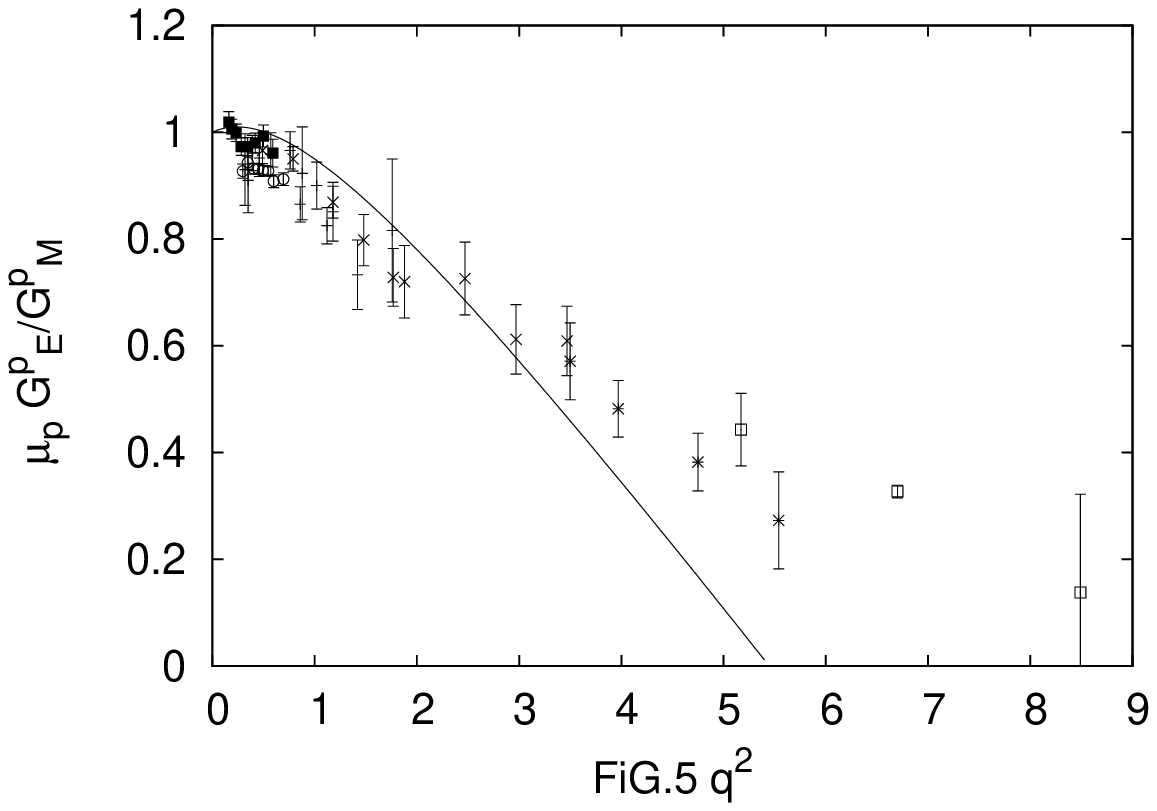}
%FIG. 5.
\end{center}
\end{figure}

The data of Fig. 5 are taken from Refs. [1,2,3, 27, 28, 29, 30]. 
The Fig. 5 shows that theoretical prediction agrees with new data as $q^2 < 5\; \textrm{GeV}^2$ and after this value of $q^2$ 
the prediction decreases faster than the data. The value of the parameter a can be increased to fit the data with larger $q^2$ better. However,
the larger a doesn't fit the data of the ratio in the range of smaller $q^2$. This is the limitation of this model. This model doesn't work well
in the region of larger $q^2$. For larger $q^2$ many new effects, like internal motion of quarks and perturbative gluons, could play roles in these physical quantities. 
On the other hand, the study shows that the assumption (76) works in the middle range of $q^2$ and it may not work for larger $q^2$. 

Fig. 6 shows the comparison of theoretical $G_E^n(q^2)$ with new data
\begin{figure}
\begin{center}
\includegraphics[width=7in, height=7in]{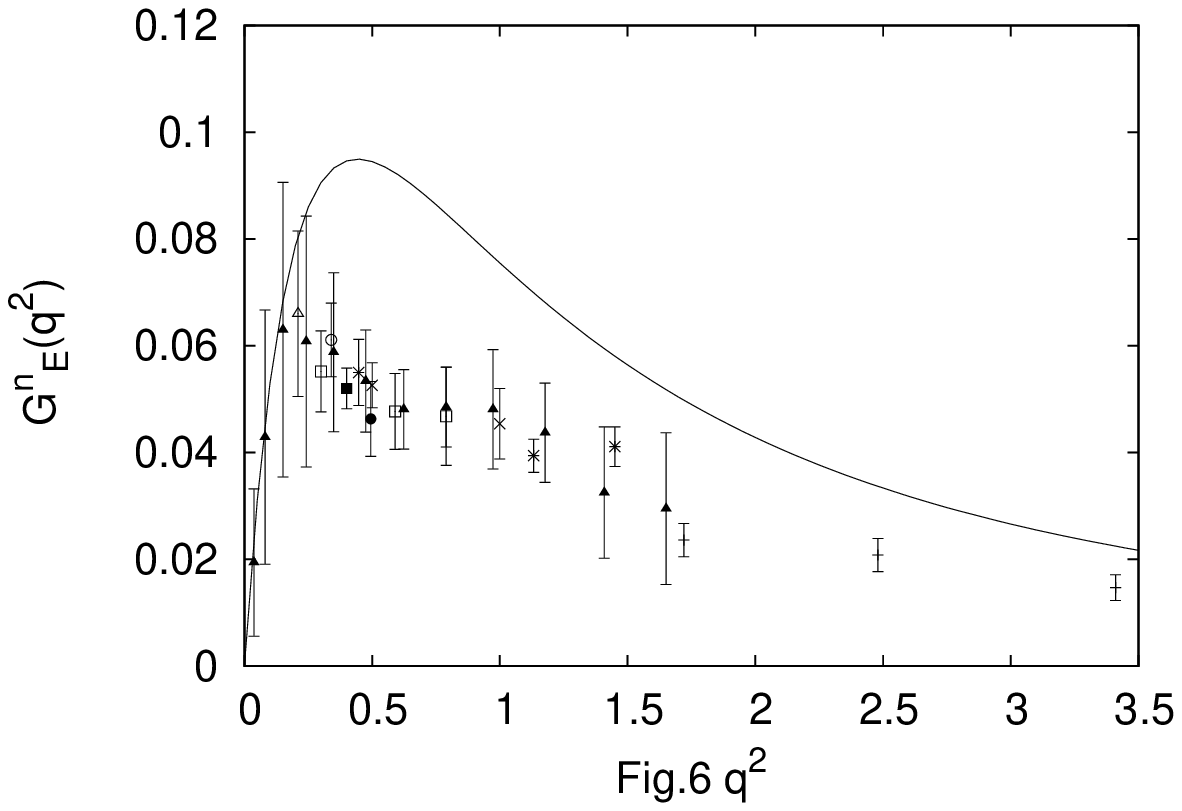}
%FIG. 6.
\end{center}
\end{figure}
The data of Fig. 6 are from Refs. [32,33,34,35,36,37,38,39,40,41,42,43,44,45].
This model predicts a nonzero and small $G_E^n(q^2)$, $|G_E^n(q^2)| < 0.1$ . 
Fig. 6 shows when $q^2 < 0.3\;\textrm{GeV}^2$ theory agrees with data. However, when $q^2 > 0.3\;\textrm{GeV}^2$ 
theoretical prediction is greater
than data by $20\%\; to\; 30\%$. Refs. [19,31] are review articles in which reviews of experiments and theoretical models 
can be found. 

%At $q^2=-4m_N^2$, there are
%\begin{eqnarray}
%G_E^p(-4m^2) &=&G_M^p(-4m^2)=0.18,  \nonumber \\
%G_E^n(-4m^2) &=&G_M^p(-4m^2)=-\frac 23G_E^p(-4m^2).
%\end{eqnarray}

\subsection{Magnetic moments of ${1\over2}^+$ baryons}

In this model the magnetic moments of the ${1\over2}^+$ baryons are obtained from Eqs. (66,77,85) as
\begin{equation}
\mu = {1\over 3} (A_2 + 5 A_1) \{ {m_p\over m} + \kappa (a + {1\over a} -1)\}.
\end{equation}
The first term of Eq. (121) is from the recoil of the whole baryon and the second term is from
the anomalous magnetic moment of quarks.
Using the parameters (107),
the magnetic moments of other
six baryons are determined (Table I) [5]
\begin{table}
\begin{center}
\caption{Magnetic moments}
\begin{tabular}{c|c|c|c|c|c|c|c|c} \hline
& $\mu _p$ & $\mu _n$ & $\mu _\Lambda $ & $\mu _{\Sigma ^{+}}$ &
$\mu _{\Sigma ^{0}}$ & $\mu _{\Sigma ^{-}}$ & $\mu _{\Xi ^0}$ &
$\mu _{\Xi ^-}$ \\ \hline theory & 2.793 &-1.862 & -0.64 & 1.74 &
0.58 & -0.57 & -0.97 & -0.51 \\& (input) & & (input) & & &&  & \\
\hline exp & 2.793 & -1.913 & -0.613 &2.458& &-1.16 &-1.250 & --0.6507 \\ & & $\pm
0.004$ && $\pm 0.010$ & &$\pm 0.025$ &$\pm 0.014 $& $\pm 0.0025$ \\ \hline
\end{tabular}
\end{center}
\end{table}
In this table the  experimental data of Ref. [46] are used. The magnetic moments of hyperons
have right signs. Hoverer, $\mu_{\Sigma^+}$ and $\mu_{\Sigma^-}$ are less than data by about 
$40\%$ to $50\%$ and $\mu_{\Xi^0}$ and $\mu_{\Xi^-}$ are less than data by about $28\%$ respectively.
 
The S-matrix element of $\Sigma ^0\rightarrow \Lambda
+\gamma $ is studied
\begin{eqnarray}
<\gamma \Lambda \mid S\mid \Sigma ^0>&=&-ie(2\pi )^4\delta
(p_i-p_f-q)\frac{e_\mu ^\lambda }{\sqrt{2\omega }}(\frac{m_\Lambda
}{E_\Lambda })^{\frac 12}\mu _{\Sigma _0\Lambda }  \nonumber \\
&&\times \frac \kappa {2m_p}\overline{u}_{\lambda ^{^{\prime
}}}(p_f)q_\nu \sigma _{\nu \mu }u_\lambda (p_i),
\end{eqnarray}
\begin{eqnarray}
 \mu _{\Sigma _0\Lambda } &=&\frac
1{2\sqrt{3}}D_3(0)\{\frac 2{a_\Lambda a_{\Sigma ^0}}-\frac
1{a_\Lambda }(1-\frac{m_p}{\kappa m_\Sigma })-\frac 1{a_{\Sigma
^0}}(1-\frac{m_p}{\kappa m_\lambda })+\frac{m_{+}^2}{2m_\Lambda
m_\Sigma }\}.
\end{eqnarray}
The dependencies of $D_1(0)$, $D_2(0)$,
$D_1^{^{\prime }}(0)$ and $D_1^{^{\prime }}(0)$ on the mass of
initial and final baryon need to be found. From Eq. (59,76)
\begin{equation}
\frac{D_2(0)}{D_2^{^{\prime }}(0)}=\frac a{a^{^{\prime }}},
\end{equation}
is obtained,
where 
\begin{equation}
a = \frac{1}{1-{m_0\over m_{\Sigma^0}}},\;\;\;a' = \frac{1}{1-{m_0\over m_{\Lambda}}}.
\end{equation}
%Eq. (59) shows when $m\longleftrightarrow m^{^{\prime}}$
On the other hand, Eq.(59) shows if $m  \longleftrightarrow  m'$ is taken, we have

\begin{eqnarray}
D_2(0)& \longleftrightarrow &D_2^{^{\prime }}(0)
\end{eqnarray}
and
\begin{eqnarray}
 D_2(0) = D_2^{^{\prime }}(0) = 1
\end{eqnarray}
when $m = m^{\prime}$. 
The general expressions of $D_2(0)$, $D_2^{^{\prime }}(0)$ which satisfy
Eqs.(124,126) can be written as
\begin{eqnarray}
D_2(0) &=&(\frac a{a^{^{\prime }}})^{\frac 12}f(m,m^{^{\prime }}),  \nonumber
\\
D_2^{^{\prime }}(0) &=&(\frac{a^{^{\prime }}}a)^{\frac 12}f(m,m^{^{\prime
}})
\end{eqnarray}
where $f(m,m^{^{\prime }})$ is a symmetric function of m, $m'$ and
\begin{equation}
f(m,m)=1.
\end{equation}
When $m\neq m^{^{\prime }}$, the deviation of $f(m,m^{^{\prime
}})$ from 1 is proportional to $(m-m^{^{\prime }})^2$. According
to Ref. [4], $f(m,m^{^{\prime }})$ is related to the effect of  Lorentz contraction. 
Possible expression is
\begin{equation}
f(m,m^{^{\prime }})=\frac{4mm^{^{\prime }}}{(m+m^{^{\prime }})^2}.
\end{equation}
For $\Sigma ^0\rightarrow \Lambda +\gamma $, the deviation of
$f(m,m^{^{\prime }})$  from 1 is only $0.1\%$. Therefore, for this decay the effect of $f(m,m^{^{\prime }}) = 1$
is taken.

\begin{eqnarray}
D_3(0) &=&\sqrt{aa^{^{\prime }}},  \nonumber \\
D_1(0) &=&\frac 1{\sqrt{aa^{^{\prime }}}}.
\end{eqnarray}
are determined too.
The magnetic moment of $\Sigma\rightarrow\Lambda$
and the decay rate are computed to be
\begin{equation}
\mu _{\Sigma ^0\Lambda }=1.053
\end{equation}
\begin{eqnarray}
\Gamma  &=&\frac \alpha 8\mu _{\Sigma ^0\Lambda }^2\frac{m_\Sigma ^3}{m_p^2}%
(1-\frac{m_\Lambda ^2}{m_\Sigma ^2})^3=3.79\times 10^{-3}MeV,  \nonumber \\
\tau  &=&\frac 1\Gamma =1.74\times 10^{-19}sec.
\end{eqnarray}
The current experimental transit magnetic moment of $\Sigma^0 \rightarrow \Lambda + \gamma$ is [46] is
\[|\mu_{\Sigma \Lambda}| = 1.61 \pm 0.08 .\]
Theoretical value of this transit magnetic moment is less than data by about $50\%$.

The SU(6) prediction of the ratio of the magnetic moments of proton and neutron is reproduced in this model and it agrees with data well. 
Inputting the $\mu_p$ and $\mu_{\Lambda}$ the magnetic moments of $\Sigma$'s, $\Xi$'s, and $\Sigma^0\rightarrow \Lambda$ are predicted.
They have right signs, but they are smaller than data.  This problem is resulted in the treatment of the flavor SU(3) symmetry breaking
in this model. In the effective EM current (48) the parameter $\kappa$ is taken to be same for u-, d-, and s-quark.
It is possible $\kappa_s$ is different from $\kappa_{u,d}$. In Eq. (49) the scalar function $M(x'_1, x'_2, x_1, x_2)$ is assumed to be 
the same for all three quarks. How to treat the the flavor SU(3) symmetry breaking is the key for the improvement of the magnetic moments of hyperons.
The study is beyond the scope of this paper. 

{\bf More discussion about the magnetic moments of baryons}\\
It is known that the effect of isospin symmetry breaking is small. The agreement between Eq. (87) and data is a good example.
We can check more relationships between the magnetic moments within the same
isospin multiple.
 
This model predicts that both the magnetic moments of $\Xi^{-}$ and $\Xi^0$ (121) are negative and
\begin{equation}
\mu_{\Xi^0} = 2 \mu_{\Xi^-}
\end{equation}
which agrees with data well. It predicts that $\mu_{\Sigma^+}$ is positive and $\mu_{\Sigma^-}$ is negative. These predictions agree with data.
However, under isospin symmetry 
\begin{equation}
\mu_{\Sigma^-} = -{1\over3} \mu_{\Sigma^+}
\end{equation}
is predicted. However,
this prediction (135) is different from data by about $30\%$ which is much larger than the effect of isospin breaking. In the limit of SU(3) symmetry 
this model predicts (121) 
\begin{equation}
\mu_{\Sigma^+} = \mu_p. 
\end{equation}
The difference between the prediction (136) and the data [46] by an reasonable $12\%$. 
%There is no reasonable theoretical explanation why the effect of isospin symmetry breaking is that large for $\Sigma^+$ and $\Sigma^-$.
The key point is how to understand the large $\mu_{\Sigma^-}$.
On the other hand, in the limit of isospin symmetry this model predicts 
\begin{equation}
\mu_{\Sigma^0} = {1\over 3} \mu_{\Sigma^+}.
\end{equation}
So far, there is no data to test Eq. (137).
%The magnetic moment of $\Sigma^+$ 
%is expressed as ()
%\begin{equation}
%\mu_{\Sigma^+} = {m_p\over m_{\Sigma^+}} + \kappa\{a_{\Sigma^+} + {1\over a_{\Sigma^+}} - 1\}
%\end{equation}
%The term ${m_p\over m_{\Sigma^+}}$ is resulted in the effect of recoil of the $\Sigma^+$ which violates the SU(3) symmetry and
%\begin{equation}
%\frac{\mu_p - 1}{\mu_{\Sigma^+} - {m_p\over m_{\Sigma^+}}} = {1.793\over 1.669} = 1.07.
%\end{equation}
Finally, it is interesting to point out that the effect of the recoil of the whole baryon plays an important rule in the magnetic moments of baryons. 
As shown above, this model works well on EM form factors of nucleon (two flavors). However,
when s-quark is involved theoretical results of the magnetic moments of hyperons have more than $30\%$ deviation from the data.
How to improve the involvement of the effects of SU(3) symmetry breaking is an important task for this model. 

\subsection{Electric and magnetic radii of nucleon}

From Eqs. (101,102) the difference of the charge and the magnetic radii of the proton is predicted as
\begin{equation}
(r^{p}_M)^2 - (r^p_E)^2 = \frac{3}{2 m^2_p}\{ a + 1 -\mu_p - {a\over \mu_p}(1+\kappa)\}.
\end{equation}
Using the values of the parameter a and $\kappa$ (107),
\begin{equation}
(r^p_M)^2 - (r^p_E)^2 = 0.0215 fm^2
\end{equation}
is obtained. This value is very sensitive to the value of the parameter a. 
This model predicts that both radii are pretty close to each other.
Eq. (139) shows that this model predicts that the $r^p_M$ is little bit greater than the $r^p_E$. 
The experimental data of the 
$r^p_M$ can be found in Refs. [16,17,18]. In review article [47] $r^p_E = 0.8775 \pm 0.0051$ fm is listed. Many more data
on $r^p_M$ can be found in Ref. [46]. 

\subsection{Electromagnetic form factors of hyperons}
It is useful to list the EM form factors of hyperons.
Both the charge and magnetic form factors of the $\Sigma^+$ are obtained from Eqs. (64,65)
respectively
\begin{eqnarray}
G^{\Sigma^+}_E (q^2) = D_2(q^2) + \tau \{D_3(q^2) + \kappa {m_{\Sigma^+}\over m_p} [ D_2(q^2) - D_1(q^2) - (1+\tau) D_3(q^2)]\},\\
G^{\Sigma^+}_M (q^2) = {m_p\over m_{\Sigma^+}} \{D_2(q^2) + \tau D_3(q^2)\} + \kappa \{ D_1(q^2) -  D_2(q^2) + (1 + \tau) D_3(q^2)\}.
\end{eqnarray}
The functions $D_{1,2,3}$ depend on the mass of the $\Sigma^+$. \(D_2(0) = 1\) leads to \(G^{\Sigma^+}_E (0) = 1\).
%It is similar to proton that the charge and magnetic radii of the $\Sigma^+$ are close to each other \(r^2_M \sim r^2_E\).

The charge and the magnetic forms of the $\Sigma^-$ are derived from Eqs. (64,65) as
\begin{eqnarray}
G^{\Sigma^-}_E(q^2) = -D_2(q^2) -{1\over3} \tau \{2 D_2(q^2) + D_3(q^2) + \kappa {m_{\Sigma^-}\over m_p}[ D_2(q^2)
- D_1(q^2) -3 (1+\tau)D_3(q^2)]\} ,\\
G^{\Sigma^-}_M(q^2) = -{1\over3}\{{m_p\over m_{\Sigma^-}} D_2(q^2) + \tau D_3(q^2) + \kappa [D_1(q^2) -  D_2(q^2) + (1+\tau) D_3(q^2)]\}.
\end{eqnarray}
Eq. (142) shows that the charge of the $\Sigma^-$ is -1.

%The study of this section is new. Only the charge radius of the $\Sigma^-$ is experimentally determined
%\begin{equation}
%r_{\Sigma^-} = 0.78 \pm 0.10 \textrm{fm}.
%\end{equation}
 
From Eqs. (64,65) the EM form factors of $\Sigma^0$ and $\Lambda$ are found 
\begin{eqnarray}
G^{\Sigma^0}_E(q^2) = {1\over3}\tau  \{ D_3(q^2) -  D_2(q^2) + \kappa {m_{\Sigma^0}\over m_p} [D_2(q^2) -  D_1(q^2)]\},\\
G^{\Sigma^0}_M(q^2) = {1\over3}\{ {m_p\over m_{\Sigma^0}} (D_2(q^2) + \tau D_3(q^2)) + \kappa [D_1(q^2) - D_2(q^2) + (1+\tau) D_3(q^2)]\},\\
G^{\Lambda}_E(q^2) = {1\over3}\tau  \{ D_2(q^2) -  D_3(q^2) + \kappa {m_{\Sigma^0}\over m_p} [D_2(q^2) -  D_1(q^2)]\},\\
G^{\Lambda}_M(q^2) = - {1\over3}\{ {m_p\over m_{\Lambda}} (D_2(q^2) + \tau D_3(q^2)) + \kappa [D_1(q^2) - D_2(q^2) + (1+\tau) D_3(q^2)]\}.
\end{eqnarray}
Eqs. (144,146) show that $G^{\Sigma^0}_E(0) = 0,\;\;G^{\Lambda}_E(0) = 0$. If ignoring the mass difference between the $\Sigma^0$ and the $\Lambda$,
this model presents
\begin{eqnarray}
G^{\Sigma^0}_M(q^2) = - G^{\Lambda}_M(q^2),\nonumber\\
\mu_{\Sigma^0} = -\mu_{\Lambda}.
\end{eqnarray}
The EM form factors of the $\Xi^0$ and the $\Xi^-$ are obtained 
\begin{eqnarray}
G^{\Xi^0}_E(q^2) = {2\over3}\tau  \{ D_2(q^2) -  D_3(q^2) + \kappa {m_{\Xi^0}\over m_p} [D_1(q^2) -  D_2(q^2)]\},\\
G^{\Xi^0}_M(q^2) = - {2\over3}\{ {m_p\over m_{\Xi^0}} (D_2(q^2) + \tau D_3(q^2))+ \kappa [D_1(q^2) - D_2(q^2) + (1+\tau) D_3(q^2)]\},\\
G^{\Xi^-}_E(q^2) = - D_2(q^2)- {1\over3}\tau \{{2 D_2(q^2) D_3(q^2) + \kappa {m_{\Xi^-}\over m_p}[D_1(q^2)-D_2(q^2)-(1+\tau}D_3(q^2)]\},\\
G^{\Xi^-}_M(q^2) = - {1\over3}\{ {m_p\over m_{\Xi^0}} (D_2(q^2) + \tau D_3(q^2))+ \kappa [D_1(q^2) - D_2(q^2) + (1+\tau) D_3(q^2)]\}.
\end{eqnarray}
Eqs. (149, 151) show that $G^{\Xi^0}_E(0) = 0$ and $G^{\Xi^-}_E(0) = -1$.
Ignoring the mass difference between $\Xi^0$ and $\Xi^-$, this model predicts
\begin{equation}
G^{\Xi^-}_M(q^2) = {1\over2} G^{\Xi^0}_M(q^2).
\end{equation} 

\section{EM transition of $p \rightarrow \Delta (1236)$}
The electromagnetic transition of $p \rightarrow \Delta (1236)$ has been studied by this model [5] and there is no any new parameter. 
Using the wave functions of proton and $\Delta$ (33,34) and the effective current (48), 
the matrix elements of EM currents of $p \rightarrow \Delta (1236)$ are obtained 
\begin{eqnarray}
<B_{\lambda ^{^{\prime }}}^{\frac 32}(p_f)^{lmn}\mid J_\mu (0)\mid
B_\lambda ^{\frac 12}(p_i)_{l_1}^{l_1^{^{\prime }}}>
= ied_{l_1jk}^{lmn}%
\varepsilon _{jk^{^{\prime }}l_1^{^{\prime }}}Q_{kk^{^{\prime
}}}\{2D_2(q^2)+\kappa [\frac{m_{+}}{m_p}D_3(q^2)+2\frac m{m_p}D_1(q^2)\nonumber \\
-\frac m{m_p}D_2(q^2)-\frac m{m_p}D_2^{^{\prime }}(q^2)]\}
\frac
1{mm^{^{\prime }}}p_\rho q_\sigma \varepsilon _{\rho \sigma \nu \mu }
\overline{\psi }_\nu ^{\lambda ^{^{\prime }}}(p')u_\lambda (p)\nonumber \\
+ie
d_{l_1jk}^{lmn}\varepsilon _{jk^{^{\prime }}l_1^{^{\prime
}}}Q_{kk^{^{\prime }}}\{D_3(q^2)-D_2(q^2)+\frac{\kappa m}{2m_p}[D_2(q^2)
+D_2^{^{\prime }}(q^2)-2D_1(q^2)]\}\nonumber \\
\frac 1{mm^{^{\prime }}}(p'_{\mu }q_\nu
-p'\cdot q\delta _{\mu \nu })\overline{\psi }_\nu ^{\lambda ^{^{\prime
}}}(p')\gamma _5u_\lambda (p)  \nonumber \\
+ie
d_{l_1jk}^{lmn}\varepsilon _{jk^{^{\prime }}l_1^{^{\prime
}}}Q_{kk^{^{\prime }}}\{D_2^{^{\prime }}(q^2)-\frac{m^{^{\prime }}}mD_2(q^2)+%
\frac{m_{-}}mD_3(q^2)\}
 \overline{\psi }_\mu ^{\lambda ^{^{\prime }}}(p')\gamma
_5u_\lambda (p).
\end{eqnarray}
m, $m'$ are the mass of $\frac 12^{+}$ baryon and $\frac
23^{+}$ baryon respectively and $m_- = m' - m$,
\[P_\mu =p_{\mu }+p'_{\mu }.\]

It is interesting to notice that the last term of Eq. (154) violates the current 
conservation. However, this term is proportional to 
\[D_2^{^{\prime }}(q^2)-\frac{m^{^{\prime }}}mD_2(q^2)+%
\frac{m_{-}}mD_3(q^2).\]
According to Eq. (62) it is equal to zero, the current conservation of 
$p \rightarrow \Delta (1236)$ (154) is guaranteed by the same condition (62) 
which guarantees the current conservation for the EM transition matrix elements of $B \rightarrow B$.

\subsection{Photoproduction of $\Delta$ resonance}
The Photoproduction of $\Delta$ resonance
is related to the transit matrix element (154)
\begin{eqnarray}
&<&\Delta _{\lambda ^{^{\prime }}}^{+}(p_f)\mid J_\mu (0)\mid p_\lambda
(p_i)>=-\frac{ie}{4\sqrt{3}}
A\frac 1{mm^{^{\prime }}}D_3(q^2)p_{i \rho} q_\sigma   \nonumber \\
&&\times \varepsilon _{\rho \sigma \nu \mu }\overline{\psi }_\nu ^{\lambda
^{^{\prime }}}(p_f)u_\lambda (p_i)-\frac{ie}{\sqrt{3}}
\frac B{mm^{^{\prime }}}D_3(q^2)  \nonumber \\
&&\times (p_{f\mu }q_\nu -p_f\cdot q\delta _{\mu \nu })\overline{\psi }_\nu
^{\lambda ^{^{\prime }}}(p_f)\gamma _5u_\lambda (p_i),
\end{eqnarray}
where
\begin{eqnarray}
A &=&\frac 2{a^{^{\prime }}}+\kappa \{1+\frac{m^{^{\prime }}}{m_p}+\frac
2{aa^{^{\prime }}}-\frac 1a-\frac 1{a^{^{\prime }}}\},  \nonumber \\
B &=&1-\frac 1{a^{^{\prime }}}+\frac \kappa 2\{\frac 1a+\frac 1{a^{^{\prime
}}}-\frac 2{aa^{^{\prime }}}\},
\end{eqnarray}
$m'$ is the mass of the $\Delta$, and
\begin{eqnarray}
a = \frac{1}{1-{m_0\over m_p}},\;\;\;a' = \frac{1}{1-{m_0\over m_{\Delta}}}\\
A=1.717,\;\;\;B=0.699.
\end{eqnarray}
It is interesting to notice that the condition (62) has been taken into account in Eq. (155) and the current conservation 
is satisfied.
Eq. (156) shows that $B = 0$ is obtained when taking $a = a'= 1$.
Therefore, the term B in Eq. (155) is directly related to the contribution of antiquark components. 
For the $\Delta^+ \rightarrow p + \gamma$ there are both 
$M1_{+}$ and $E1+$ amplitudes. 
%\subsection{Photoproduction of the $\Delta$ resonance}
The S matrix element of $\gamma p\rightarrow \Delta^+ \rightarrow \pi N$ is written as
\begin{eqnarray}
<\pi N\mid S\mid \gamma p>&=&-i(2\pi )^4\delta (p_\gamma
+p_i-p_\pi -p_N)\sum_{\lambda ^{^{\prime }}}<\pi N\mid T\mid
\Delta _{\lambda ^{^{\prime }}}^{+}(p_f)>  \nonumber \\ &&\times
<\Delta _{\lambda ^{^{\prime }}}^{+}(p_f)\mid T\mid \gamma
p>\frac{E_\Delta }{m_\Delta }\frac 1{W-m_\Delta +\frac i2\Gamma
(W)},
\end{eqnarray}
where W is the mass of the final state, $\Gamma (W)$ is the total
width of the strong decay of $\Delta (1236)$. The calculation is
done in the rest frame of $\Delta (1236)$. $<\pi N\mid T\mid
\Delta _{\lambda ^{^{\prime }}}^{+}(p_f)>$ is the amplitude of the
strong decay of $\Delta (1236)$
\begin{eqnarray}
<\pi N\mid T\mid \Delta _{\lambda ^{^{\prime }}}^{+}(p_f)>=
g(W)\frac{p_{\pi \mu }}{m_N}\overline{u}(p_N)\psi
_\mu ^{\lambda ^{^{\prime }}}.
\end{eqnarray}
The electric transition amplitude in Eq.(159) is expressed as
\begin{eqnarray}
<\Delta _{\lambda ^{^{\prime }}}\mid T\mid \gamma p>=-
e_\mu <\Delta _{\lambda ^{^{\prime }}}\mid J_\mu (0)\mid p>.
\end{eqnarray}
By using following equation
\begin{eqnarray}
\sum_{\lambda ^{^{\prime }}}\psi _\mu ^{\lambda ^{^{\prime }}}\overline{\psi
}_{\mu ^{^{\prime }}}^{\lambda ^{^{\prime }}} &=&\frac 13(1+\gamma
_0)\{\delta _{\mu \mu ^{^{\prime }}}+\frac 12\gamma _5\gamma _j\varepsilon
_{j\mu \mu ^{^{\prime }}}\}
\end{eqnarray}
where \(j,\mu ,\mu ^{^{\prime }} = 1, 2, 3 \).
Using Eq.(155), it is obtained
\begin{eqnarray}
\sum_{\lambda ^{^{\prime }}}\overline{u}_\gamma (p_N)\psi _\mu ^{\lambda
^{^{\prime }}} &<&\Delta _{\lambda ^{^{\prime }}}\mid J_\nu (0)\mid
p_\lambda >p_{\pi \mu }e_\nu   \nonumber \\
&=&\frac{eD_3(0)}{24\sqrt{3}m_N^2m_\Delta }\{\frac{m_N(m_N+E_N)}{E_i(m_N+E_i)%
}\}^{\frac 12}\overline{u}_\gamma \{[A(m_N+m_\Delta )^2+B(m_\Delta
^2-m_N^2)] \nonumber \\ &&\times [2{\bf k\cdot (e\cdot p_\pi}
)+i{\bf \sigma \cdot ep_\pi \cdot k}-i{\bf \sigma \cdot kp_\pi
\cdot e}] \nonumber \\ &&-3iB(m_\Delta ^2-m_N^2)({\bf \sigma \cdot
ep_\pi \cdot k}+{\bf \sigma \cdot kp_\pi \cdot e})\}u_\lambda ,
\end{eqnarray}
where $E_i$ is the energy of the initial proton, k is the energy
of the photon, and $D_3(0)$ is given by Eq. (131). The amplitudes of the magnetic and electric transitions
are obtained by comparing with the photo production amplitudes in Ref. [48]
\begin{eqnarray}
M1+ &=&\frac{eD_3(0)}{96\sqrt{3}\pi m_N^2m_\Delta }\{\frac{m_N+E_N}{m_\Delta
E_i(m+E_i)}\}^{\frac 12}\frac{g(W)p_\pi k}{W-m_\Delta +\frac i2\Gamma (W)}
\nonumber \\
&&\times \{A(m_N+m_\Delta )^2+B(m_\Delta^2-m_N^2)\}, \\
E1+ &=&-\frac{eD_3(0)}{96\sqrt{3}\pi m_N^2m_\Delta }\{\frac{m_N+E_N}{%
m_\Delta E_i(m+E_i)}\}^{\frac 12}\frac{g(W)p_\pi k}{W-m_\Delta +\frac
i2\Gamma (W)}  \nonumber \\
&&\times B(m_\Delta ^2-m_N^2),
\end{eqnarray}
where W is the total energy of the photon and the proton in the frame of center of mass and
g(W) is the amplitude of $\Delta\rightarrow \pi + N$,
$p_\pi$ and k are the momenta of the pion and the photon respectively. This model predicts a nonzero amplitude of electric quadrupole $E1+$.
Inputting $\Gamma = 0.12 \textrm{GeV}$ at $W = m_\Delta$,  $g = 1.66$ is determined
The $M1+$ amplitude at $W = m_\Delta$ is calculated to be 
\[M1+ = 47.3\times10^{-3}/m_{\pi^+}.\]
In Ref. [49] the $\gamma^* p\rightarrow \Delta$ reaction at low $q^2$ is measured. At $q^2 = 0.06\; \textrm{GeV}^2$ the average value
of the $M1+$ amplitude 
\[M1+ = 40.33 \pm 0.27 \pm 0.57 \pm 0.61\; 10^{-3}/m_{\pi^+}\]
is presented.
Many effective theories [50] are used to fit the data. 
The theoretical result presented above [5] is at $q^2 = 0\; \textrm{GeV}^2$. In order to compare with the data at $q^2 = 0.06\; \textrm{GeV}^2$ [49]
the correction by the form factor $D_3(q^2) = \sqrt{a a'} D_2(q^2)$ (112) has to be taken into account. The correction is 0.82.
Therefore, this model [5] predicts 
\[M1+ = 38.8 10^{-3}/m_{\pi^+}\]
at the $q^2 = 0.06\; \textrm{GeV}^2$. This value is in agreement with the average value of Ref. [49] within the errors.  
No new parameter is taken in this prediction.

As mentioned above in the rest frame the wave functions of nucleon and $\Delta$ resonance are in s-wave only. Eq. (165) shows that 
\[E1+ \propto B,\;\;and\;\;E1+ \propto \; m^2_\Delta - m^2_N.\]
Nonzero B comes from the effects of antiquarks and $m^2_\Delta - m^2_N$ is resulted in the effect of recoil or the effect of Lorentz contraction
in the moving frame. The Lorentz contraction makes that the nucleon or $\Delta$ in moving frame contain components of many partial waves.
Nonzero $E1+$ predicted in this model is resulted in both the antiquark components and the recoil effects.
It is interesting to notice that in the rest frame both the proton and the $\Delta$ are spherical. This model provides a new mechanism for the $E1+$
amplitude in the process $ \Delta\rightarrow p + \gamma$.
   
At the peak $W = m_\Delta$
\begin{equation}
\frac{E1+}{M1+}=\frac{-B(m_\Delta - m_N)}{A(m_\Delta +m_N)+B(m_\Delta -m_N)}%
=-5.4\;\%.
\end{equation}
In the paper [5] following early data of this ratio are quoted\\
-0.045 [51], -0.073 [52],
-0.024 [53].\\
In Ref. [46] the newer estimated value of this ratio is presented as
\[-0.025 \pm 0.005\]
for the process $\Delta \rightarrow N + \gamma$. However, the absolute value of the ratio $\frac{E1+}{M1+}$ at the pole
is determined as
\[0.065 \pm0.007\; [54],\;\;and\;\;0.058\; [55].\]
This model predicts a negative and small $\frac{E1+}{M1+}$ at the pole by this mechanism. 

\subsection{$\Delta \rightarrow N + \gamma$ decay}
Using Eq. (155), 
the width of the $\Delta
^{+}(1236)\rightarrow p+\gamma $ is derived
\begin{eqnarray}
\Gamma _\gamma  &=&\frac{k^2}{2\pi }\frac{m_N}{m_\Delta }\{\mid
A_{\frac 32}\mid ^2+\mid A_{\frac 12}\mid ^2\},  \\
A_{\frac 32} &=&-\frac{eD_3(0)(m_\Delta +m_N)(m_\Delta ^2-m_N^2)}{8\sqrt{6}%
(m_Nm_\Delta )^{3/2}}\{A+2B\frac{m_\Delta -m_N}{m_\Delta +m_N)}\}  \nonumber
\\
&=&-0.21\;\textrm{GeV}^{-\frac 12},  \nonumber \\
A_{\frac 12} &=&-\frac{eD_3(0)(m_\Delta +m_N)(m_\Delta ^2-m_N^2)}{24\sqrt{2}%
(m_Nm_\Delta )^{3/2}}\{A-2B\frac{m_\Delta -m_N}{m_\Delta +m_N)}\}  \nonumber
\\
&=&-0.10\;\textrm{GeV}^{-\frac 12}.
\end{eqnarray}
The experimental data
are [54]
\begin{equation}
A_{\frac 32} = -0.24\;\textrm{GeV}^{-\frac 12},\;\;A_{\frac 12} = -0.14\;\textrm{GeV}^{-\frac 12}.
\end{equation}
The new data are [46]
\begin{equation}
A_{\frac 32} = -0.250 \pm 0.008\; \textrm{GeV}^{-\frac 12},\;\;A_{\frac 12} = -0.135 \pm 0.006\; \textrm{GeV}^{-\frac 12}.
\end{equation}
The decay width is computed to be
\begin{equation}
\Gamma _{\Delta\rightarrow N\gamma} =0.64\; \textrm{MeV},
\end{equation}
The experimental data [56] is 0.65 $\textrm{MeV}$. The new data [46] is (0.63 - 0.78) $\textrm{MeV}$. 
Theoretical results agree with data pretty well.

\subsection{The electromagnetic form factors of
$p\rightarrow \Delta ^{+}(1236)$}

The differential cross section of the
\begin{center}
$e+p\rightarrow e+\Delta ^{+}(1236)$ \\ \hspace*{0.9in}
$\hookrightarrow N+\pi$
\end{center}
is expressed as
\begin{equation}
\frac 1{\Gamma _t}\frac{d^2\sigma }{d\Omega dE^{^{\prime }}}=\sigma
_T+\varepsilon \sigma _S.
\end{equation}
where $E'$ is the energy of the outgoing electron, $\sigma_T$ and $\sigma_\sigma$ are the cross sections of 
the transverse and longitudinal photons respectively, and $\varepsilon$ is the polarization parameter. Using of the Eq. (155)
and the equation
\begin{eqnarray}
\sum_\lambda \psi _\mu ^\lambda (p)\overline{\psi }_{\mu ^{^{\prime
}}}^\lambda (p) &=&\frac 12(1-\frac i{m^{^{\prime }}}\widehat{p})\{\delta
_{\mu \mu ^{^{\prime }}}+\frac 23\frac{p_\mu p_{\mu ^{^{\prime }}}}{%
m^{^{\prime }2}}-\frac 13\gamma _\mu \gamma _{\mu ^{^{\prime }}}  \nonumber
\\
&&-\frac i{3m^{^{\prime }}}(p_\mu \gamma _{\mu ^{^{\prime }}}-p_{\mu
^{^{\prime }}}\gamma _\mu )\},
\end{eqnarray}
it is obtained
\begin{eqnarray}
\sigma _T &=&\frac{m\alpha q^{*2}}{m_\Delta(W^2-m^2)}\frac{\Gamma (W)}{%
(W-m_\Delta)^2+\frac 14\Gamma ^2(W)}\frac{D_3^2(q^2)}{18m^2}%
\{A^2(q^2+m_{+}^2)  \nonumber \\
&&+2AB(m^2_\Delta-m^2-q^2)+4B^2(q^2+m_{-}^2)(1-\frac{q^2}{q^{*2}})\}, \\
\sigma _S &=&\frac{m\alpha q^{*2}}{m_\Delta(W^2-m^2)}\frac{\Gamma (W)}{%
(W-m_\Delta)^2+\frac 14\Gamma ^2(W)}\frac{2D_3^2(q^2)}{9m^2}%
B^2(q^2+m_{+}^2)\frac{q^2}{q^{*2}},
\end{eqnarray}
where
\begin{equation}
 W^2=-(p_i+p_e-p_{e^{^{\prime }}})^2,\;\;q^{*2}=q^2+\frac
1{4m^2_\Delta}(m^2_\Delta-m^2-q^2)^2,
\end{equation}
$p_i$ is the momentum of the proton, m is the mass of nucleon, and $m_+ = m_\Delta + m$, and $m_- = m_\Delta - m$.a
Eq. (175) shows that the cross section of the longitudinal photon $\sigma_S \propto B^2$ and B is originated in the effect of antiquark components of the proton and the $\Delta$.
Therefore, the cross section  $\sigma_S$ is the consequence of the antiquark components in this model.
The ratio of $\sigma _S$ and $\sigma _T$ is obtained
\begin{eqnarray}
R &=&\frac{\sigma _S}{\sigma _T}=4B^2(q^2+m_{-}^2)\frac{q^2}{q^{*2}}%
/\{(Am_{+}+Bm_{-})^2  \nonumber \\
&&+(A-B)^2q^2+3B^2(q^2+m^2)-4B^2(q^2+m_{-}^2)\frac{q^2}{q^{*2}}\}.
\end{eqnarray}
The behavior of R is expressed as
\begin{eqnarray}
q^2&=&0,\;\;\;R=0;  \nonumber \\
q^2&\rightarrow&\infty,\;\;\;R\sim \frac 1{q^2}\rightarrow 0.
\end{eqnarray}
In the range of $q^2>3\; \textrm{GeV}^2$, $R \sim 0.27$.

According to the definition of multiples [57], the magnetic transition
form factor $G_{M1+}^2(q^2)$, the electric quadrupole transition form factor
$G_{E1+}^2(q^2)$ and the Coulomb transition form factor $G_{S1+}^2(q^2)$ are found
\begin{eqnarray}
G_{M1+}^2(q^2) &=&\frac{D_3^2(q^2)}{18m^2}%
\{(Am_{+}+Bm_{-})^2+(A-B)^2 q^2-B^2(q^2+m_{-}^2)\frac{q^2}{q^{*2}}\}, \\
G_{E1+}^2(q^2) &=&\frac{D_3^2(q^2)}{18m^2}B^2(q^2+m_{-}^2)(1-\frac{q^2}{%
q^{*2}}), \\
G_{S1+}^2(q^2) &=&\frac{D_3^2(q^2)}{18m^2}B^2(q^2+m_{-}^2).
\end{eqnarray}
Eqs. (179-181) show that besides the magnetic form factor $G_{M1+}(q^2)$ this model predicts
nonzero $E1+,\;S1+$ form factors $G_{E1+}(q^2)$ and $G_{S1+}(q^2)$. There is a relationship between $G_{E1+}(q^2)$ and $G_{S1+}(q^2)$
\[G^2_{E1+}(q^2) = (1-\frac{q^2}{ q^{*2}})G^2_{S1+}(q^2).\]
Both the $G_{E1+}(q^2)$ and the $G_{S1+}(q^2)$ are proportional to B. 
Therefore, these two form factors are from the contributions of the antiquark components of the nucleon.

The differential cross
section (172) is expressed as
\begin{eqnarray}
\frac 1{\Gamma _t}\frac{d^2\sigma }{dE^{^{\prime }}d\Omega } &=&\frac{%
m\alpha q^{*2}}{m^{^{\prime }}(W^2-m^2)}\{G_{M1+}^2(q^2)+3G_{E1+}^2(q^2)
\nonumber \\
&&+4\varepsilon G_{S1+}^2(q^2)\frac{q^2}{q^{*2}}\}\frac{\Gamma (W)}{%
(W-m^{^{\prime }})^2+\frac 14\Gamma ^2(W)}.
\end{eqnarray}

From Eq.(179), the transit magnetic moment of $p\rightarrow
\Delta ^{+}(1236)$ is derived
\begin{equation}
\mu_{p\rightarrow\Delta} =G_{M1+}(0)=\frac{D_3(0)}{3\sqrt{2}m}(Am_{+}+Bm_{-})=1.23\frac{2\sqrt{2%
}}3\mu _p.
\end{equation}
The data are
\begin{equation}
1.22\frac{2\sqrt{2}}3\mu _p^{[53]},\;\;\;1.28\frac{2\sqrt{2}}3\mu
_p^{[56]}.  \nonumber
\end{equation}
Taking $a = a^\prime = 1$ and $m = m^\prime$ in Eq. (183), 
the prediction of SU(6), $\mu_{p\rightarrow\Delta} = \frac{2\sqrt{2}}3\mu _p$, 
is revealed. The correction factor 1.23 in Eq. (183)
is from the effects of recoil and antiquarks in this model. Theory agrees with the data.
%Lorentz contraction effect is considered in Eq.(84), for
%$p\rightarrow \Delta ^{+}(1236)$
%\begin{equation}
%f(m,m^{^{\prime }})=0.98.
%\end{equation}
%The theoretical value of the magnetic
%transition moment is 
%\[\mu_{p\rightarrow \Delta} = 1.24\frac{2\sqrt{2}}3\mu _p.\]

The expression
\begin{equation}
\sigma _T^R=\frac{m\alpha q^{*2}}{m^{^{\prime }}(W^2-m^2)}\frac{\Gamma (W)}{%
(W-m^{^{\prime }})^2+\frac 14\Gamma ^2(W)}G^{*2}_M(q^2)
\end{equation}
has been used to determine $G_M^{*2}(q^2)$. $G^*_M$ is obtained from Eq.(174) that
\begin{eqnarray}
G^{*2}_M(q^2) &=&G_{M1+}^2(q^2)+3G_{E1+}^2(q^2)  \nonumber \\
&=&\frac{D_3^2(q^2)}{18m^2}%
\{(Am_{+}+Bm_{-})^2+(A-B)^2q^2+B^2(q^2+m_{-}^2)(3-4\frac{q^2}{q^{*2}})\}.
\end{eqnarray}
The $D_3(q^2)$ for $p\rightarrow \Delta
^{+}(1236)$ is expressed as
\begin{equation}
D_3(q^2)=\frac{4mm^{^{\prime }}\sqrt{aa^{^{\prime }}}}{(m+m^{^{\prime }})^2}%
(1+2.39\frac{q^2}{4m^2})^{-1}(1+\frac{q^2}{0.71})^{-2}.
\end{equation}
The $G_{M1+}^2(q^2)$ form factor dominates the $G_M^{*2}(q^2)$.
Eq. (186) shows that
\[G^*_M(q^2) \propto D_3(q^2)\]
and $D_3(q^2)$ has triple poles. Therefore, this model predicts the form factor $G^*_M(q^2)$ decreases with $q^2$ faster than 
the $G^p_M(q^2)$ which is in form of dipole.
Comparisons with experimental data
are shown in Fig.7 and Fig.8 (a).
 
The data
for Fig.7 comes from Ref.[26] and that for Fig.8 are from Ref.[61].
\begin{figure}
\begin{center}
\includegraphics[width=7in, height=7in]{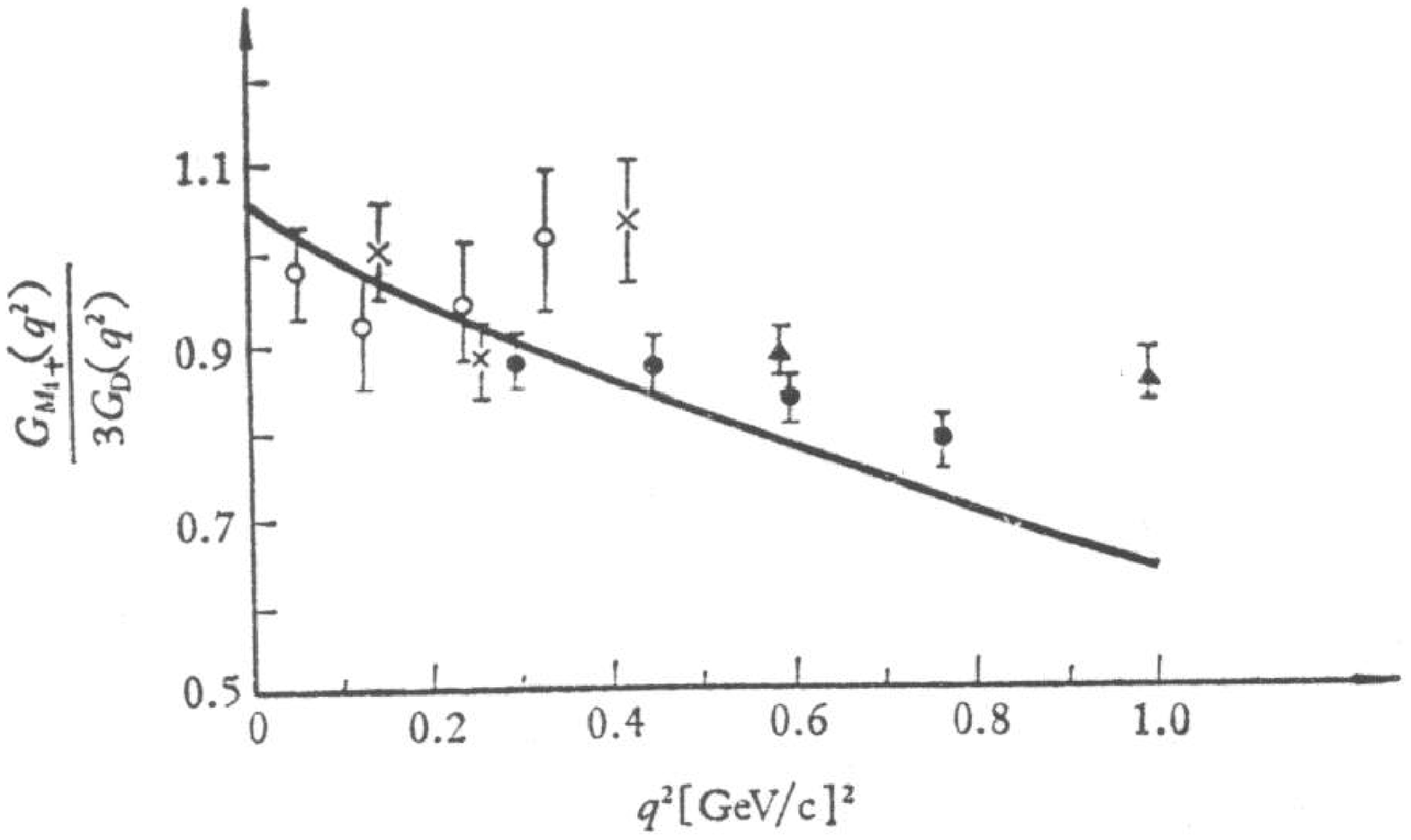}
Fig. 7
\end{center}
\end{figure}

\begin{figure}
\begin{center}
\includegraphics[width=7in, height=7in]{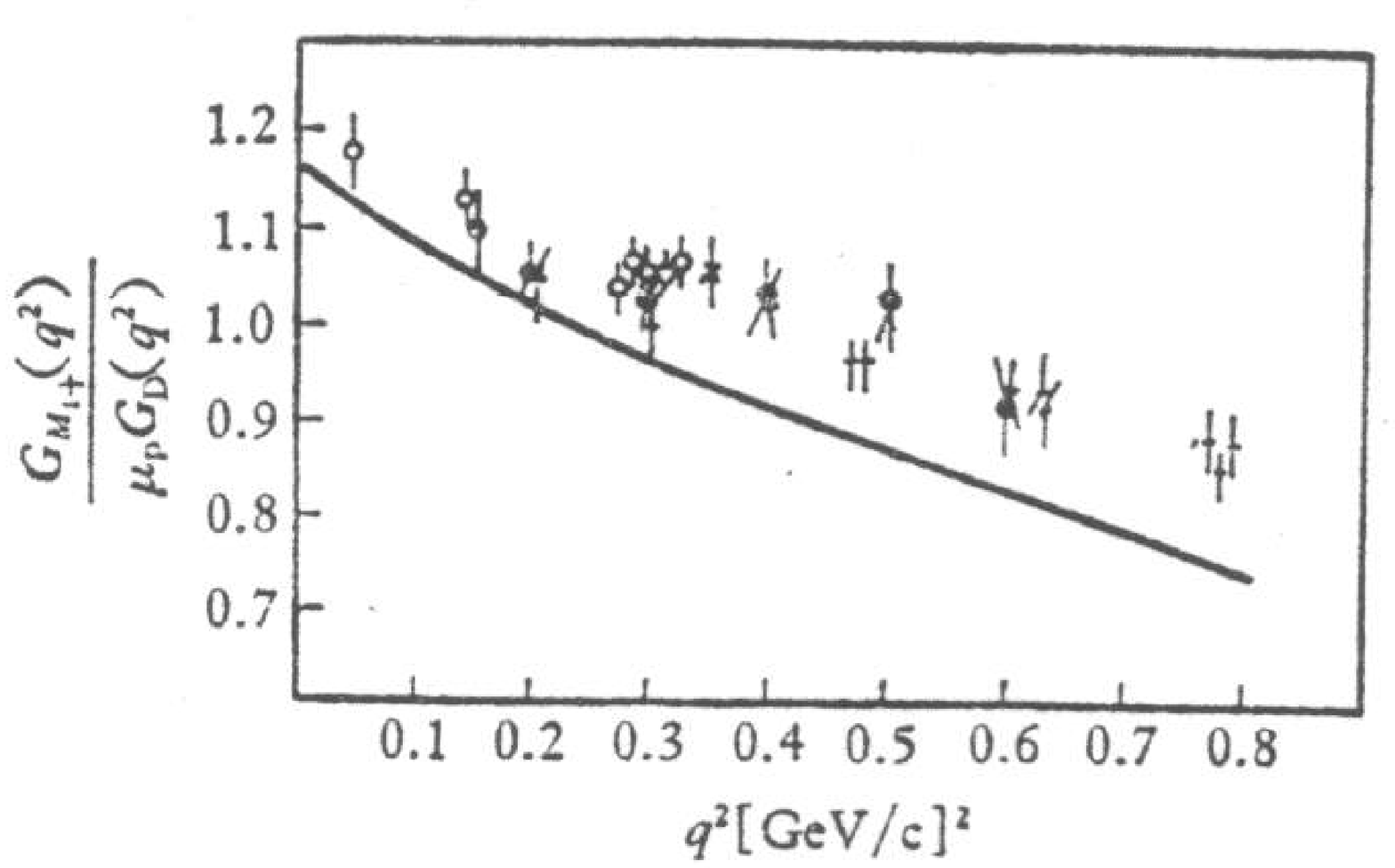}
Fig. 8 (a)
\end{center}
\end{figure}

These two figures show that: \\
1) this model predicts that $G^*_M(q^2)$ decreases faster than $G_D(q^2)$ or $G^p_M(q^2)$;\\
2) theoretical prediction of $G^*_M(q^2)$ agrees with data in the region of low $q^2$;\\
3) in the larger region theoretical $G^*_M(q^2)$ decreases faster than data. \\
There are new measurements of the form factor $G^*_M(q^2)$ [63,64,65]
in the region of larger $q^2$. These new data show that the $G^*_M(q^2)$ indeed decreases faster
than $G^p_M(q^2)$ ($\sim \mu_p G_D(q^2)$). However, theoretical prediction of $G^*_M(q^2)$ decreases
faster than these new data too. SU(6) symmetry breaking is the possible reason. As shown in the transit magnetic moment (183)
the effect of SU(6) symmetry breaking by the mass difference of nucleon and $\Delta$ is about $23\%$.
Therefore,
in order to improve the behavior of the $G^*_M(q^2)$ in the region of larger $q^2$
the effects of SU(6) symmetry breaking must be taken into account.
In the expression of the $D_3(q^2)$ (187) the factor
\[1/(1 + 2.39 q^2/(4 m^2_p))/(1 + q^2/0.71)^2\]
is determined from the magnetic form factor of the proton. 
The $G^p_M(q^2)$ the quantity with dimension of $mass$ is proportional to  $m_p$.
Based on SU(6) symmetry Eq. (187) has been applied to the magnetic form factor $G_{M1+}(q^2)$.
In order to include the effect of the SU(6) symmetry breaking it is natural to include $m_\Delta$
in this factor of Eq. (187). In this paper following possibility
\[m_p \rightarrow {1\over2}(m_p + m_\Delta)\]
is tried. The quantity $0.71 \textrm{GeV}^2$ can be rewritten as
\[0.71 \rightarrow 0.71 (m_p + m_\Delta)^2/(4 m^2_p)\]
and the $2.39/(4m^2_p)$ in Eq. (187) can be replaced as
\[2.39/(m_p + m_\Delta)^2.\]

The comparison between this scheme and the data [63,64,65] is presented in Fig. 8(b). 

\begin{figure}
\begin{center}
\includegraphics[width=7in, height=7in]{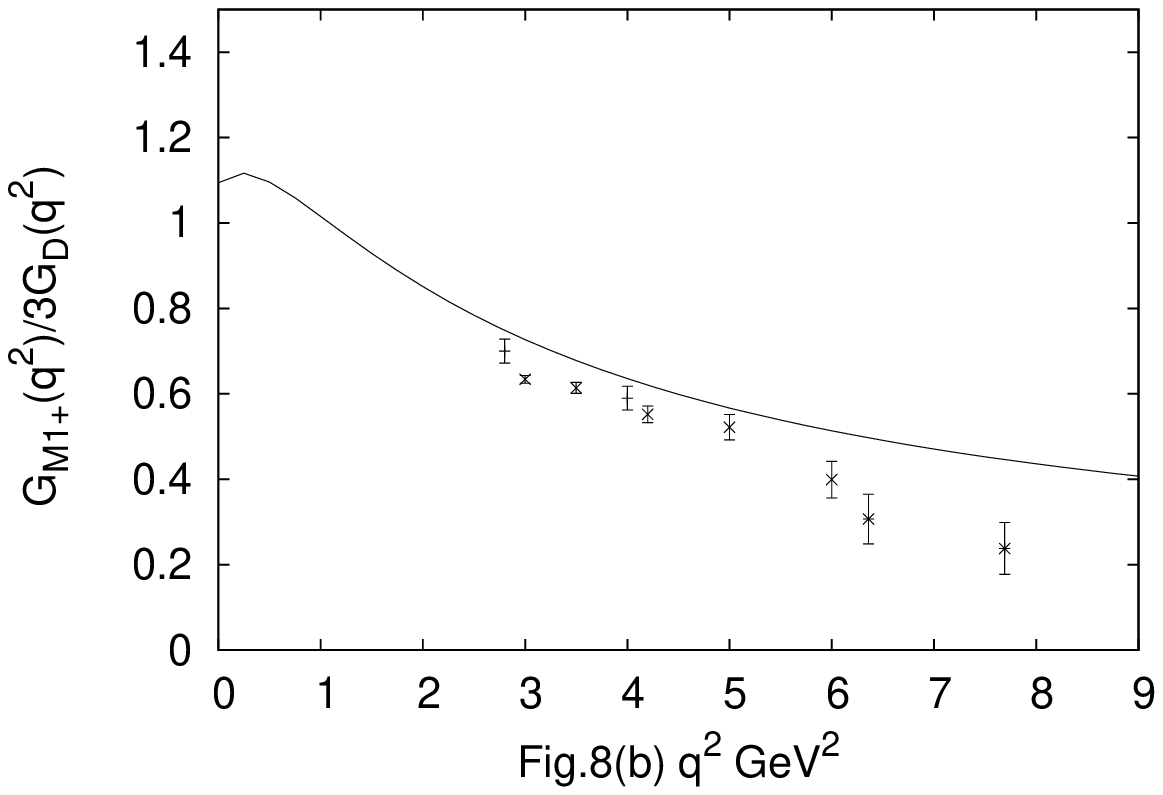}
%Fig. 8(b).
\end{center}
\end{figure}

Fig. 8(b) shows that the theoretical results of the $G_{M1+}(q^2)$ are indeed improved. 
However, when $q^2 > 6\; \textrm{GeV}^2$ theoretical $G_{M1+}(q^2)/(3 G_D(q^2))$ decreases
still faster than current data. 
As mentioned above that this model may not be working well for large $q^2$.

At the peak of the $\Delta$ resonance the electric
multiple moments are obtained from Eq.(180,181)
\begin{eqnarray}
E1+ = G_{E1+}(0) = -\frac{D_3(0)}{3\sqrt{2}m}Bm_{-}=-0.17,\\
S1+ = G_{S1+}(0) = -\frac{D_3(0)}{3\sqrt{2}m}Bm_{-}=-0.17.
\end{eqnarray}
This model predicts small and negative E1+, S1+, and
\begin{eqnarray}
S1+ &=&E1+,  \nonumber \\
R_{SM} = \frac{S1+}{\mu_{p\rightarrow \Delta}}  &=& -5.4\%,\\
R_{EM} = \frac{E1+}{\mu_{p\rightarrow \Delta}}  &=& -5.4\%.
\end{eqnarray}
The data [26] is
\begin{equation}
\frac{S1+}{\mu_{p\rightarrow\Delta}} = (-5\pm3)\%.
\end{equation}
Theoretical result agree with this experimental data.
There are many new measurements on the ratios of $\frac{E1+}{\mu_{p\rightarrow \Delta}}$
and $\frac{S1+}{\mu_{p\rightarrow \Delta}}$ in different regions of $q^2$ [66-74].
All the new data show both the $R_{EM}$ and $R_{SM}$ are negative and small, which are 
the predictions of this model. The value of the $R_{SM}$ predicted by this model is compatible
with these new data. However, the value of the $R_{EM}$ predicted by this model is about half 
the value measured. 

The cross section $\sigma_S$ (175) is calculated in Ref. [5].
Taking
\begin{equation}
W=m^{^{\prime }}=1.236\;GeV,\;\;\;\Gamma (m^{^{\prime }})=0.12\;GeV,
\end{equation}
we obtain
\begin{equation}
\sigma _S=48.4q^2(q^2+0.0888)(1+0.679q^2)^{-2}(1+\frac{q^2}{0.71}%
)^{-4}\times 10^{-28}cm^2.
\end{equation}
Comparison with the data [62] is shown in Fig.9.
%\begin{figure}
%\begin{center}
%\includegraphics[width=7in, height=7in]{fig6.ps}
%FIG. 8.
%\end{center}
%\end{figure}
\begin{figure}
\begin{center}
\includegraphics[width=7in, height=7in]{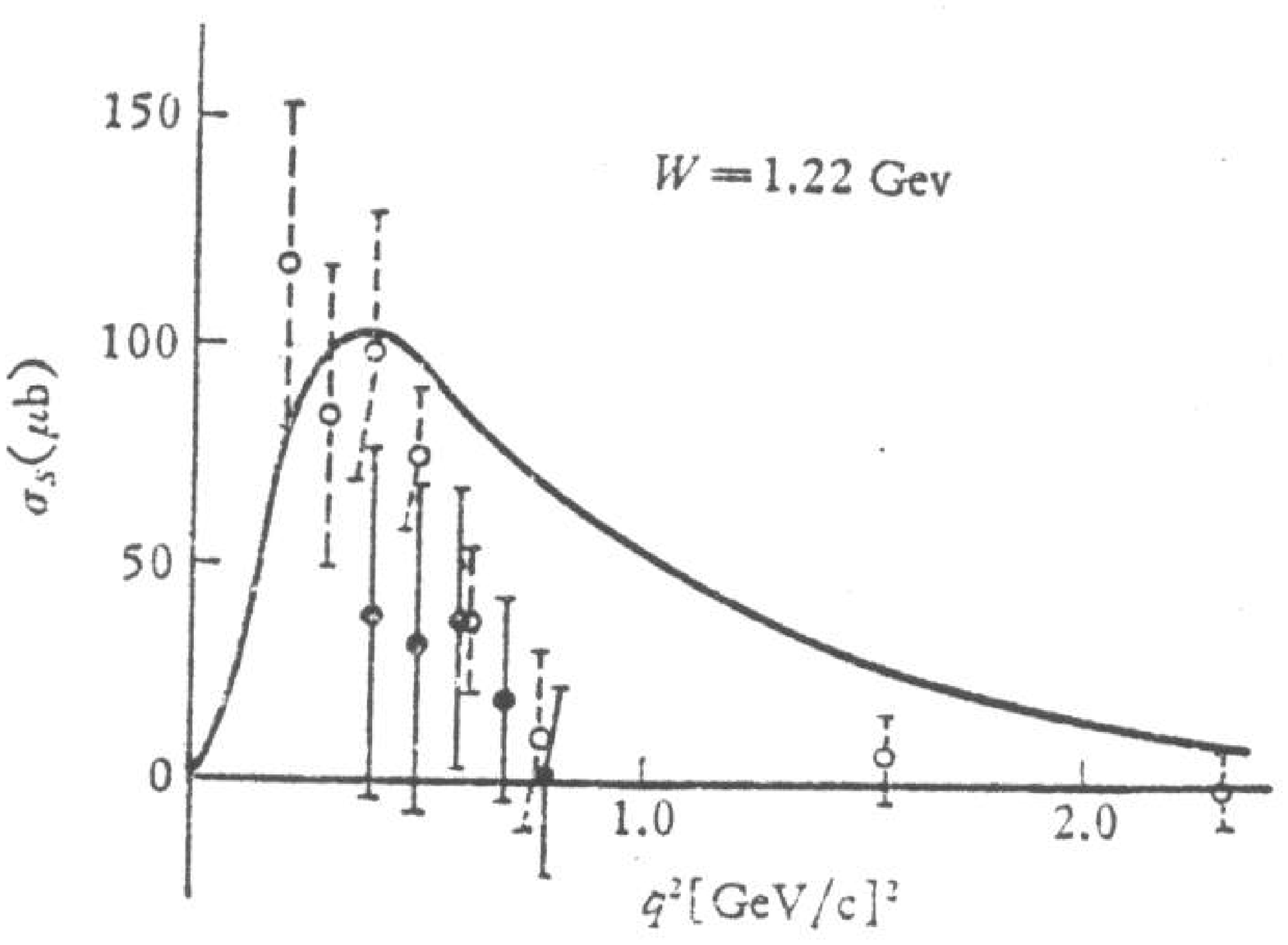}
Fig. 9.
\end{center}
\end{figure}

\section{The axial-vector and pseudoscalar form factors of baryons}
This model has been applied to study the weak semileptonic decays of baryons and the charged quasielastic neutrino reactions of 
baryons in Ref. [6] and the review of this study is presented
in this section.

The effective Lagrangian of weak interactions of charged currents between quark and lepton is
\begin{equation}
{\cal L}_w=\frac{G}{\sqrt{2}}\bar{l}(x)(1+\gamma_5)\gamma_\mu\nu J^W_\mu + h.c..
\end{equation}
where $J^W_\mu$ is the charged weak currents of quarks. The effects of strong interactions must be taken into account. 
In $J^W_\mu$ there are both vector and axial-vector parts. Because of CVC the vector part is well determined from Eq. (48). It is known that 
the $G_A$ of the $\beta$ decay of the neutron cannot be determined by SU(6) symmetry, therefore, a new parameter $\lambda$ has to be introduced.
The effective charged current of quarks are expressed as [6]
\begin{equation}
J^W_\mu=\bar{\psi}(x)Q_W\{\gamma_\mu+\lambda\gamma_\mu\gamma_5+\frac{\kappa}{2m_N}
\sigma_{\mu\nu}q_\nu\}\psi(x),
\end{equation}
where \(Q_w=cos\theta\lambda_{12}+sin\theta\lambda_{13}\), $\theta$ is Cabbibo
angle, $\lambda$ is a new parameter, and the parameter $\kappa$ has been determined already (107). 
\subsection{Weak matrix elements of ${1\over2}^+$ baryons} 
Using the baryon wave functions (33,35,46,47), the effective Lagrangian of weak interaction (196), and 
the expression of the matrix elements similar to (49), 
the transition matrix elements of charged weak currents are calculated
\begin{eqnarray}
<B^{{1\over2}c'}(p_f)|J^W_\mu(0)|B^{{1\over2}c}(p_i)> = -{1\over2}\int d^4 x_1 
d^4 x_2 M(x_1,x_2, x'_2,x_1)\bar{B}^{{1\over2}c',i'j'k'_1}_{\alpha\beta\gamma,ijk}
(x_1,x_2,0)^{l'_1}_{l_1}Q^{k'_1 k'_2}_{Wk_1 k_2}\nonumber \\
\{\gamma_\mu+\frac{\kappa}{2m_p}
\sigma_{\mu\nu}q_\mu+
+\lambda\gamma_5\gamma_\mu\}_{\gamma\gamma'}B^{{1\over2}c,k'_2 j'i'}_{\gamma'
\beta\alpha,k_1 ji}(0,x_2,x_1)^{l'_2}_{l_2}
=\bar{u}_{c^\prime}(p_f)\{A_1 I_1+A_2 I_2\}u_c(p_i),
\end{eqnarray}
where
\begin{equation}
A_1 = Tr\bar{B} Q_w B,\;\;\;A_2 = Tr\bar{B} B Q_w,
\end{equation}
and
\begin{eqnarray}
I_1={1\over6}\{[(4-{m'\over m})D_2(q^2)+(4-{m\over m'})D'_2(q^2)
+5(\frac{q^2+m^2_-}{2mm'}+\frac{\kappa m_+ q^2}{10mm' m_p})D_3 (q^2)]\gamma_\mu\nonumber \\
+[10D_1(q^2)-(5-\frac{2m_p}{\kappa m})D_2 (q^2)-(5-\frac{2m_p}{\kappa m'})D'_2(q^2)\nonumber \\
+\frac{3q^2+5m^2_+}{2mm'}D_3(q^2)]\frac{\kappa}{2m_p}\sigma_{\mu\nu}q_\nu
+5\lambda[D_2(q^2)+D'_2(q^2) + \frac{q^2+m^2_-}{2mm'}D_3(q^2)]\gamma_\mu\gamma_5\nonumber \\
+i(1+\frac{\kappa m+}{2m_p})\frac{m_-}{mm'}D_3(q^2)q_\mu-i \lambda[{1\over m}D_2(q^2)+
{1\over m'}D'_2(q^2)-\frac{m_+}{mm'}D_3(q^2)]q_\mu\gamma_5\},
\end{eqnarray}
\begin{eqnarray}
I_2={1\over6}\{[-(1+2{m'\over m})D_2(q^2)-(1+2{m\over m'})D'_2(q^2)+(\frac{q^2+m^2_-}
{2mm'}+\frac{\kappa m_+ q^2}{mm'm_p}D_3(q^2)]\gamma_\mu\nonumber \\
+[2D_1(q^2)-(1-{4m_p\over\kappa m})D_2(q^2)-(1-{4m_p\over\kappa m'})D'_2(q^2)\nonumber \\
+\frac{m^2_+-3q^2}{2mm'}D_3(q^2)]{\kappa\over2m_p}\sigma_{\mu\nu}q_\nu
+\lambda[D_2(q^2)+D'_2(q^2)+\frac{q^2+m^2_-}{2mm'}D_3(q^2)]\gamma_\mu\gamma_5\nonumber \\
+2i(1+\frac{\kappa m_+}{2m_p}){m_-\over mm'}D_3(q^2)q_\mu-2i\lambda[{1\over m}D_2(q^2)
{1\over m'}D'_2(q^2)-\frac{m_+}{mm'}D_3(q^2)]q_\mu\gamma_5\},
\end{eqnarray}
where \(m_{\pm} = m' \pm m\), m and $m^\prime$ are the masses of the initial and final baryons respectively.
The functions $D_1(q^2),\;D_2(q^2),\;D'_2(q^2),\;D_3(q^2)$ are defined by Eq. (59).

It is necessary to point out that
the condition of the current conservation (62) prohibits the appearance of 
the second class current, 
\[\bar{u}\gamma_5 q_\nu \sigma_{\mu\nu}u,\] 
in Eq. (199,200). 

The amplitude (197) can be expressed as
\begin{eqnarray}
<B^{{1\over2}c'}(p_f)|J^W_\mu(0)|B^{{1\over2}c}(p_i)> = b D_1(q^2) \bar{u}(p_f)_{c'}\nonumber\\
\{f_V(q^2) \gamma_\mu + f_T(q^2) \sigma_{\mu\nu} q_\nu + if_S(q^2) q_\mu\nonumber \\
+ g_A(q^2) \gamma_\mu \gamma_5 +i g_P(q^2) q_\mu \gamma_5\}u(p_i)_c,
\end{eqnarray}
where b is a coefficients and
\begin{equation}
D_1(q^2) = \frac{1}{\sqrt{aa'}(1+2.39\tau)(1+{q^2\over 0.71})^2}.
\end{equation}
All these form factors of the weak
matrix elements (201) are predicted by this model and 
b, $f_V,\;f_T,\;f_S,\;g_A,\;g_P$ are listed in Table 2.
 
\begin{table}[h]
\begin{center}
\caption{Form factors of weak interactions}
\begin{tabular}{|c|c|c|c|c|c|c|} \hline
process&b&$f_V$&$f_T$&$f_S$&$g_A$&$g_P$\\ \hline
$n\rightarrow p$&${1\over6}$& & & & &  \\
$\Xi^-\rightarrow \Sigma^0$&${1\over6\sqrt{2}}$&$(5-{m_+\over m})a+(5
-{m_+\over m'})a^\prime$ &$\mu[10-(5-{1\over\mu m})a-(5-{1\over\mu m'})a^\prime]$
&S
&5A&-P\\
$\Xi^0\rightarrow \Sigma^+$&-${1\over6}$
&+$(5\zeta_-+\frac{\mu m_+
q^2}{mm'})aa'$
&+$(5\zeta_+-\frac{q^2}{mm'})aa^\prime]$
& & &  \\ \hline
$\Xi^-\rightarrow\Xi^0$&-${1\over6}$&$(1-{2m_+\over m})a+(1-{2m_+\over m'})a'$
&$\mu[2-(1-{2\over\mu m})a-(1-{2\over\mu m'})a'$&2S&A&-2P\\ 
%\hline
$\Sigma^-\rightarrow n$&${1\over6}$&+$(\zeta_-+{2\mu m_+ q^2\over mm'})aa'$
&$+(\zeta^+ - {2q^2\over mm'})aa']$ & & & \\ \hline
$\Sigma^-\rightarrow\Sigma^0$&$-{1\over 6\sqrt{2}}$&$(4+{m_+\over m})a
+(4+{m_+\over m'})a'$&$\mu[8-(4+{1\over\mu m})a-(4+{1\over\mu m'})a'$&-S&4A&P\\
 & &+$(4\zeta_--{\mu m_+ q^2\over mm'})aa'$&$+(4\zeta_++{q^2\over mm'})aa']$
& & &  \\ \hline
$\Sigma^+\rightarrow\Lambda$&${1\over\sqrt{6}}$&${\mu' q^2\over 2mm'}aa'$
&$\mu[2-(1-{1\over2\mu m})a-(1-{1\over2\mu m'})a'$&${S\over2}$
&A&-${P\over 2}$ \\
$\Sigma^-\rightarrow\Lambda$&${1\over\sqrt{6}}$& &+$\frac{aa'm^2_+}{mm'}]$
& & &  \\ \hline
$\Lambda\rightarrow p$&-${3\over2\sqrt{6}}$&$a+a'+aa'\zeta_-$
&$\mu[2-(a+a'-aa'\zeta_+)]$&0&A&0\\ \hline
$\Xi^-\rightarrow \Lambda$&${1\over2\sqrt{6}}$&$(1+{m_+\over m})a+(1+{m_+\over m'})a'$
&$\mu[2-(1+{1\over\mu m})a-(1+{1\over\mu m'})a'$&-S&A&P\\
 & &+$(\zeta_--{\mu m_+ q^2\over mm'})aa'$&+$(\zeta_++\frac{q^2}{mm'})aa']$
& & & \\ \hline
\end{tabular}
\end{center}
where $\mu={\kappa\over2m_p},\;\;$$\mu'=1+\mu m_+,\;\;$$\zeta_\pm=\frac{q^2+m^2_\pm}
{2mm'},\;\;$
$S={\mu' m_-\over mm'}aa'\;\;$$A=\lambda(a+a'+aa'\zeta_-),\;\;$$P=\lambda({a\over m}
+{a'\over m'}-{aa'm_+\over mm'})$
\end{table}

\subsection{Axial-vector form factor of nucleon}
Using the Table 2, the axial-vector form factor of $n\rightarrow p$ is obtained
\begin{eqnarray}
G_A(q^2) = {5\over6} \lambda D_1(q^2) \{a + a' +\zeta_- a a'\},
\end{eqnarray}
where the definition of the $\zeta_-$ can be found in Tab. 2.
Ignoring the mass difference of proton and neutron and using the definitions of $D_{2,3}(q^2)$,
Eq. (203) is expressed as
\begin{equation}
G_A(q^2)=G_A(0) \{D_2(q^2)+\tau D_3(q^2)\},
\end{equation}
where
\begin{equation}
G_A(0) = {5\over3}\lambda.
\end{equation}
Eq.(205) is the result of SU(6). 

After ignoring the mass difference of proton and neutron, the vector form factor of $n\rightarrow p$ 
is found from Tab. 2
\begin{eqnarray}
f_V = {1\over6}\{ 6a + \tau (10 + 4 \kappa) a^2 \} D_1(q^2),\\
f_V = D_2(q^2) + {1\over3} \tau (5 + 2 \kappa) D_3(q^2).
\end{eqnarray}
Therefore, 
\begin{equation}
f_V(0) = 1.
\end{equation}
In Ref. [6]
inputting ${G_A\over G_V} = 1.242$ (where $G_V = f_V$) the parameter $\lambda$ is found to be
\begin{equation}
\lambda = 0.745.
\end{equation}
The new data is ${G_A\over G_V} = 1.2701 \pm 0.0025$ [46] and $\lambda = 0.762$ is determined.
The difference is about $2\%$.

From Eqs.(52,80,81), the Dirac form factor $F_1(q^2)$ of proton is found to be
\begin{equation}
F_1(q^2) = D_2(q^2) + \tau D_3(q^2).
\end{equation}
Therefore, this model predicts
\begin{eqnarray}
G_A(q^2) = {5\over 3} \lambda F_1(q^2),\\
F^p_1(q^2) = {1\over 1 + \tau }\{G^p_E(q^2) + \tau G^p_M(q^2)\}.
\end{eqnarray}
Following the notation in literature, the the axial-vector form factor is rewritten as
\begin{equation}
{5\over 3} \lambda G_A(q^2).
\end{equation}
%The matrix element of nucleon transation of $n\rightarrow p$ is expressed as
%\begin{equation}
%<p(p_f)|j^W_\mu(0)|n(p_i)>={1\over3}\bar{u}_{c'}(p_f)\{5G^p_M(q^2)\gamma_\mu
%+5\lambda G_A(q^2)\gamma_\mu\gamma_5+{i\over m}P_\mu J(q^2)\}u_c(p_i),
%\end{equation}
%where \(P_\mu=p_{i\mu}+p_{f\mu}\),  
%$G^p_M(q^2)$ is shown in Eq.().  
%Using Table 2 the form factors
%are determined
%\begin{eqnarray}
%G_A(q^2) = F^p_1(q^2) = D_2(q^2)+\tau D_3(q^2),\\
%&&J(q^2) = D_2(q^2)+{5\over2}\kappa\{D_1(q^2)-D_2(q^2)+(1+{5\over3}\tau)D_3(q^2)\}.

In the study of $\nu + N$ scattering the axial-vector form factor is taken
as a form of dipole
\begin{equation}
G_A(q^2) = \frac{1}{(1 + {q^2\over M^2_A})^2}.
\end{equation}
The relationship between parameter $M_A$ and the charge radius of proton (80) is predicted 
\begin{equation}
{1\over M^2_A} = {1\over12} <r^2_E> - (\mu_p - 1){1\over 8m^2_p},
\end{equation}
where $<r^2_E>$ is the charge radius of proton squared.
Using Eq. (139) and inputting $<r^2_M> = (0.777 \pm 0.013 \pm 0.010)^2 fm^2$, it is predicted
\begin{equation}
M_A = 1.002\; \textrm{GeV}.
\end{equation}
On the other hand, if ignoring the contribution of the antiquarks, as mentioned above, $a = 1$
should be taken and 
\begin{equation}
G_A(q^2)=(1+{q^2\over0.71})^{-2} = {1\over \mu_p} G^p_M(q^2).
\end{equation}
The form factor $G_A(q^2)$, indeed, takes the form of dipole and $M_A = 0.84 \textrm{GeV}$.
Comparing with the data, this $G_A(q^2)$ decreases too fast.
Using the parametrization (77), the axial-vector form factor of nucleon, $G_A(q^2)$, is expressed as
\begin{equation}
G_A(q^2) = \frac{1 + 4.5 \tau}{(1 +2.39 \tau)(1 + {q^2\over 0.71})^2}.
\end{equation}
The axial-vector form factor predicted by this model is not in the form of dipole. However, because of the factor
\[\frac{1 + 4.5 \tau}{1 +2.39 \tau}\]
the $G_A(q^2)$ (218) decreases with $q^2$ slower than $(1+{q^2\over0.71})^{-2}$ and an equivalent $M_A$ is determined (216).
The comparison between the expression (218) predicted by this model and the form of dipole (214) 
with $M_A = 1.002\; \textrm{GeV}$ is shown in Fig. 10 .

\begin{figure}
\begin{center}
\includegraphics[width=7in, height=7in]{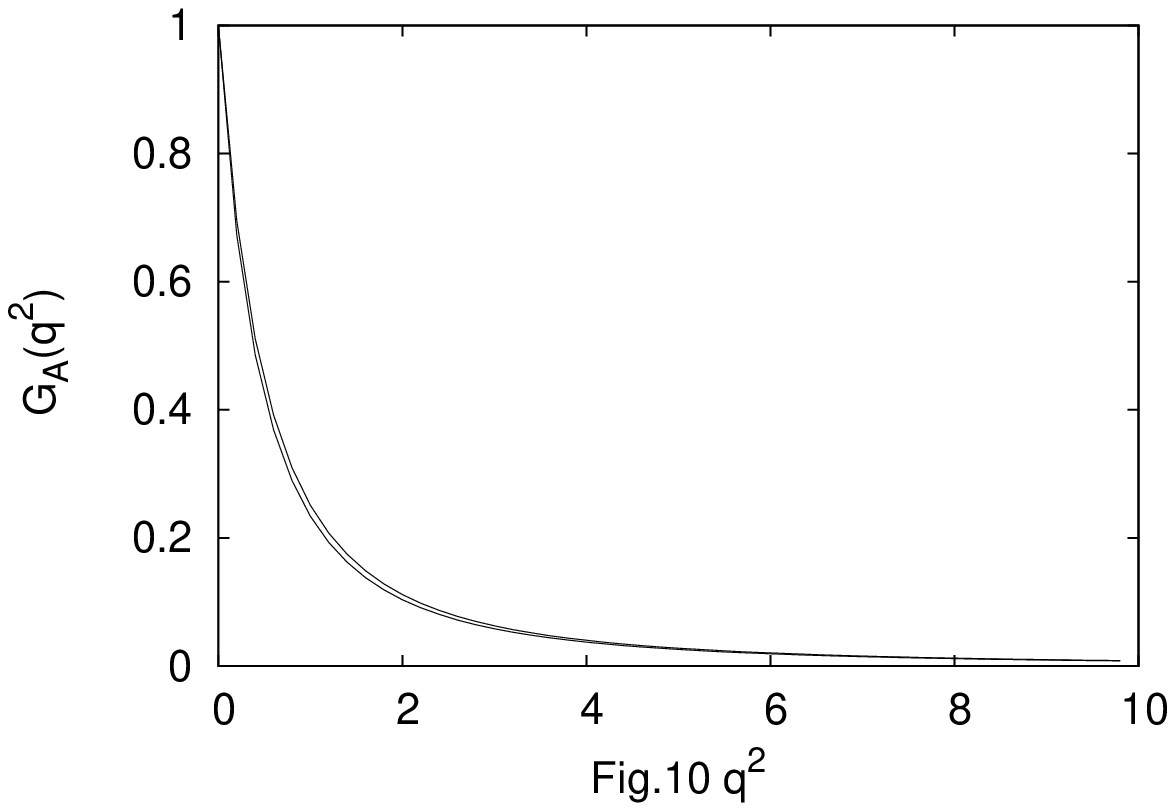}
%Fig. 10.
\end{center}
\end{figure}
The Fig. 10 shows that in a wide range of $q^2$ these two expressions of $G_A(q^2)$ are almost identical.
There are many measurements of the $M_A$. In Ref. [18] two groups of the experimental values of the $M_A$
are presented. One group from quasi-elastic neutrino and antineutrino scattering experiments and the resulting world
average is 
\begin{equation}
M_A = (1.026 \pm 0.021) \textrm{GeV}.
\end{equation}
The other determinations of the $G_A(q^2)$ are based on the analysis of charged pion electroproduction off proton and the world average 
is
\begin{equation}
M_A = (1.069 \pm 0.016)\textrm{GeV}.
\end{equation}
In Ref. [79] 
\begin{equation}
M_A^{world-average} = 1.014 \pm 0.014 \textrm{GeV}
\end{equation}
is presented. In Ref. [87] a list of the value of $M_A$ used by many experiments of $\nu_\mu + N$ and $\bar{\nu}_\mu + N$ is presented. 
The prediction of the $M_A$ (216) by this model is in good agreement with these values (219,220,221)
within the experimental errors.

\subsection{Pseudoscalar form factor of $n\rightarrow p$}
Ignoring the mass difference between neutron and proton, the pseudoscalar form factor of $n\rightarrow p$ is predicted by this model (Table 2)
(after the $g_P$ form factor in Eq. (201) is rewritten as 
$ g_P{q_\mu\over m} \gamma_5$)  
\begin{equation}
g_P = {1\over 3}\lambda (a - 1) D_2(q^2).
\end{equation}
If $a = 1$ is taken, $g_P = 0$ is revealed. Therefore, in this model the pseudoscalar form factor of nucleon is resulted 
in antiquark components of nucleon. For very small $q^2$
\begin{equation}
g_P(0) = {1\over3} \lambda (a - 1) = 0.872.
\end{equation}
Based on the work presented in Refs.[81,82] and systematic chiral expansion of low energy QCD Green function the authors of Ref. [78] 
obtain
\begin{equation}
g_P = 8.26 \pm 0.16.
\end{equation}
In Ref. [95] a measurement of the $g_P$ from low energy pion electroproduction has been reported as
\[g_P(q^2) = 0.082 \pm 0.018\;m_N/\textrm{MeV},\;\;at\;\; q^2 = 0.012\;\textrm{GeV}^2.\]
It is claimed that the measurements are consistent with the pion pole dominance in the region of small $q^2$
\[g_P(q^2) \sim (1 - q^2/m^2_\pi).\]
Using this form factor, it is obtained 
\[g_P(0) =  2.33 \pm 0.51 .\]

The value of the $g_P$ (223) obtained in this model is smaller than these two values obtained in Refs. [78, 95]. 
On the other hand, the $g_P(q^2)$ obtained in this model (222) in the region of small $q^2$ is expressed as
\[1 - q^2/0.286 .\]
The constant 0.286 is more than 10 times of the $m^2_\pi$. Therefore, the pion pole of the $g_P(q^2)$ is not revealed from this model.
There is an issue about satisfaction of PCAC in this model. On the other hand,
in this model $g_P$ is resulted in antiquark components of nucleon. The antiquark components of nucleon can come from pion clouds.
It is interesting to study the relationship between the factor, a - 1 (223) and the pion clouds. 
This study is beyond the scope of this paper. Phenomenologically,
the contributions of the $g_P$ to the semileptonic decays and the cross sections of the quasielastic neutrino scatterings are small.
   
\subsection{Semileptonic decays of baryons} 
For the semileptonic decays of ${1\over2}^+$ baryons the transfer momenta are very small and only the $g_A(0)$
and the $f_V(0)$ contribute to the decays. After inputting the value of the $\lambda$, the $g_A(0)$ is obtained and and the $f_V(0)$ 
is found from Table 2. 
The Cabbibo angle is chosen to be 
\[\theta=0.22.\]
The branching ratios are of these semileptonic decays are predicted. The results and comparisons 
with data [46] are shown in Table 3.

%In the calculation of the decay rates the mass difference between proton and neutron is neglected.
%Under this condition the scalar and pseudoscalar terms are ignored.
%Using Eq.(),
%\begin{eqnarray}
%\lefteqn{G_A(q^2)=(1+a\tau)D_2(q^2),}\\
%&&J(q^2)=\{1+{5\over2}(\mu_p-1)+{3\over2}\kappa a\tau\}D_2(q^2).
%\end{eqnarray}
%$D_2(q^2)$ is shown in Eq.().

%Theoretical results of the ${G_A\over G_V}$ and the branching ratios of semileptonic 
%decays of baryons are obtained(Table2).
\newpage
\begin{table}[h]
\begin{center}
\caption{Theoretical values of R and ${G_A\over G_V}$}
\begin{tabular}{|c|c|c|c|c|} \hline
Process&BR(theory)&BR(exp.)&${G_A\over G_V}(th)$&${G_A\over G_V}(exp)$\\ \hline
$n\rightarrow pe^-\bar{\nu}$&$0.892\times10^3$s&$(.8801\pm0.0011)\times10^3 s$
&input&1.25\\ \hline
$\Sigma^+\rightarrow\Lambda e^+\nu$&$1.92\times10^{-5}$&$(2.0\pm0.5)\times10^{-5}$&0(here is ${G_V\over G_A}$)&
\\ \hline
$\Sigma^-\rightarrow\Lambda e^-\bar{\nu}$&$0.59\times10^{-4}$&$(0.573\pm0.027)
\times10^{-4}$&0(here is ${G_V\over G_A})$&$0.01\pm0.10$(${G_V\over G_A}$)\\ \hline
$\Sigma^-\rightarrow\Sigma^0 e^-\bar{\nu}$&$1.43\times10^{-10}$&&0.50&\\ \hline
$\Xi^-\rightarrow\Xi^0 e^-\bar{\nu}$&$2.32\times10^{-10}$&$<2.3\times10^{-3}$&-0.25 &\\ \hline
$\Lambda\rightarrow pe^-\bar{\nu}$&$8.79\times10^{-4}$&$(8.32\pm0.14)\times10^{-4}$&0.75
&$0.718\pm0.015$\\ \hline
$\Lambda\rightarrow p\mu^-\bar{\nu}$&$1.51\times10^{-4}$&$(1.57\pm0.35)\times10^{-4}$&0.75&$0.718\pm0.015$\\ \hline
$\Sigma^-\rightarrow ne^-\bar{\nu}$&$1.01\times10^{-3}$&$(1.017\pm0.034)\times10^{-3}$
&-0.25&-$(0.34\pm0.017)$\\ \hline
$\Sigma^-\rightarrow n\mu^-\bar{\nu}$&$0.48\times10^{-3}$&$(0.45\pm0.04)\times10^{-3}$
&-0.25& \\ \hline
$\Xi^-\rightarrow\Lambda e^-\bar{\nu}$&$0.52\times10^{-3}$&$(0.563\pm0.031)\times10^{-3}$
&0.25&$0.25\pm0.05$\\ \hline
$\Xi^-\rightarrow\Lambda\mu^-\bar{\nu}$&$0.15\times10^{-3}$&$(0.35^{+0.35}_{-0.22})\times10^{-4}$&0.25
&\\ \hline
$\Xi^-\rightarrow\Sigma^0 e^-\bar{\nu}$&$0.42\times10^{-4}$&$(0.87\pm0.17)
\times10^{-4}$&
0.92& \\ \hline
$\Xi^-\rightarrow\Sigma^0\mu^-\bar{\nu}$&$0.54\times10^{-6}$&$<0.8\times10^{-3}$&0.92
& \\ \hline
$\Xi^0\rightarrow\Sigma^+ e^-\bar{\nu}$&$2.5\times10^{-4}$&$(2.53\pm0.08)\times10^{-4}$&
1.24&\\ \hline
$\Xi^0\rightarrow\Sigma^+\mu^-\bar{\nu}$&$0.22\times10^{-5}$&$<(4.6^{+1.8}_{-1.4})\times10^{-6}$&1.24
& \\ \hline
\end{tabular}
\end{center}
\end{table}
\newpage 
Table 3 shows that theoretical predictions of the branching ratios and ${G_A\over G_V}$ (or  ${G_V\over G_A}$ in some cases)
are in good agreement with data.

It is interesting to notice (see Table 2) that the vector form factors of  $\Sigma^{\pm}\rightarrow \Lambda$ 
are expressed as
\begin{equation}
G_V(q^2) = {q^2\over 2\sqrt{6}m_\Lambda m_\Sigma}(1 + \kappa\frac{m_\Lambda + m_{\Sigma}}{2 m_p}a_\Lambda a_\Sigma) D_1(q^2),
\end{equation}
where $a_\Lambda = (1 - {m_0\over m_\Lambda})^{-1}$, and  $a_\Sigma = (1 - {m_0\over m_\Sigma})^{-1}$.
Therefore, at $q^2 = 0$ the vector form factors for both processes are equal to zero 
\begin{equation}
{G_V\over G_A}= 0. 
\end{equation}
This prediction originates in the condition of current 
conservation (62). In 70's the experimental data was $-0.37\pm0.20$. The newer data
is $0.01\pm0.10$ [46]. Theory agrees with data very well.
As a matter of fact in both  $\Sigma^{\pm}\rightarrow \Lambda + e^+(e^-) + \nu (\bar{\nu})$ decays
transfer momenta are very small and the maximum of $q^2$ are $0.55\times 10^{-2}\;\textrm{GeV}^2,\;0.67\times 10^{-2}\;\textrm{GeV}^2$
respectively. For the $q^2_{max}$ 
\[G_V(q^2_{max}) = 0.164 \times 10^{-2},\;\;0.199\times 10^{-2}\]
respectively for both decays.
Therefore, for both decays
\begin{equation}
{G_V\over G_A} \sim 0.
\end{equation}
These predictions are in good agreements with data.

The coefficient between lepton and neutrino, $\alpha _{e\nu}$, and the asymmetric parameters, $\alpha_e,\;\alpha_\nu,\;\alpha_B$,
are predicted by this model (see Table 4).
\begin{table}[h]
\begin{center}
\caption{Theoretical values of coefficients}
\begin{tabular}{|c|c|c|c|c|} \hline
process&$\alpha_{e\nu}$&$\alpha_e$&$\alpha_\nu$&$\alpha_B$\\ \hline
$\Sigma^+\rightarrow \Lambda$&-0.40&-0.70&0.68&0.06\\ \hline
$\Sigma^-\rightarrow \Lambda$&-0.40&-0.70&0.68&0.06\\ \hline
$\Sigma^-\rightarrow\Sigma^0$&0.43&0.28&0.85&-0.71\\ \hline
$\Xi^-\rightarrow\Xi^0$&0.789&-0.52&-0.31&).53\\ \hline
$\Lambda\rightarrow p$&0.0058&0.031&0.97&-0.61 \\ \hline
$\Sigma^-\rightarrow n$&0.56&-0.58&-0.46&0.45\\ \hline
$\Xi^-\rightarrow \Lambda$&0.63&0.26&0.50&-0.50\\ \hline
$\Xi^-\rightarrow\Sigma^0$&-0.35&-0.39&0.92&-0.30\\ \hline
$\Xi^0\rightarrow\Sigma^+$&-0.20&-0.17&0.99&-0.48\\ \hline
\end{tabular}
\end{center}
\end{table}
In Table 5 the related data [76] of the process $\Lambda\rightarrow p e^- \bar{\nu}$
are listed.
%\end{document}

\begin{table}[h]
\begin{center}
\caption{Experimental values of coefficients of $\Lambda\rightarrow p e^- \bar{\nu}$}
\begin{tabular}{|c|c|c|c|c|} \hline
Lab&$\alpha_{e\nu}$&$\alpha_e$&$\alpha_\nu$&$\alpha_p$\\ \hline
Argonne& $-0.08 \pm 0.10$ & $0.09 \pm 0.11$ & $0.75 \pm 0.11$ & $-0.55 \pm 0.11$ \\ \hline
CERN& $-0.07\pm0.09$&$0.15\pm0.09$&$0.89\pm0.08$&$-0.52\pm0.08$\\ \hline
\end{tabular}
\end{center}
\end{table}
The experimental value of $\alpha_e$ of the 
$\Sigma^-\rightarrow n e^- \bar{\nu}$ 
\[\alpha_e = -0.26 \pm 0.37\]
can be found in Ref. [77]. Theoretical results of these coefficients are compatible with the existing data.
In Ref. [87] it has been reported
\[\alpha_{e\nu} = -0.27 \pm 0.013,\;\; G_A(0)/G_V(0) = 0.731 \pm 0.016\]
for the decay $\Lambda \rightarrow p + e^- + \bar{\nu}$.
In Ref. [88] for the decay $\Sigma^- \rightarrow n + e^- + \bar{\nu}$ following quantities have been measured
\[\alpha_e = -0.519 \pm 0.104,\;\;
\alpha_n = 0.509 \pm 0.102,\;\;\alpha_\nu = -0.230 \pm 0.061.\] 
Theoretical values of these quantities (see Tab. 4) are compatible with the measured values.

\section{Charged quasielastic reactions of neutrino and nucleon}
\subsection{$\Delta S=0$ scatterings of neutrino (antineutrino) and nucleon}
There are two $\Delta S=0$ charged quasielastic reactions
\[\nu_\mu+n\rightarrow p+\mu^-,\;\;\;\bar{\nu}_\mu+p\rightarrow n+\mu^+.\]
AS shown in Eqs. (218,219) the $G_A(q^2)$ and the $M_A$ are all predicted in this model.
The vector form factors of these two processes are determined completely and
the cross sections of them can be calculated without any new parameter.
Therefore, the cross sections of these scattering processes are the predictions of this model.\\
{\bf $\nu_\mu+n\rightarrow p+\mu^-$}\\
Using Eq. (201) and ignoring the mass difference between proton and neutron, the matrix element of the charged weak quark current can be written 
\begin{equation}
<p(p_f)|j^W_\mu(0)|n(p_i)>={1\over3}\bar{u}_{c'}(p_f)\{5G^p_M(q^2)\gamma_\mu
+5\lambda G_A(q^2)\gamma_\mu\gamma_5+{i\over m}P_\mu J(q^2)\}u_c(p_i),
\end{equation}
where \(P_\mu=p_{i\mu}+p_{f\mu}\), the form factors $G^p_M(q^2)$ and
$G_A(q^2)$ are shown in Eqs. (95,218) respectively,
\begin{eqnarray}
J(q^2)=D_2(q^2)+{5\over2}\kappa\{D_1(q^2)-D_2(q^2)+(1+{5\over3}\tau)D_3(q^2)\}\nonumber \\
= \{1+{5\over2}(\mu_p-1)+{3\over2}\kappa\; a\;\tau\}D_2(q^2).
\end{eqnarray}
$D_2(q^2)$ is shown in Eq. (112).

The cross section of $\nu_\mu+n\rightarrow p+\mu^-$ is derived as
\begin{equation}
\frac{d\sigma}{dq^2}=\frac{G^2}{9\pi x}cos^2\theta\{{\tau\over x}W_1
+[{x\over2}-\tau(1+{1\over2x})]W_2-{\tau\over x}(x-\tau)W_3\},
\end{equation}
where \(x={E_\nu\over m}\), $E_\nu$ is the energy of neutrino, 
\begin{eqnarray}
\lefteqn{W_1=25\{\tau G^p_M(q^2)^2+(1+\tau)\lambda^2 G^2_A(q^2)\},}\nonumber \\
&&W_2=\{5G^p_M(q^2)-2J(q^2)\}^2+25\lambda^2G^2_A(q^2)+4\tau J^2(q^2),\nonumber \\
&&W_3=-50\lambda G_A(q^2)G^p_M(q^2).
\end{eqnarray}
Using
\begin{equation}
q^2=\frac{4m^2 x^2 sin^2{\alpha\over2}}{1+2xsin^2{\alpha\over2}}
\end{equation}
($\alpha$ is the scattering angle of muon), 
the limits of $q^2$ are found
\begin{equation}
0\leq q^2 \leq\frac{4m^2 x^2}{1+2x}.
\end{equation}
Integrating over $q^2$, the total cross section as a function of $E_\nu$ is obtained.
The comparison between theory and experimental data [83] is shown in Fig. 11. 
\begin{figure}
\begin{center}
\includegraphics[width=7in, height=7in]{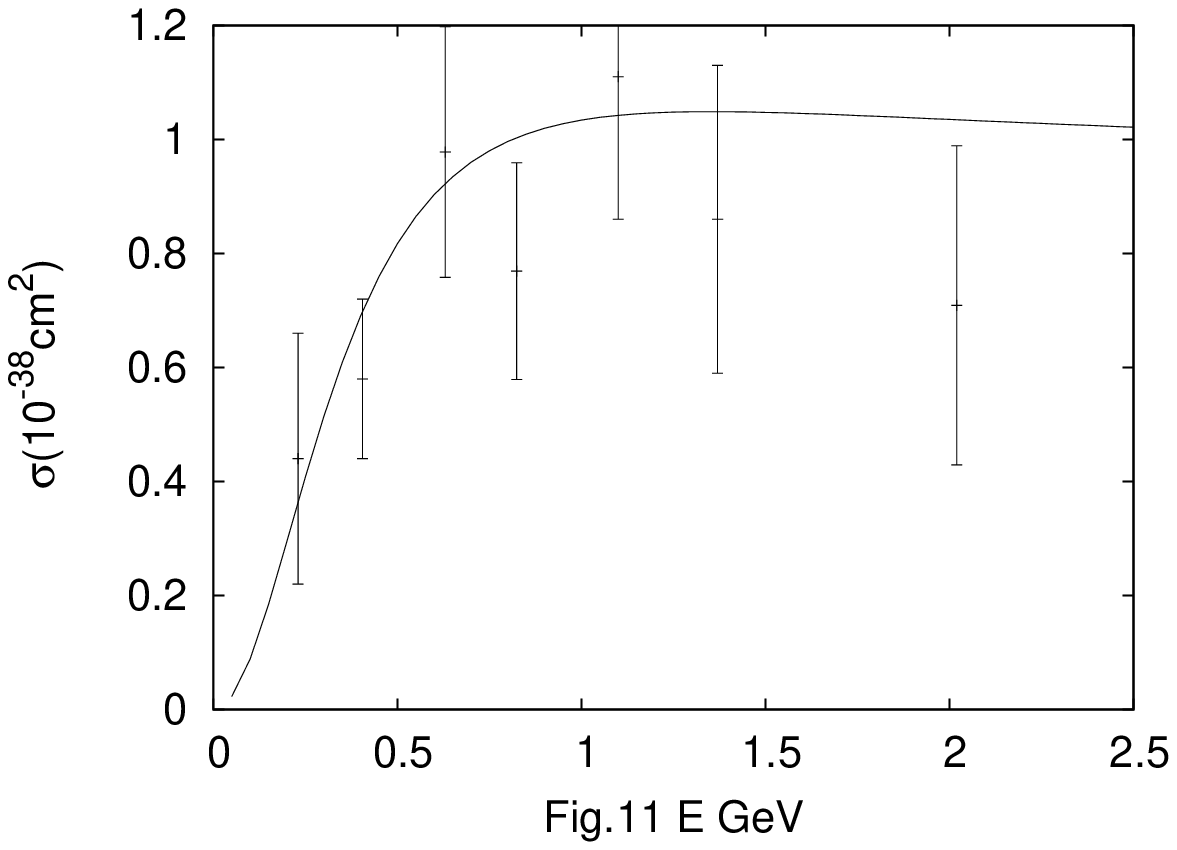}
%FIG. 10.
\end{center}
\end{figure}

The comparison with newer data is shown in Fig. 12
\begin{figure}
\begin{center}
\includegraphics[width=7in, height=7in]{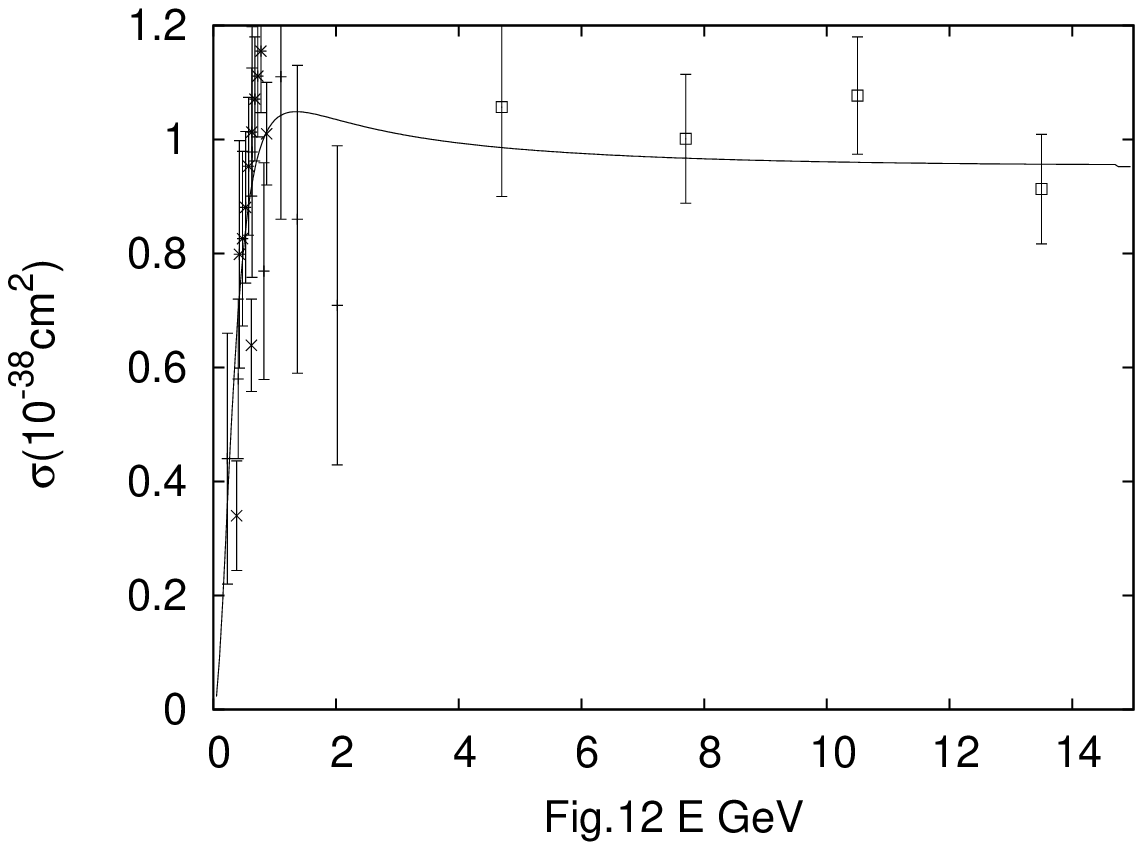}
%FIG. 11.
\end{center}
\end{figure}

The experimental data of Fig. 12 are taken from Refs. [84,85,86].
When $E_\nu > 15 \textrm{GeV}$ theoretical $\sigma(\nu_\mu + n \rightarrow p + \mu^-)$ 
is very flat (see below).
In this calculation there is no new adjustable parameter. Theory
agrees with data very well.

{\bf  $\bar{\nu}_\mu + p\rightarrow n + \mu^+$}\\
The cross section of this process is obtained from Eq. (230) by changing the minus sign
associated with the $W_3$ to plus 
\begin{equation}
\frac{d\sigma}{dq^2}=\frac{G^2}{9\pi x}cos^2\theta\{{\tau\over x}W_1
+[{x\over2}-\tau(1+{1\over2x})]W_2+{\tau\over x}(x-\tau)W_3\},
\end{equation}
where $W_{1,2,3}$ are shown in Eqs. (231). 
The theoretical values of the cross section is shown in Fig. 13. 
\begin{figure}
\begin{center}
\includegraphics[width=7in, height=7in]{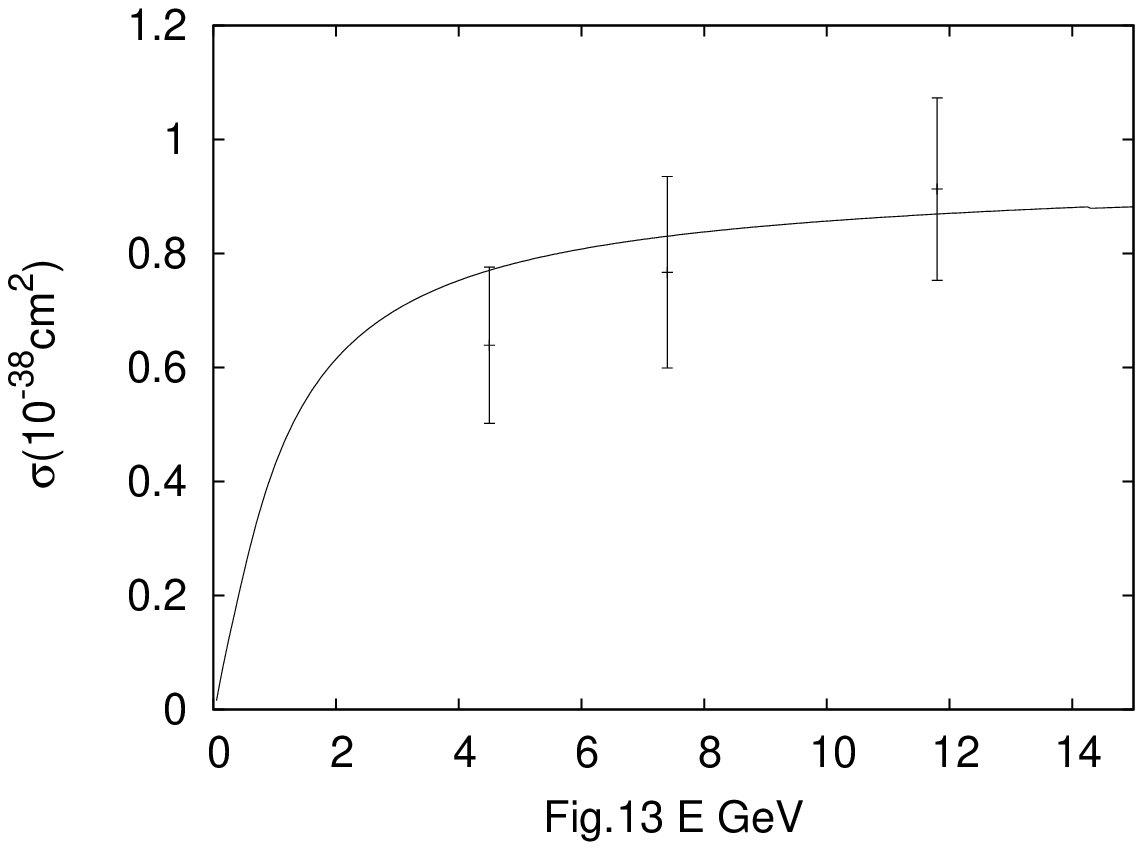}
%FIG. 12.
\end{center}
\end{figure}
The data is taken from Ref. [86]
Theory agrees with the data well.
When $E_{\nu} \rightarrow \infty $ $W_{1,3}$ make no contributions to the cross section.
Therefore, 
\begin{equation}
\lim_{E_\nu\rightarrow \infty} \sigma(\bar{\nu}_\mu + p\rightarrow n+\mu^+) = \lim_{E_\nu\rightarrow \infty} \sigma(\nu + n \rightarrow p + \mu)
= 0.89 \times 10^{-38} cm^2.
\end{equation}

\subsection{$\Delta S=1$ quasielastic reactions of neutrino and nucleon}

{\bf  $\bar{\nu}_\mu+p\rightarrow\Lambda+\mu^+$}\\
The decays $\Lambda \rightarrow p + e^- + \bar{\nu}_e,\;p + \mu^- + \bar{\nu}_\mu$
have been studied and the results are shown in Table 2. Theory agrees with data.
The matrix element of the weak current of this process 
is obtained from Eq. (201) and Table 2
\begin{equation}
<\Lambda(p_f)|J^W_\mu(0)|p(p_i)>={3\over2\sqrt{6}}
\bar{u}_{c'}(p_f)\{G^\Lambda_V(q^2)\gamma_\mu+\lambda G^\Lambda_A(q^2)\gamma_\mu\gamma_5
+{i\over m_+}
G(q^2)P_\mu\}u_c(p_i),
\end{equation}
where \(m_+ = m_\Lambda + m_p\),
\begin{eqnarray}
G^\Lambda_A(q^2)=D_2(q^2)+D'_2(q^2)+\frac{q^2+(m_\Lambda-m_p)^2}
{2m_p m_\Lambda}D_3(q^2),\nonumber \\
G(q^2)=\frac{\kappa(m_\Lambda+m_p)}{2m_p}\{2D_1(q^2)-D_2(q^2)-D'_2(q^2)
+\frac{q^2+(m_\Lambda+m_p)^2}{2m_p m_\Lambda}D_3(q^2)\},\nonumber \\
G^\Lambda_V(q^2)=G^\Lambda_A(q^2)+G(q^2).
\end{eqnarray}
According to Eq.(77), there are
\begin{equation}
D_2(q^2)=a_p D_1(q^2),\;\;\;D'_2(q^2)=a_\Lambda D_1(q^2),\;\;\;
D_3(q^2)=a_p a_\Lambda D_1(q^2),
\end{equation}
where $a_p$ and $a_\Lambda$ are obtained from Eq.(79). Using Eq. (238), we have
\begin{eqnarray}
\lefteqn{G^\Lambda_A(q^2)=D_1(q^2)\{a_p+a_\Lambda+\frac{q^2+(m_\Lambda-m_p)^2}
{2m_p m_\Lambda}a_p a_\Lambda\},}\nonumber \\
&&G(q^2)=\frac{\kappa(m_\Lambda+m_p)}{2m_p}D_1(q^2)\{2-a_p-a_\Lambda+
\frac{q^2+(m_\Lambda+m_p)^2}{2m_p m_\Lambda}a_p a_\Lambda\}.
\end{eqnarray}
All the form factors of this process are determined.

The experimental estimation of the parameter $M_A$ of the axial-vector form factor
(239)  is $0.6\pm0.2\;\textrm{GeV}$ [83]. Eq.(239) shows that the axial-vector form  
factor $G^\Lambda_A(q^2)$ decreases faster than the axial-vector form factor (218) of 
$\nu_\mu+n\rightarrow p + \mu^-$. Comparing with the dipole form factor, for $\Lambda$
production 
\[M_A=0.865\; \textrm{GeV}.\]
is determined.
It is compatible with the experimental estimation.

The cross section of $\bar{\nu}_\mu+p\rightarrow\Lambda+\mu^+$ is 
written as
\begin{equation}
\frac{d\sigma}{d q^2}=\frac{3G^2 sin^2\theta}{8\pi m^2_p x^2}\{\tau W_1+[
x({x\over2}-\frac{m^2_\Lambda-m^2_p}{4m^2_p})-\tau({1\over2}+x)]W_2
-\tau(\tau-x+\frac{m^2_\Lambda-m^2_p}{4m^2_p})W_3\},
\end{equation}
where $x = {E_\nu\over m_p}$, 
\begin{eqnarray}
\lefteqn{W_1={1\over4m^2_p}\{[q^2+(m_\Lambda-m_p)^2]G^\Lambda_V(q^2)^2
+\lambda^2[q^2+(
m_\Lambda+m_p)^2]G^\Lambda_A(q^2)^2\},}\nonumber \\
&&W_2=(1+\lambda^2)[G^\Lambda_A(q^2)^2+\frac{q^2}{(m_\Lambda+m_p)^2}G^2(q^2)],
\nonumber \\
&&W_3=-2\lambda G^\Lambda_A(q^2) G^\Lambda_V(q^2).
\end{eqnarray}
$q^2$ is expressed as
\begin{equation}
q^2=\frac{2x(2m^2_p x+m^2_p-m^2_\Lambda)sin^2{\alpha\over2}}{1+2xsin^2{\alpha\over2}}.
\end{equation}
The limits of $q^2$ is determined as
\begin{equation}
0\leq q^2 \leq\frac{4m^2_p x(x-\frac{m^2_\Lambda-m^2_p}{2m^2_p})}{1+2x}.
\end{equation}
The lower limit of the energy of neutrino
is 
\begin{equation}
E_\nu > \frac{m^2_\Lambda - m^2_p}{2 m_p}.
\end{equation}
The total cross section is obtained and shown in Fig.14. 
In the calculation there is no new adjustable parameter. In the range of neutrino 
energies from 0.9-3.3 GeV the experimental value of the cross section [83] is
\begin{equation}
\sigma = (1.3^{+0.9}_{-0.7})\times 10^{-40}\;cm^2/proton.
\end{equation}

\begin{figure}
\begin{center}
\includegraphics[width=7in, height=7in]{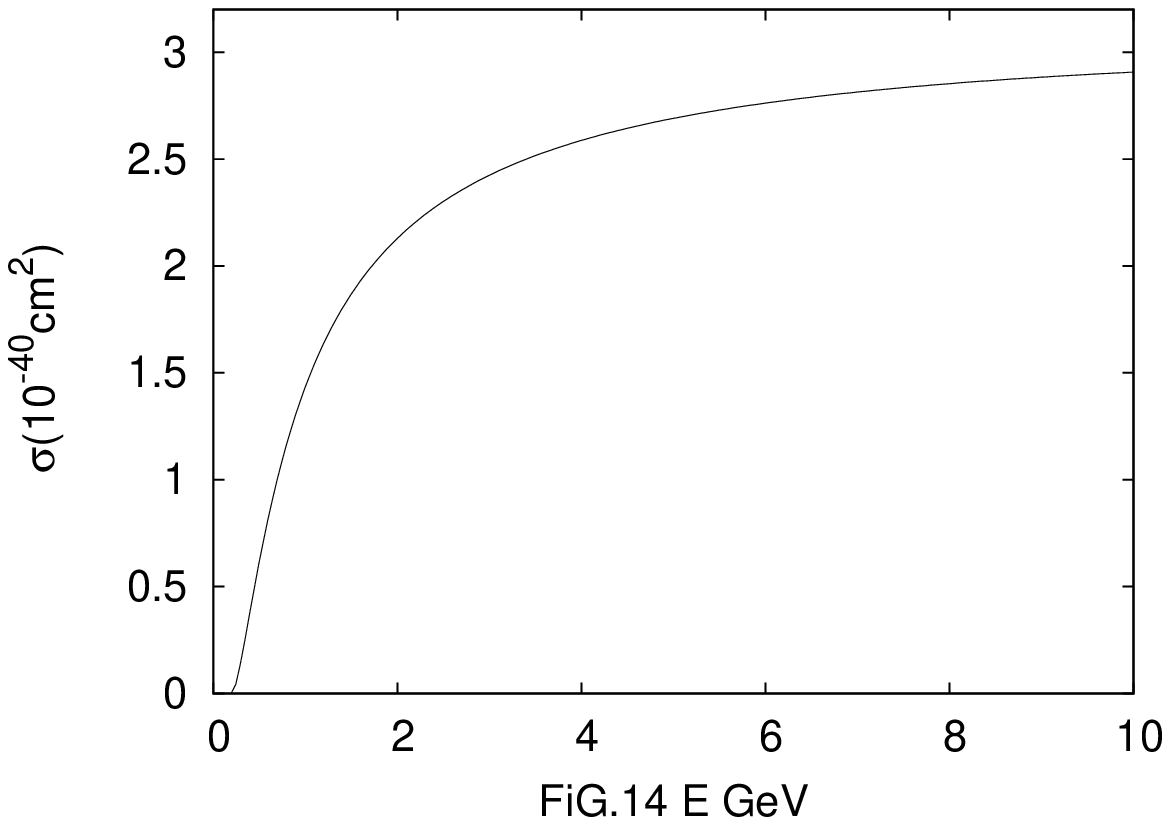}
%FIG. 14.
\end{center}
\end{figure}
It is at the order of $10^{-40} cm^2$ which is smaller than the cross section
of $\nu_\mu + n \rightarrow p +\mu^-$ by two order of magnitude. The reason is that this cross section
is $\propto sin^2\theta$. 

When $ E_\nu \rightarrow \infty$ the cross section approaches a constant
\begin{equation}
\lim_{E_\nu \rightarrow \infty}\sigma = 2.69 \times ^{-40}\; cm^2.
\end{equation}
Theory is compatible with data.

{\bf $\bar{\nu}_\mu+p\rightarrow\Sigma^0+\mu^+$}\\
The matrix element of this process is obtained from Eq. (201)
\begin{equation}
<\Sigma^0_{c'}(p_f)|J^W_\mu(0)|p_c(p_i)>=\frac{3}{2\sqrt{6}} 
\bar{u}_{c'}(p_f)\{G^\Sigma_V(q^2)\gamma_\mu+\lambda G^\Sigma_A(q^2)
\gamma_\mu\gamma_5+{i\over m_\Sigma}G^{\Sigma}(q^2)P_\mu\}u_c(p_i),
\end{equation}
where
\begin{eqnarray}
G^\Sigma_V(q^2)=\frac{1}{3\sqrt{3}}\{D_2(q^2)+D'_2(q^2)+
\frac{q^2+(m_\Sigma-m_p)^2}{2m_p m_\Sigma}D_3(q^2)\nonumber \\
+\frac{\kappa(m_\Sigma+m_p)}{2m_p}[2D_1(q^2)-D_2(q^2)-D'_2(q^2)
+\frac{q^2+(m_\Sigma-m_p)^2}{2m_p m_\Sigma}D_3(q^2)]\},\nonumber \\
= (0.988 + 1.6 \tau)/(1 + 2.39 \tau)/(1 + 4.96 \tau)^2,\\
G^{\Sigma}_{A}(q^2)={1\over 3\sqrt{3}}\{D_2(q^2)+D'_2(q^2)+
\frac{q^2+(m_\Sigma-m_p)^2}{2m_p m_\Sigma}D_3(q^2)\}\nonumber \\
= (0.42 + 1.03 \tau))/(1 + 2.39 \tau)/(1 + 4.96 \tau)^2,\\
G^\Sigma(q^2)={2\over 3\sqrt{3}}\{D'_2(q^2)+{m_\Sigma\over m_p}D_2(q^2)
-\frac{\kappa q^2}{2m^2_p}D_3(q^2)+\frac{\kappa m_\Sigma}{4m_p}[2D_1(q^2)-D_2(q^2)
-D'_2(q^2)\nonumber \\
+\frac{q^2+(m_\Sigma-m_p)^2}{2m_p m_\Sigma}D_3(q^2)]\}\nonumber \\
= (1.26 - 0.95\tau)/(1 + 2.39 \tau)/(1 + 4.96 \tau)^2.
\end{eqnarray}

Using the substitution $m_\Lambda\rightarrow m_\Sigma$ in Eq. (240), the cross section
of $\bar{\nu_\mu}+p\rightarrow\Sigma^0+\mu^+$ is obtained and 
the $W_{1,2,3}$ of this process are defined as
\begin{eqnarray}
\lefteqn{W_1={1\over4m^2_p}\{[q^2+(m_\Sigma-m_p)^2]G^\Sigma_V(q^2)^2+\lambda^2[q^2+(
m_\Sigma+m_p)^2]G^\Sigma_A(q^2)^2\},}\nonumber \\
&&W_2=[\frac{m_\Sigma+m_p}{m_\Sigma}G^\Sigma q^2)-G^\Sigma_V(q^2)]^2+\lambda^2
G^\Sigma_A(q^2)^2+{q^2\over m^2_\Sigma}G^\Sigma(q^2)^2,
\nonumber \\
&&W_3=-2\lambda G^\Sigma_A(q^2) G^\Sigma_V(q^2).
\end{eqnarray}
There is no new adjustable parameter. 
When $E_\nu\rightarrow \infty$ the cross section approaches a constant
\begin{equation}
\lim_{E_\nu\rightarrow \infty} \sigma = 0.38 \times 10^{-40}\; cm^2.
\end{equation}

The numerical results are shown in Fig.15.
\begin{figure}
\begin{center}
\includegraphics[width=7in, height=7in]{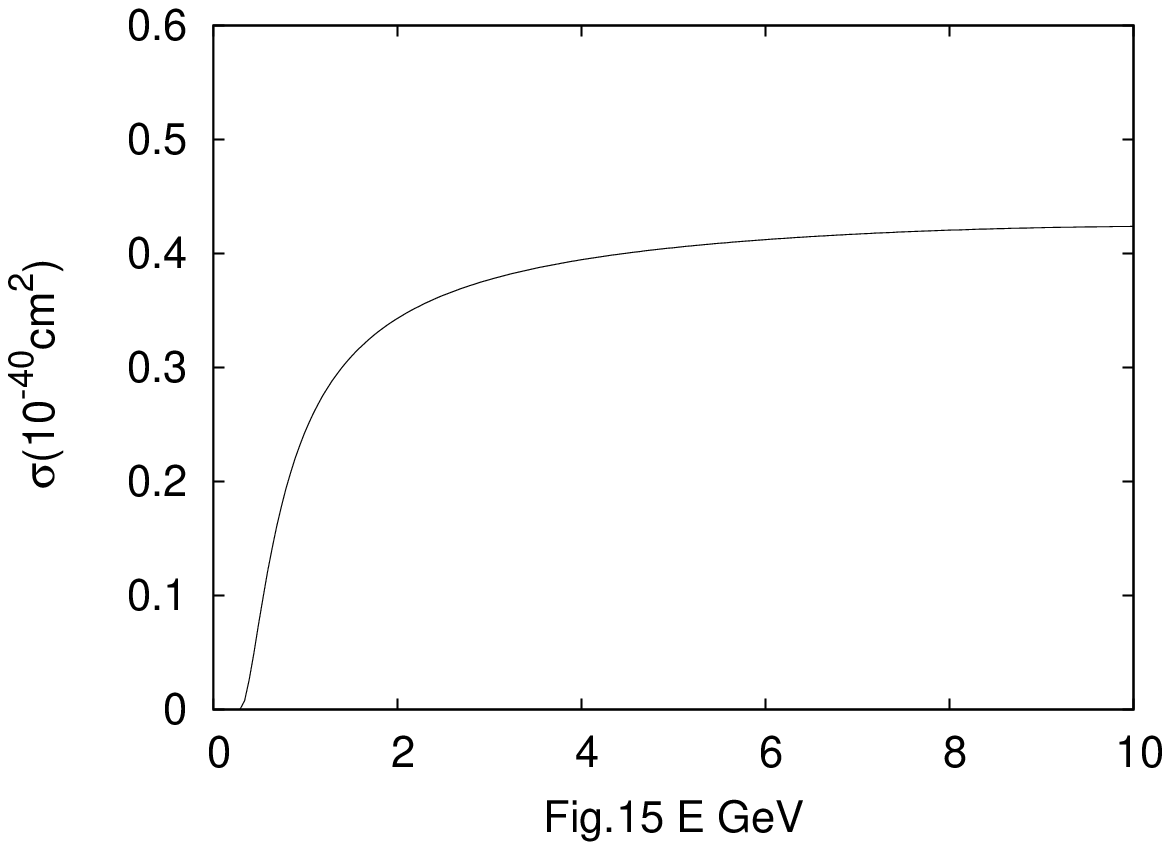}
%FIG. 15.
\end{center}
\end{figure}
Comparing Fig.15 with Fig.14, it is found
\[ \sigma(\bar{\nu}_\mu+p\rightarrow\Sigma^0+\mu^+)\sim 
{1\over 6}\sigma(\bar{\nu}_\mu+p\rightarrow\Lambda+\mu^+).\]

In the experiment of quasielastic hyperon production by antineutrino [83] 10 $\Lambda$ and
2 $\Sigma^{0}$ are found. Theory is compatible with data.

\section{$\nu_\mu + p\rightarrow\Delta^{++} + \mu^-$ scattering}
Using the same approach, the $\nu_\mu+p\rightarrow\Delta^{++} + \mu^-$ scattering process has been studied in Ref. [7].
In this section the review of the study done in Ref. [7] is presented.
\subsection{Form factors of $p\rightarrow \Delta^{++}$}
The S-matrix element of neutrino production of $\Delta$ is written as
\begin{equation}
<\mu^-\Delta^{++}|S|\nu_\mu p>=-i(2\pi)^4\delta^4(p_i+p_\nu-p_f-p_\mu)
{G\over\sqrt{2}}cos\theta<\mu^-|j_\mu(0)|\nu_\mu><\Delta^{++}|J^W_\mu(0)|p>.
\end{equation}
The hadronic matrix element is expressed as
\begin{eqnarray}
<\Delta^{++}|J^W_\mu(0)|p>=-{i\over 2}\int d^4x'd^4y'd^4xd^4y\bar{B}^{{3\over2}
\lambda'}_{\alpha\beta\gamma}(x',y',0)^{i'j'k'}_{ij1,111}M(y',x',x,y)\nonumber \\
\{\gamma_\mu(1+\lambda\gamma_5)+{\kappa\over2m_p}\sigma_{\mu\nu}q_\nu\}
_{\gamma\gamma'}B^{{1\over2}\lambda}_{\gamma'\beta\alpha}(0,y,x)^{k'j'i',3}_{2ji,1}.
\end{eqnarray}
Using the wave functions(33,46,47), the matrix element (254) is expressed as 
\begin{eqnarray}
\lefteqn{<\Delta^{++}|J^W_\mu(0)|p>={1\over4}D_3(q^2)
\bar{\psi}^{\lambda'}_{\sigma}(p')\{{A\over mm'}P_\rho q_\nu\epsilon_{\rho\nu\sigma\mu}
+{4B\over mm'}(p'_\mu q_\sigma-p'\cdot q\delta_{\mu\nu})\gamma_5}\nonumber \\
&&+4\lambda[C\delta_{\sigma\mu}-{1\over mm'a'}(p'\cdot q\delta_{\sigma\mu}
-p'_\mu q_\sigma)]+{2\lambda D\over mm'}p_\rho q_\nu\epsilon_{\rho\nu\sigma\mu}
\gamma_5\}u_\lambda(p),
\end{eqnarray}
where m, E, $m'$, $E'$ are the masses and energies of proton and $\Delta$ respectively,
\begin{eqnarray}
\lefteqn{A={2\over a'}+\kappa\{{2\over aa'}-{1\over a}-{1\over a'}+1+{m'\over m}\},}
\nonumber \\
&&B=1-{1\over a'}+{\kappa\over 2}({1\over a}+{1\over a'}-{2\over aa'}),\nonumber \\
&&C={1\over a}+{m'\over m}{1\over a'},\nonumber \\
&&D=1-{1\over a'},
\end{eqnarray}
where a and $a'$ are the proportional constants of proton and $\Delta$ respectively,
\(P_\mu=p_\mu+p'_\mu\), $D_3(q^2)$ is given by Eq. (187), A and B are the same as Eqs. (156).
Eq.(256) shows that the coefficients B and D are resulted in the effects of antiquarks of nucleon. 
The conservation of the vector current is satisfied.
  
According to Ref. [7], there are 8 form factors 
\begin{eqnarray}
T\equiv {G\over \sqrt{6}}<\Delta^{++}|J^W_\mu(0)|p>
<\mu^-|j_\mu(0)|\nu_\mu>\nonumber \\
={G\over\sqrt{2}}cos\theta\bar{\psi}(p')_\alpha\{-[{1\over m}G^V_3(q^2)\gamma_\mu
+{1\over m^2}G^V_4(q^2)p'_\mu+{1\over m^2}G^V_5(q^2)p_\mu]\gamma_5F^{\mu\nu}
\nonumber \\
G^V_6(q^2)j^\alpha\gamma_5-[{1\over m}G^A_3(q^2)\gamma_\mu+{1\over m^2}G^A_4(q^2)
p'_\mu]F^{\mu\nu}+G^A_5(q^2)j^\alpha+{1\over m^2}G^A_6(q^2)q_\alpha q\cdot j\}u(p),
\end{eqnarray}
where 
\begin{eqnarray}
\lefteqn{j_\alpha=<\mu^-|j_\alpha(0)|\nu>,}\nonumber \\
&&F^{\mu\nu}=q_\mu j_\nu-q_\nu j_\mu,
\end{eqnarray}
p and $p^\prime$ are the momenta of the proton and the $\Delta$ respectively,
$q = p - p^\prime$.
Comparing with Eq. (255), it is obtained
\begin{eqnarray}
\lefteqn{G^V_3(q^2)=\frac{A}{2\sqrt{3}}D_3(q^2),}\\ 
&&G^V_4(q^2)=-{1\over2\sqrt{3}}{m\over m'}(A-2B)D_3(q^2), \\
&&G^V_5(q^2)= G^V_6(q^2)=0, \\
&&G^A_3(q^2)=\frac{\lambda D}{\sqrt{3}}D_3(q^2),
\\
&&G^A_4(q^2)=-\frac{\lambda}{\sqrt{3}}({1\over a'}-D){m\over m'}D_3(q^2),\\
&&G^A_5(q^2)={1\over\sqrt{3}}\lambda CD_3(q^2),\\
&&G^A_6(q^2)=0.
\end{eqnarray}
Three axial-vector form factors are predicted. The $G^A_5(q^2)$ has been mentioned in Refs. [85,86]. 
$G^A_5(0)$ is computed
\begin{equation}
G^A_5(0)={1\over \sqrt{3}} \lambda C \sqrt{a a'} = 1.09.
\end{equation}
In Ref.[85] many values of $G^A_5(0)$ have been listed. Using PCAC, 
\[G^A_5(0) = \frac{g_\Delta f_\pi}{2\sqrt{3} M} = 1.2\]
is obtained [85] in the limit $m_\pi\rightarrow 0$ and 
\[G^A_5(0) = 0.84,\;1.07,\;1.9,\]
are presented too from different approaches [85]. From Eq. (264) the $G^A_5(q^2)$ is expressed as
\begin{equation}
G^A_5(q^2) = G^A_5(0) (1 + 2.39 \tau)^{-1}(1 + {q^2\over 0.71})^{-2}.
\end{equation}
In Refs. [85,86] a dipole expression 
\begin{equation}
G^A_5(q^2) = G^A_5(0)(1 + {q^2\over M^2_A})^{-2}
\end{equation}
has been applied to fit the data of the cross section of $\nu + p\rightarrow \mu^- + \Delta^{++}$
and the parameter $M_A$ are determined to be
\[ M_A = 0.92 \pm 0.14\; \textrm{GeV},\;\;M_A = 0.84 \pm 0.15 \;\textrm{GeV},\]
and 
\[M_A = 0.98\; \textrm{GeV},\;\;0.95\; \textrm{GeV}\]
respectively.
Obviously, unlike the form of dipole [85,86] the axial-vector form factor of $p\rightarrow \Delta^{++}$ (267) predicted in this model 
takes the form of triple pole and it is very different from the one of the dipole (268).
However, when $q^2$ is small the parameter $M_A$ is determined from Eq. (267) to be
\begin{equation}
M_A = 0.76\; \textrm{GeV}.
\end{equation}
Besides the $G^A_5(q^2)$ this model predicts other two axial-vector form factors for the transition
$p\rightarrow \Delta^{++}$ 
\begin{eqnarray}
G^A_3(q^2) = -0.8434 (1 + 2.39 \tau)^{-1}(1 + {q^2\over 0.71})^{-2},\\
G^A_4(q^2) = 0.1965 (1 + 2.39 \tau)^{-1}(1 + {q^2\over 0.71})^{-2}.
\end{eqnarray}
Therefore, in this model the $G^A_3(q^2)$ and the $G^A_5(q^2)$ are the two major axial-vector form factors 
for
the transition $p\rightarrow \Delta^{++}$. The two vector form factors of this process (259,260) are
determined to be
\begin{eqnarray}
G^V_3(q^2) = 1.645 (1 + 2.39 \tau)^{-1}(1 + {q^2\over 0.71})^{-2},\\
G^V_4(q^2) = -0.2323 (1 + 2.39 \tau)^{-1}(1 + {q^2\over 0.71})^{-2},\\
G^V_4(q^2) =-0.141 G^V_3(q^2).
\end{eqnarray}
The $G^V_3(q^2)$ is the major vector form factor of this process and it is the 
magnetic form factor. This model predicts that two major axial-vector and one vector form factors 
contribute to the process $\nu + p\rightarrow \mu^- + \Delta^{++}$. These three form factors are
in the forms of triple poles.

\subsection{Cross section}
Using Eq.(255), the differential cross section is written as
\begin{equation}
\frac{d^2\sigma}{d\Omega dE'}=\frac{G^2}{512\pi^3}cos^2\theta E^{'2}
\tau_{\mu\nu}W_{\mu\nu}\frac{\Gamma(m')}{(m'-m_\Delta)^2+{1\over4}\Gamma^2(\Delta)},
\end{equation}
where
\begin{eqnarray}
\lefteqn{\tau_{\mu\nu}W_{\mu\nu}=\frac{2}{3mm'\epsilon\epsilon'}D^2_3(q^2)
\{4m^2 q^2 W_1+[4m\epsilon(2m\epsilon+m^2-m^{'2})-2mq^2(m+2\epsilon)]W_2}\nonumber \\
&&-q^2(q^2-4m\epsilon+m^{'2}-m^2)W_3\}\\
%+{1\over2}q^2 m^2_\mu W_4+pq m^2_\mu W_5\},\\
&&W_1={1\over m^2}(q^2+m^2_+)(4\lambda^2 C^{'2}+{A^2\over m^2}q^{*2})+{4\over m^2}
(q^2+m^2_-)[B^2\frac{p'q)^2}{m^2 m^{'2}}+\frac{\lambda^2 D^2}{m^2}q^{*2}]
\nonumber \\
&&+4{q^{*2}\over m^2}(AB{p'q\over m^2}-2\lambda^2 C' D{m'\over m}),\nonumber \\
&&W_2=\frac{q^2+m^2_+}{mm'}\{4\lambda^2[{m\over m'}C^{'2}+{2C'\over a'}
{p'q\over m^{'2}}+{q^{*2}\over mm' a'}]+{A^2\over mm'}q^2\}\nonumber \\
&&+\frac{4q^2}{m^2 m^{'2}}(q^2+m^2_-)(B^2+\lambda^2 D^2)+{4q^2\over mm'}
[AB{p'1\over mm'}-2\lambda^2 C'D],\nonumber \\
&&W_3=8\lambda A(\frac{q^2+M^2_+}{mm'}C'-{q^{*2}\over m^2}D)+16\lambda B
{p'q\over mm'}(C'-D\frac{q^2+m^2_-}{mm'}),\nonumber \\
%&&W_4=\frac{q^2+m^2_+}{mm'}\{{A^2\over mm'}p^2+4\lambda^2[{m\over m'}C^{'2}
%+{2C'\over a'}({p'q\over m^{'2}}-2)+\frac{q^{*2}}{mm' a^{'2}}]\}\nonumber \\
%&&+4\frac{q^2+m^2_-}{mm'}\{\lambda^2 D^2{p^2\over mm'}+4B^2({q^2\over mm'}
%+{4p'q\over mm'})\}\nonumber \\
%&&+4{p^2\over mm'}(AB{p'q\over mm'}-2\lambda^2C'D),\nonumber \\
%&&W_5=-\frac{q^2+m^2_+}{mm'}\{4\lambda^2[{m\over m'}C^{'2}
%+{2C'\over a'}({p'q\over m^{'2}}-1)+\frac{q^{*2}}{mm' a^{'2}}]+A^2{p'q\over mm'}\}
%\nonumber \\
%&&-4\frac{q^2+m^2_-}{mm'}\{\lambda^2 D^2{p'q\over mm'}+B^2(2{p'q\over mm'}
%+{q^2\over mm'}\}\nonumber \\
%&&-4{p'q\over mm'}(AB{p'q\over mm'}-2\lambda^2 C' D),
\end{eqnarray}
$\epsilon$ is the energy of neutrino, \(m^{'2}=(p+p_\nu-p_\mu)^2\), $\Gamma(m')$ 
is the decay width of $\Delta$. 
Using Eq. (257), the cross sections and the differential cross section of $\nu + p\rightarrow \mu^- + \Delta^{++}$
are calculated. The comparison between theoretical results and the experimental data are shown in 
Fig. 16, 17, 18, 19.

\begin{figure}
\begin{center}
\includegraphics[width=7in, height=7in]{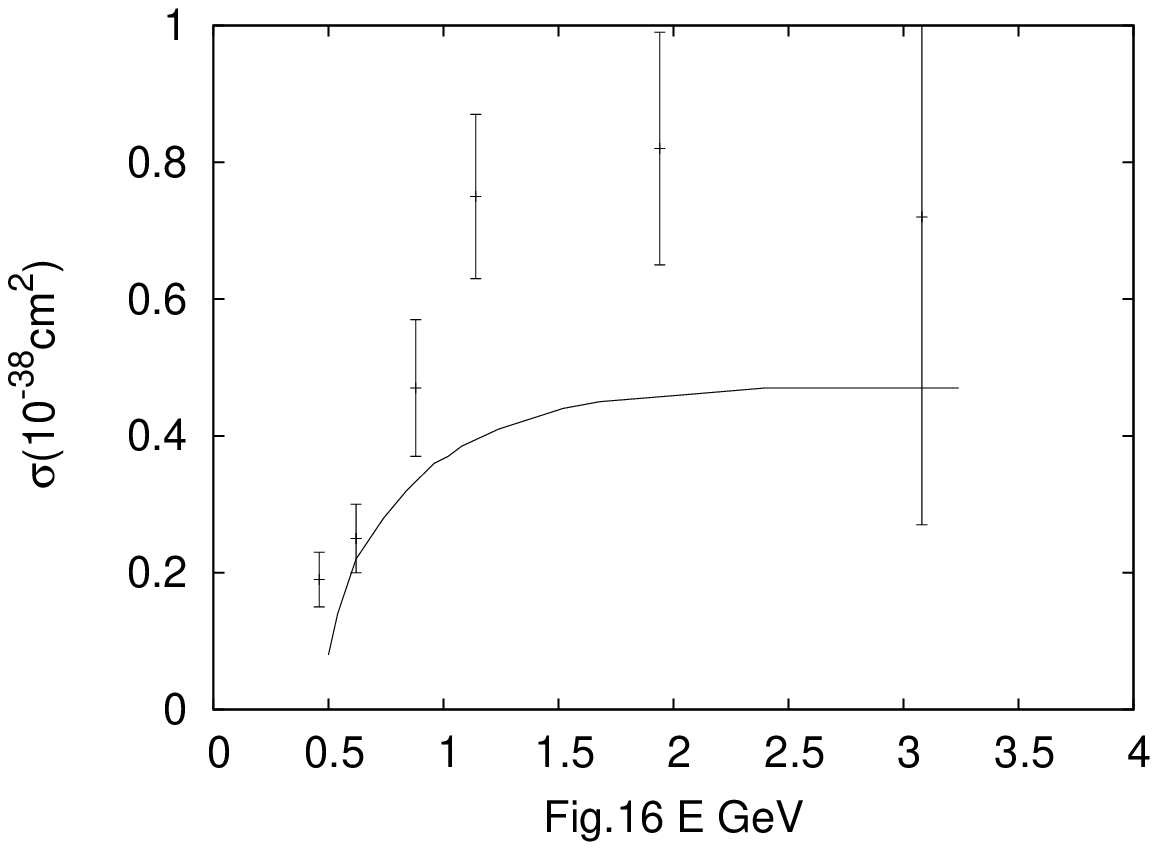}
%FIG. 16.
\end{center}
\end{figure}

\begin{figure}
\begin{center}
\includegraphics[width=7in, height=7in]{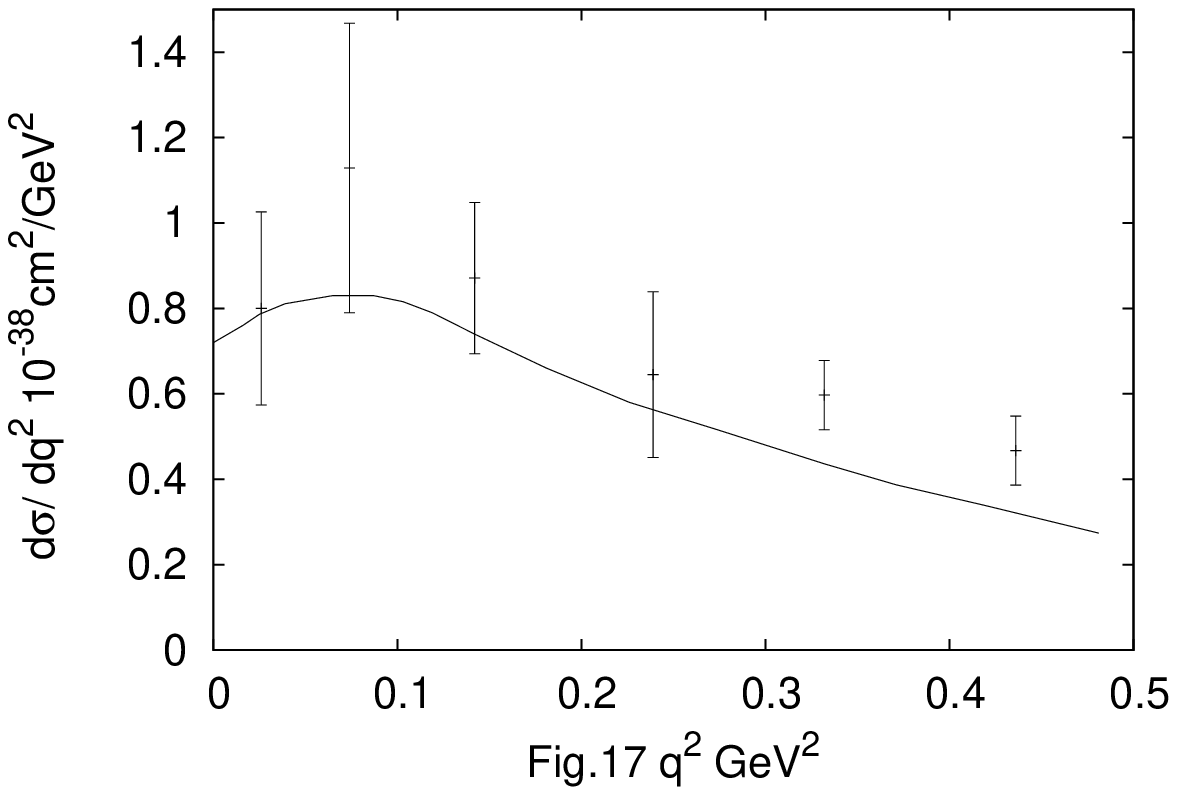}
%FIG. 17.
\end{center}
\end{figure}

The data of Fig. 16 and Fig. 17 are taken from Ref. [85]. 

Newer experimental data [86] of $\sigma(\nu + p\rightarrow \mu^- + \Delta^{++}(p + \pi^+))$ 
and ${d\sigma\over dq^2}$ are shown in Fig. 18 and Fig. 19. 
\begin{figure}
\begin{center}
\includegraphics[width=7in, height=7in]{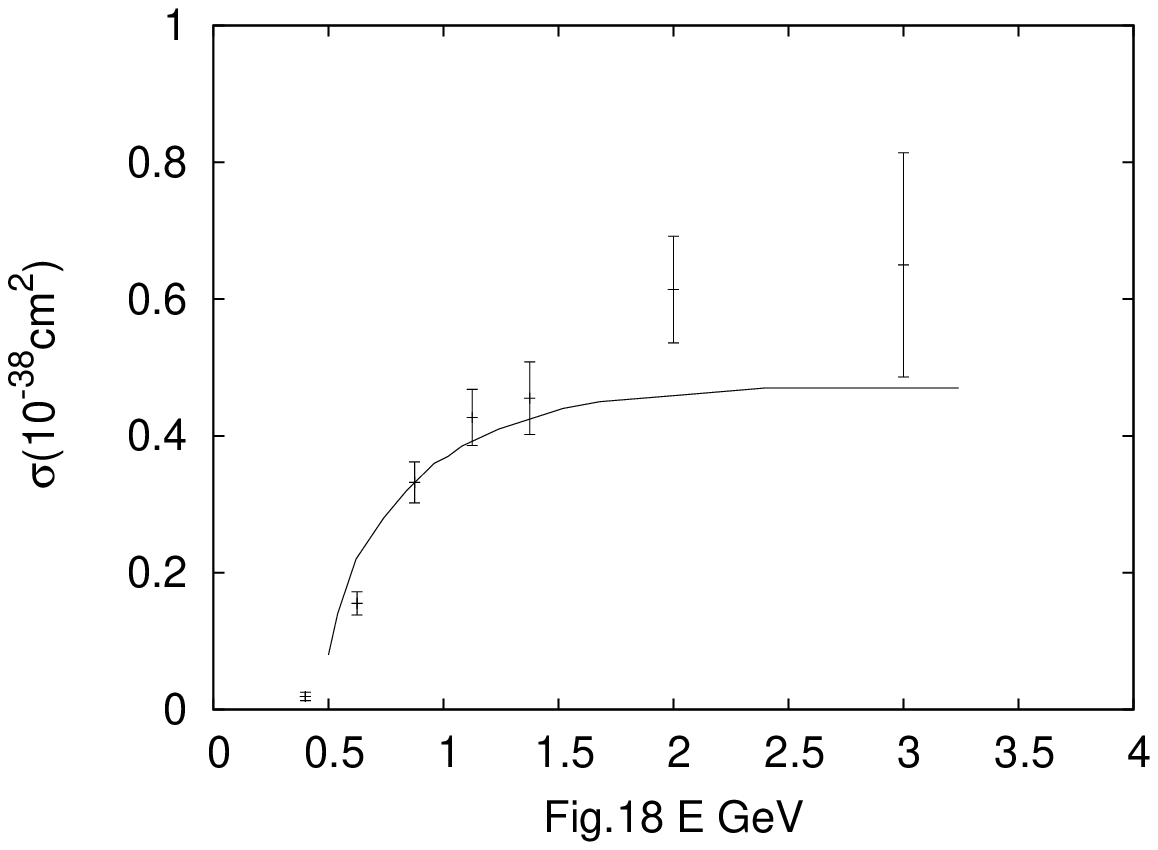}
%FIG. 18.
\end{center}
\end{figure}
\begin{figure}
\begin{center}
\includegraphics[width=7in, height=7in]{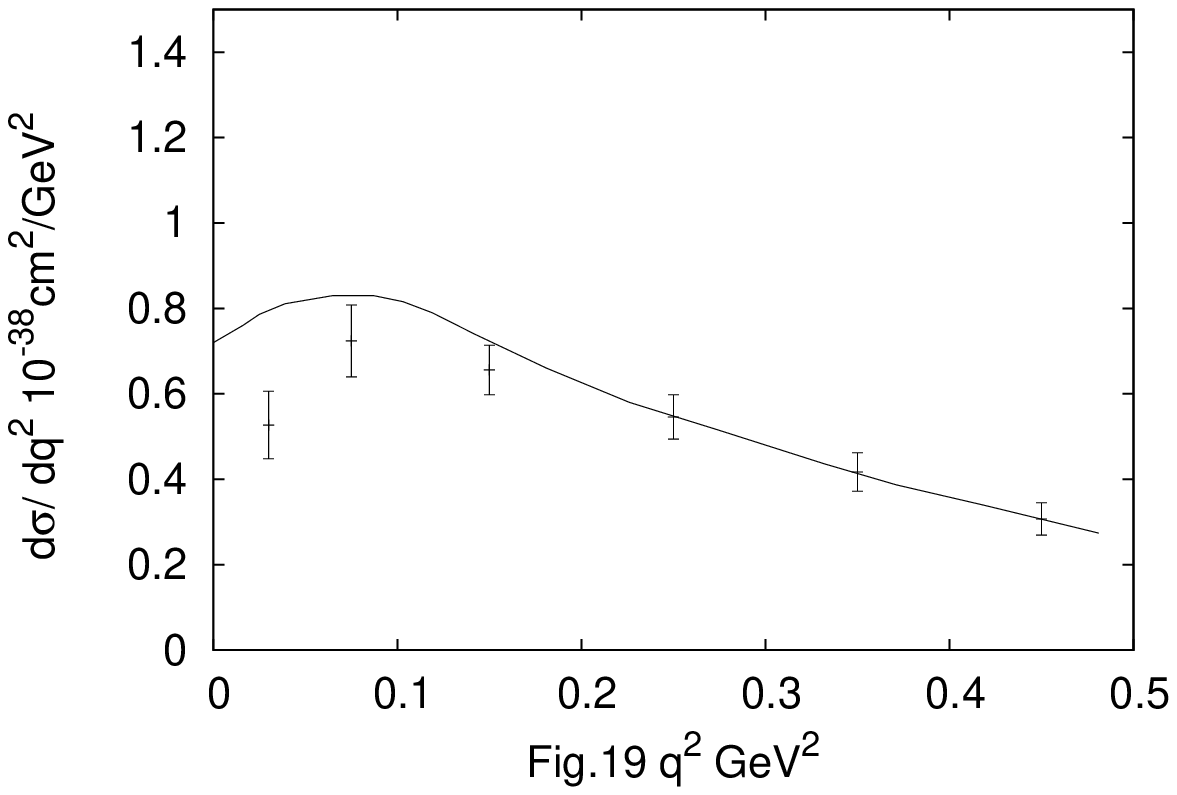}
%FIG. 19.
\end{center}
\end{figure}
The data used in Fig. 18 and Fig. 19 [86] are the cross sections of $\nu + p \rightarrow 
\mu^- p + \pi^+$ when $M_{N\pi} < 1.4\; \textrm{GeV}$. The background has not been subtracted.
Fig. 16, 17, 18, 19 show that theoretical predictions are compatible with data.

\subsection{Density matrix}
When pions produced from the decay of the $\Delta$ are measured the differential cross
section is written as
\begin{eqnarray}
\lefteqn{\frac{d^4\sigma}{dq^2 dm^{'2} d\Omega_\pi}={1\over 4\pi}\frac{d^2\sigma}
{dq^2 dm^{'2}}\{Y_{00}-{2\sqrt{5}}(\rho_{33}-{1\over2})Y_{20}}\nonumber \\
&&+{4\over\sqrt{10}}\rho_{31}ReY_{21}-{4\sqrt{10}}\rho_{3-1}ReY_{22}\},
\end{eqnarray}
$\frac{d^2\sigma}{dq^2 dm{'2}}$ is derived from Eq.(257). The results are shown in 
Table 6.
\begin{table}[h]
\begin{center}
\caption{Density matrix elements}
\begin{tabular}{|c|c|c|c|} \hline
 &$\rho_{33}$&$\rho_{3-1}$&$\rho_{31}$
\\ \hline
Experimental value [85]&$0.58\pm0.09$&$-0.24\pm0.11$&$-0.18\pm0.11$ \\  \hline
Experimental value [86]&$0.661\pm0.036$&$-0.088\pm0.042$&$-0.107\pm0.040$ \\  \hline
Theoretical value &0.95&-0.01&-0.26\\ \hline
\end{tabular}
\end{center}
\end{table}

\section{$\sigma$ term of nucleon}
In this model there are two Lorentz invariant functions, $f_1$ and $f_2$, in the 
wave functions. It is shown in section 1 that the difference between $f_1$ and $f_2$ 
represents the antiquark spinors. If \(f_1=f_2\) the effects of the antiquark 
spinors disappear and theoretical results disagree with data.
As shown in this paper the contribution of the antiquarks plays essential role in
the EM and weak form factors of nucleon or the structure of nucleon.

%In Ref. [14] the antiquark content of proton is defined as Eq. (42).
%\begin{equation}
%\bar{q}_i={1\over2}<p|\bar{\psi}_i\psi_i-\bar{\psi}_i\gamma_0\psi_i|p>.
%\end{equation}
%This definition (42) has been used to study the quark sea of nucleon and the 
%$\sigma$ term of nucleon. Eq. (42) can be used to argue 
%the existence of the contents of
%$\bar{u}$ and $\bar{d}$ in a proton.
 
%From the expressions of the wave functions of nucleon (33,34) if \(f_1 = f_2\)
%is taken only the quark components contribute.  
%The matrix $\gamma_0$ doesn't affects the wave function in the rest frame, therefore,
%\[<p|\bar{\psi}_i\psi_i|p>=<p|\bar{\psi}_i\gamma_0\psi_i|p>\]
%the antiquark content of nucleon is zero. However, if \(f_1\neq f_2\)
%antiquark spinors are part of the wave functions. When $\gamma_0$ acts on antiquark
%spinors a minus sign is generated. Then
%\[<p|\bar{\psi}_i\psi_i|p>\neq<p|\bar{\psi}_i\gamma_0\psi_i|p>\]
%and according to Eq. (42), $\bar{q}_i \neq 0$. Of course, meson clouds
%^contribute to the antiquark contents of nucleon. 

The $\sigma$ term of nucleon is calculated
\begin{equation}
\sigma = <p|m_u\bar{u}u+m_d\bar{d}d|p> = (a+{1\over a}-1)(2m_u+m_d)=3.73\; ( 2m_u + m_d).
\end{equation}
There is a wide range for $m_u$ and $m_d$ [46]: $m_u = 2.3 ^{+0.7}_{-0.5}\; \textrm{MeV}$,
$m_d = 4.8 ^{+0.7}_{-0.3}\; \textrm{MeV}$.
The $\sigma$ term of nucleon is determined to be
\begin{equation}
\sigma = 30.2 - 42.9\; \textrm{MeV}.
\end{equation}
If $a = 1$ is taken,
\[\sigma = 2 m_u + m_d = 8.1 - 11.5\;\textrm{MeV}\]
is obtained. This value of the $\sigma$ term is much smaller than the one (280) and much smaller than other
results.

In Ref. [88] $\sigma \sim 45 \;\textrm{MeV}$ is determined. In Refs. [92,94] the
$\sigma$ term determined by Lattice calculations are listed
\[15 - 25\; \textrm{MeV},\;40 - 60\; \textrm{MeV},\;50 \pm 3\; \textrm{MeV},\;45 - 55\; \textrm{MeV},\;
18 \pm 5\; \textrm{MeV},\;49\pm 3\; \textrm{MeV}.\]
In Ref. [14] $\sigma = 37 \pm 8 \pm 6 \;\textrm{MeV}$ is reported. $\sigma = 52 \pm 3 \pm 8\; \textrm{MeV}$
is obtained in Ref. [93]. The value of the $\sigma$ term obtained in this paper is compatible with
these values. 
\section{Antiquark components of nucleon} 
As mentioned in this paper that the antiquark spinors $v_{\pm}$ (30) of baryons play very important role in understanding
the structure of baryon. Eq. (42) shows that the antiquark spinors contribute to the density of the antiquarks of nucleon.

In Ref. [14] the antiquark content of proton is defined as Eq. (42).
%\begin{equation}
%\bar{q}_i={1\over2}<p|\bar{\psi}_i\psi_i-\bar{\psi}_i\gamma_0\psi_i|p>.
%\end{equation}
This definition (42) has been used to study the quark sea of nucleon and the
$\sigma$ term of nucleon. Eq. (42) can be used to argue
the existence of the contents of
$\bar{u}$ and $\bar{d}$ in a proton.

From the expressions of the wave functions of nucleon (33,34) if \(f_1 = f_2\)
is taken only the quark components contribute.
The matrix $\gamma_0$ doesn't affects the wave function in the rest frame, therefore,
\[<p|\bar{\psi}_i\psi_i|p>=<p|\bar{\psi}_i\gamma_0\psi_i|p>\]
the antiquark content of nucleon is zero. However, if \(f_1\neq f_2\)
antiquark spinors are part of the wave functions. When $\gamma_0$ acts on antiquark
spinors a minus sign is generated. Then
\[<p|\bar{\psi}_i\psi_i|p>\neq<p|\bar{\psi}_i\gamma_0\psi_i|p>\]
and according to Eq. (42), $\bar{q}_i \neq 0$. Of course, meson clouds
contribute to the antiquark contents of nucleon.

Using the wave functions (33,34), the antiquark densities 
\begin{eqnarray}
\bar{u} = {1\over2} <p|\bar{u}(1 - \gamma_0) u|p>,\nonumber\\
\bar{d} = {1\over2} <p|\bar{d}(1 - \gamma_0) d|p>,\nonumber\\
\bar{s} = {1\over2} <p|\bar{s}(1 - \gamma_0) s|p>
\end{eqnarray}
are calculated 
\begin{eqnarray}
\bar{u} = {2\over a}(a - 1)^2,\nonumber \\
\bar{d} = {1\over a}(a - 1)^2,\nonumber \\
\bar{s} = 0.
\end{eqnarray}
In this model there is no strange quark component. Therefore, $\bar{s} = 0$ is natural. The antiquark density, $\bar{u}$ and $\bar{d}$, vanish when $a = 1$ 
as expected. Obviously, there are other contributors for the density of the antiquarks of nucleon.
Eq. (281) shows the contribution of the antiquark spinors of nucleon to the density of antiquark of proton only.
It is very interesting to explore the dynamics of nonzero antiquark density. This investigation is beyond the scope of this paper.

\section{Summary}
A review of an approach of the study of EM and weak structure of nucleon done in Refs. [4,5,6,7] in 70's is presented. 
In these study a nucleon in the rest frame is spherical. 
The wave functions of ${1\over2}^+$ and ${3\over2}^+$ 
in the frame of center of mass are SU(6) symmetric and are boosted to moving frame
by Lorentz transformation. Antiquark components (spinors of antiquarks) are naturally cooperated in these wave functions.
These wave functions, effective Lagrangian of electromagnetic and weak interactions [1] have been
applied to study the electromagnetic and weak form factors of nucleons and $p\rightarrow\Delta$.
There are three inputs: $G^p_M(q^2)$, $\mu_p$, and $\mu_\Lambda$.
A new expression of
$R = \frac{\mu_p G^p_E(q^2)}{G^p_{M}(q^2)}$ is obtained. In small region of $q^2$  the ratio is flat and $R \sim 1$ and it agrees with data.
As $q^2$ increases the ratio decreases and is consistent with data when $q^2 < 5\;\textrm{GeV}^2$. The ratio predicted by this model  
decreases faster than the data when $q^2 > 5\; \textrm{GeV}^2$.
Nonzero small electric form factor $G^n_E(q^2)$ is predicted and agrees with data when $q^2 < 0.3\;\textrm{GeV}^2$. When $q^2 > 0.3\;\textrm{GeV}^2$ 
$G^n_E(q^2)$ is larger than the experimental data. In this model the $G^n_E(q^2)$
is resulted in the contribution of the antiquark components. The magnetic moments of hyperons 
predicted by this model have the right signs but smaller than data by about $30\%$. The SU(3) symmetry breaking must be taken into account.
The transit magnetic moment $\mu_{p\rightarrow \Delta}$ predicted by this model agrees well with data.
Two helicity amplitudes of $\Delta \rightarrow N + \gamma$ are predicted. The the magnetic transition is dominated 
and $E1+$ moment contributes about $5\%$ of the magnetic amplitude. The magnetic amplitude agrees with data.
The helicity amplitudes and the decay rate predicted are in good agreement with data.
For the process $e + p \rightarrow \Delta^+ + e$ three form factors, $G_{M1+},\;G_{E1+},\;G_{S1+}$, are determined.
The $G_{M1+}$ decreases faster with $q^2$ than $G^p_M(q^2)$ and agrees with data in small $q^2$ region. 
In the region of
larger $q^2$ the $G_{M1+}$ predicted decreases faster. In the form factor $G_{M1+}$ $SU(6)$ symmetry is applied and the
${1\over \mu_p}G^p_M(q^2) = \frac{1}{(1 + {q^2\over 0.71})^2}$ is inputted to $G_{M1+}$
and the mass difference between proton and $\Delta$ is ignored. Phenomenologically, when the effect of the physical mass of the $\Delta$
is taken into account the $G_{M1+}$ agrees with data when $q^2 < 5\; \textrm{GeV}^2$. 
Therefore, the SU(6) symmetry breaking effect plays an important role in the behavior of the $G_{M1+}$ at larger $q^2$.
This model presents a new picture for the structure of nucleon. Both the negative and small $E1+(q^2)$ and the $S1+(q^2)$ form factors
are resulted in both the contribution of antiquark components and Lorentz contraction. In the rest frame both the proton and the $\Delta$ 
are spherical. 
The $S1+(q^2)$ form factor is consistent
with data and the $E1+(q^2)$ is small and about twice of the data. This model doesn't work well in the range of large $q^2$. 
In the range of large $q^2$ many new physical effects like internal motions of quarks and perturbative gluons should play roles.
The assumption $f_2 = a f_1$ may not work well for large $q^2$ too.
This is the limitation of this model.

The new wave functions of baryons are applied to study the weak interactions of baryons. There is an additional parameter $\lambda$ for
the axial-vector currents of quarks [1], which is determined by inputting $G_A(0)$ of the $\beta$ decay of nucleon. 
The axial-vector constants of hyperons, $G_A$, are predicted. The theoretical predictions of the semi-leptonic decays of 
neutron and all hyperons are in good agreement with data.
The axial-vector form factor $G_A(q^2)$ of nucleon is predicted to be $G_A(q^2) = F^p_1(q^2)$. It is not in the form of dipole exactly. However, numerical calculation shows that 
the $G_A(q^2)$ predicted is in good agreement with the form of dipole with $M_A = 1.002 \textrm{GeV}$. 
It is interesting to notice that the pseudoscalar form factor of nucleon is caused by the antiquark components.
The theoretical predictions of the 
cross sections of $\nu + n \rightarrow p + \mu$,
$\bar{\nu} + p \rightarrow n + \mu^+$ agree with data well. The $\Delta S = 1$ quasielastic neutrino scattering 
$\bar{\nu} + p \rightarrow \Lambda + \mu^+$ and $\bar{\nu} + p \rightarrow \ + \mu^+$ are predicted. 
This approach has been applied to study $\nu + p\rightarrow \Delta^{++} + \mu^-$ without new parameters. There are three vector form factors 
and three axial-vector form factors. One vector and two axial-vector form factors play dominant roles. 
The form factors, especially the axial-vector form factors are not in the form of dipoles, but in the form of tripoles.
The cross section, differential cross
section, and density matrix elements are predicted. Theory are consistent with data.

The study presented in this review paper shows that antiquark components of nucleon play an essential role in understanding nucleon structure. 

\newpage
{\bf Appendix}\\

{\bf Appendix I $\;\;$Flavor wave functions of baryons}
The flavor wave functions for ${1\over2}^+$ baryons:\\
\(p,\;\; B = \left(\begin{array}{ccc}
                        0\;\;0\;\;1\\
                        0\;\;0\;\;0\\ 
                        0\;\;0\;\;0
                        \end{array}
                        \right) ,\;\;n,\;\; B = \left(\begin{array}{ccc}
                        0\;\;0\;\;0\\
                        0\;\;0\;\;1\\
                        0\;\;0\;\;0
                        \end{array}
                        \right),\) \\
\(\Sigma^+,\;\;B = \left(\begin{array}{ccc}
                        0\;\;1\;\;0\\
                        0\;\;0\;\;0\\
                        0\;\;0\;\;0
                        \end{array}
                        \right) ,\;\;\Sigma^-,\;\; B = \left(\begin{array}{ccc}
                        0\;\;0\;\;0\\
                        1\;\;0\;\;0\\
                        0\;\;0\;\;0
                        \end{array}
                        \right) ,\)\\
\(\Sigma^0,\;\;B = {1\over\sqrt{2}}\left(\begin{array}{ccc}
                        1\;\;0\;\;0\\
                        0\;\;-1\;\;0\\
                        0\;\;0\;\;0
                        \end{array}
                        \right) ,\;\;\Lambda,\;\; B = {1\over\sqrt{6}}\left(\begin{array}{ccc}
                        1\;\;0\;\;0\\
                        0\;\;1\;\;0\\
                        0\;\;0\;\;-2
                        \end{array}
                        \right) ,\)\\
\(\Xi^-,\;\;B = \left(\begin{array}{ccc}
                        0\;\;0\;\;0\\
                        0\;\;0\;\;0\\
                        1\;\;0\;\;0
                        \end{array}
                        \right) ,\;\;\Xi^0,\;\; B = \left(\begin{array}{ccc}
                        0\;\;0\;\;0\\
                        0\;\;0\;\;0\\
                        0\;\;1\;\;0
                        \end{array}
                        \right) ,\) \\
The flavor wave functions of ${3\over2}^+$ baryons are defined as
\[d^{111}_{111}=\Delta^{++},\;\;\;d^{112}_{112}={1\over\sqrt{3}}\Delta^{+},\;\;
d^{122}_{122}={1\over\sqrt{3}}\Delta^{0},\;\;
d^{222}_{222}=\Delta^{-},\]
\[d^{113}_{113}={1\over\sqrt{3}}\Sigma^{*+},\;\;d^{123}_{123}={1\over\sqrt{6}}\Sigma^{*0},
\;\;d^{223}_{223}={1\over\sqrt{3}}\Sigma^{*-},\]
\[d^{133}_{133}={1\over\sqrt{3}}\Xi^{*0},\;\;d^{233}_{233}={1\over\sqrt{3}}\Xi^{*-},
\;\;d^{333}_{333}=\Omega^-.\]
{\bf Appendix II $\;\;$Permutation operators}\\
The four operators of permutation are expressed as
\begin{eqnarray}
\lefteqn{O_1={1\over6}\{2e+2(12)-(13)-(32)-(123)-(132)\}}\\
&&O_2={1\over6}\{2e-2(12)+(13)+(32)-(123)-(132)\}\\
&&Y_s={1\over6}\{e+(12)+(13)+(32)+(123)+(132)\}\\
&&Y_a={1\over6}\{e-(12)-(13)-(32)+(123)+(132)\}
\end{eqnarray}
The $O_1$ and $O_2$ projectors satisfy
\begin{eqnarray}
\lefteqn{O_1\cdot O_1=O_1}\nonumber \\
&&O_2\cdot O_2=O_2\nonumber \\
&&O_1\cdot O_2=0.
\end{eqnarray}
{\bf Appendix III $\;\;$Wave functions of excited baryons}\\
As mentioned in the paper the electromagnetic and weak
interactions of baryons which are in s-wave in the rest frame have been studied by this approach reasonably well. This approach can be extended to 
study the EM and neutrino( antineutrino) productions of the low-lying excited
baryons. 
The effective electromagnetic currents and the weak currents are the same as Eqs. (48,196). The wave functions of 
excited baryons can be constructed in the same way. 
The wave functions of low lying excited baryons have been constructed in Ref. [4] and are 
presented in this section of Appendix.
In the rest frame of the center of mass the $O(3)\times SU(6)$ symmetry is assumed for these states.
In the rest frame O(3) symmetry determines the part of orbital angular momentum of baryon and  
the SU(6) symmetry determines the parts of the spin and the flavor. The flavor, spin, and the orbital wave functions 
of the baryon is totally symmetric. Of course the color part is antisymmetric.
According to SU(6) symmetry, baryons can be classified as $\underline{56}$, $\underline{70}$, $\underline{20}$
states. These states are decomposed as states of spin ($S={1\over2}$ or $S={3\over2}$)
and flavor ( octet, decuplet, or singlet)
\begin{eqnarray}
\underline{56} = ({3\over2}, 10) + ({1\over2}, 8),\nonumber \\
\underline{70} = ({3\over2}, 8) + ({1\over2}, 8) + ({1\over2}, 10) + ({1\over2}, 1), \nonumber \\
\underline{20} = ({1\over2}, 8) + ({3\over2}, 1).
\end{eqnarray}
The wave functions of baryons of the $\underline{56}$ in s-wave have been studied and presented in this paper. 
The wave functions in p wave and d waves [4] are constructed below.
Baryon is a system of three quarks. There are two independent relative coordinates $x = x_1 - x_2$, $y = {1\over2}(x_1 + x_2) - x_3$.
Therefore, in the rest frame of the baryon there are two relative orbital angular momentum.\\
{\bf  (1s1p) wave functions of baryon}\\
In the rest frame there are two spacial wave functions 
\begin{eqnarray}
f_1(x, y) x_j,\;\;f_2(x,y) y_j,
\end{eqnarray}
where $j = 1, 2, 3$, 
$x_j$ and $y_j$ have $O_2$ or $O_1$ symmetry respectively
\begin{eqnarray}
O_2 x_j = x_j,\;\;O_1 y_j = y_j,
\end{eqnarray}
$f_{1,2}(x,y)$ are new Lorentz invariant functions which are total symmetric in $x_1,\;x_2,\;x_3$. The parity of the baryon of p-wave is negative and they are $\underline{20}-plet$.
The spacial, the spin, and the flavor wave function must be total symmetric.
The wave functions are constructed 
\begin{enumerate}
\item $S = {1\over 2}$ and Octet
\begin{eqnarray}
B^{JM}_{\alpha\beta\gamma,ijk}(x,y)^m_l = {1\over2} \epsilon_{i'j'k'}\sum_{\lambda,c}C^{JM}_{1\lambda{1\over2}c}
x\cdot e^\lambda(p)\{\Gamma^{{1\over2}}_{\gamma\beta,\alpha}(x,y,p)_c \epsilon_{jkm}\delta_{il}\nonumber\\
+ \Gamma^{{1\over2}}_{\alpha\gamma,\beta}(x,y,p)_c \epsilon_{ikm}\delta_{jl}\} 
+ {1\over3} \epsilon_{i'j'k'}
\sum_{\lambda,c}C^{JM}_{1\lambda{1\over2}c}y\cdot e^\lambda(p) \nonumber \\
\{\Gamma^{{1\over2}}_{\gamma\beta,\alpha}(x,y,p)_c (\epsilon_{jim}\delta_{kl}+\epsilon_{kim}\delta_{jl})+
\Gamma^{{1\over2}}_{\alpha\gamma,\beta}(x,y,p)_c (\epsilon_{jim}\delta_{kl}+\epsilon_{jkm}\delta_{il})\}.
\end{eqnarray}
\item $S = {3\over 2}$ and Octet\\
\begin{eqnarray}
B^{JM}_{\alpha\beta\gamma,ijk}(x,y)^m_l =  \epsilon_{i'j'k'}\sum_{\lambda,c}C^{JM}_{1\lambda{3\over2}c}
\Gamma^{{3\over2}}_{\alpha\beta\gamma}(x,y,p)_c\{{1\over2}
x\cdot e^\lambda(p)\nonumber \\
\epsilon_{ijm}\delta_{kl} + {1\over3} y\cdot  e^\lambda(p)(\epsilon_{jim}\delta_{kl} + \epsilon_{jkm}\delta_{il})\}
\end{eqnarray}
\item $S = {1\over 2}$ and Decuplet\\
\begin{eqnarray}
B^{JM, lmn}_{\alpha\beta\gamma,ijk}(x,y) = {1\over2} \epsilon_{i'j'k'} d^{lmn}_{ijk}\sum_{\lambda,c}
C^{JM}_{1\lambda{1\over2}c} \{x\cdot e^\lambda(p)\Gamma^{{1\over2}}_{\alpha\beta,\gamma}(x,y,p)_c  \nonumber \\
+ {2\over3} y\cdot e^\lambda(p) [\Gamma^{{1\over2}}_{\gamma\alpha,\beta}(x,y,p)_c
+ \Gamma^{{1\over2}}_{\gamma\beta,\alpha}(x,y,p)_c]\}
\end{eqnarray}
\item $S = {1\over 2}$ and Singlet\\
\begin{eqnarray}
B^{JM}_{\alpha\beta\gamma}(x,y) = \epsilon_{i'j'k'}\epsilon_{ijk}\sum_{\lambda,c}
C^{JM}_{1\lambda{1\over2}c}\{{1\over2}x\cdot e^\lambda(p)[\Gamma^{{1\over2}}_{\beta\gamma,\alpha}(x,y,p)_c\nonumber \\
+ \Gamma^{{1\over2}}_{\alpha\gamma,\beta}(x,y,p)_c] +  y\cdot e^\lambda(p)\Gamma^{{1\over2}}_{\alpha\beta,\gamma}(x,y,p)_c\}
\end{eqnarray}
\end{enumerate}
{\bf  (1p1p) and (1s1d) wave functions of baryons}\\
The parity of those states are positive. For states of (1p1p) there are three orbital angular momentum
\[L = 2, 1, 0.\] 
Their spacial wave functions and property of symmetry of these states are
\begin{eqnarray}
L = 2,\;\;f(x,y)\sum_{\lambda_1, \lambda_2} C^{(2\lambda}_{1\lambda_1,1\lambda_2}x\cdot e^{\lambda_1}(p) y\cdot e^{\lambda_2}(p)\;\;\;\;
O_2 \nonumber \\
L = 1,\;\;f(x,y)\sum_{\lambda_1, \lambda_2} C^{(1\lambda}_{1\lambda_1,1\lambda_2}x\cdot e^{\lambda_1}(p) y\cdot e^{\lambda_2}(p)\;\;\;\;
Y_a\nonumber \\
L = 0,\;\;f(x,y)\{x\cdot y+{1\over m^2}p\cdot x p\cdot y\},\;\;\;\; O_2.
\end{eqnarray}
The spacial wave functions of the (1s1d) states are $L = 2$ and
\begin{eqnarray}
f_1(x,y)C^{(2\lambda}_{1\lambda_1,1\lambda_2}x\cdot e^\lambda_1(p) x\cdot e^\lambda_2(p),\;\;\;\;\;Y_s,\;\;O_1,\nonumber \\
f_2(x,y)C^{(2\lambda}_{1\lambda_1,1\lambda_2}y\cdot e^\lambda_1(p) y\cdot e^\lambda_2(p),\;\;\;\;\;Y_s,\;\;O_1.
\end{eqnarray}

The classification of these states are\\
\[(1s1d)\;\; are\;\; \underline{56}-plet,\]
\[the\;\; L = 2\;\; states\;\; of\; (1s1d)\;\; and\;\;(1p1p)\;\; are\; \;\underline{70}-plet,\]
\[the \;\;L = 1\;\; of\;\;(1p1p)\; \;are\;\; \underline{20}-plet,\]
\[the \;\;L = 0\;\; of\;\;(1p1p)\; \;are\;\; \underline{70}-plet.\]
The complete wave functions of these states are constructed as
\begin{enumerate}
\item $L = 2$ $\underline{56}$-plet\\
{\bf Octet}\\
\begin{eqnarray}
B^{JM}_{\alpha\beta\gamma,ijk}(x,y)^m_l =  \epsilon_{i'j'k'}\sum_{\lambda,c}C^{JM}_{2\lambda{1\over2}c}C^{2\lambda}_{1\lambda_1 1\lambda_2}
\{x\cdot e^{\lambda_1}(p) x\cdot e^{\lambda_2}(p) \nonumber \\
+ {4\over3} y\cdot e^{\lambda_1}(p) y\cdot e^{\lambda_2}(p)\}
\{\Gamma^{{1\over2}}_{\alpha\beta,\gamma}(x,y,p)_c(\epsilon_{ijm}\delta_{kl} + \epsilon_{ikm}\delta_{jl})
\nonumber \\
+ \Gamma^{{1\over2}}_{\beta\gamma,\alpha}(x,y,p)_c(\epsilon_{jkm}\delta_{il} + \epsilon_{ikm}\delta_{jl})\}.
\end{eqnarray}
{\bf Decuplet}\\
\begin{eqnarray}
B^{JM, lmn}_{\alpha\beta\gamma,ijk}(x,y) = \epsilon_{i'j'k'} d^{lmn}_{ijk}\sum_{\lambda1,\lambda_2}
C^{JM}_{2\lambda{3\over2}c}
C^{2\lambda}_{1\lambda_1,1\lambda_2}\{x\cdot e^{\lambda_1}(p) x\cdot e^{\lambda_2}(p) \nonumber \\
+ {4\over3} y\cdot e^{\lambda_1}(p) y\cdot e^{\lambda_2}(p)\}
\Gamma^{{3\over2}}_{\alpha\beta\gamma}(x,y,p)_c.
\end{eqnarray}
\item $L = 0$ $\underline{56}$-plet\\
The wave functions can be obtained by replacing
\[\sum_{\lambda1,\lambda_2}C^{JM}_{2\lambda{3\over2}c}\{x\cdot e^{\lambda_1}(p) x\cdot e^{\lambda_2}(p) 
+ {4\over3} y\cdot e^\lambda_1(p) y\cdot e^\lambda_2(p)\}\]
of Eqs. (293,294) by
\[x^2 + {4\over3} y^2 + {1\over m^2}\{(p\cdot x)^2 + {4\over 3} (p\cdot y)^2\}.\]
These states are the radial excitations of the ground states of the $\underline{56}$-plet.
\item $L = 0$ $\underline{70}$-plet
{\bf  $S = {1\over2}$ Octet}\\
\begin{eqnarray}
B^{{1\over2}\lambda}_{\alpha\beta\gamma,ijk}(x,y)^m_l = {1\over2} \epsilon_{i'j'k'}
(x\cdot y + {1\over m^2} p\cdot x p\cdot y)
\{\Gamma^{{1\over2}}_{\alpha\gamma,\beta}(x,y,p)_\lambda \epsilon_{ikm}\delta_{jl}\nonumber \\
+ \Gamma^{{1\over2}}_{\gamma\beta,\alpha}(x,y,p)_\lambda \epsilon_{jkm}\delta_{il}\}
+ {1\over 8}\epsilon_{i'j'k'}\{x^2 - {4\over3} y^2 + {1\over m^2}((p\cdot x)^2 - 
{4\over3} (p\cdot y)^2)\}\nonumber \\
\{\Gamma^{{1\over2}}_{\gamma\beta,\alpha}(x,y,p)_\lambda(\epsilon_{jim}\delta_{kl} + \epsilon_{kim}\delta_{jl})
+ \Gamma^{{1\over2}}_{\gamma\alpha,\beta}(x,y,p)_\lambda(\epsilon_{kjm}\delta_{il} + \epsilon_{ijm}\delta_{kl})\}.
\end{eqnarray}
{\bf $S = {3\over2}$ Octet}\\
\begin{eqnarray}
B^{{3\over2}\lambda}_{\alpha\beta\gamma,ijk}(x,y)^m_l = {1\over2} \epsilon_{i'j'k'}\Gamma^{{3\over2}}_{\gamma\beta\alpha}(x,y,p)_\lambda
\{(x\cdot y + {1\over m^2} p\cdot x p\cdot y)\epsilon_{ijm}\delta_{kl}\nonumber \\
+ {1\over4} [x^2 - {4\over 3} y^2 + {1\over m^2} (p\cdot x)^2 - {4\over 3}{1\over m^2} 
(p\cdot y)^2] (\epsilon_{kjm}\delta_{il} + \epsilon_{kim}\delta_{jl})\}.
\end{eqnarray}
{\bf $S = {1\over2}$ Decuplet}
\begin{eqnarray}
B^{{1\over2}\lambda, lmn}_{\alpha\beta\gamma,ijk}(x,y) = \epsilon_{i'j'k'} d^{lmn}_{ijk}
(x\cdot y + {1\over m^2}p\cdot x p\cdot y)\nonumber \\
\{ \Gamma^{{1\over2}}_{\gamma\beta,\alpha}(x,y,p)_\lambda + \Gamma^{{1\over2}}_{\gamma\alpha,\beta}(x,y,p)_\lambda\}.
\end{eqnarray}
{\bf $S = {1\over2}$ Singlet}\\
\begin{eqnarray}
B^{{1\over2}\lambda}_{\alpha\beta\gamma}(x,y) = \epsilon_{i'j'k'}\epsilon_{ijk} (x\cdot y + 
{1\over m^2}p\cdot x p\cdot y)\nonumber \\
\{\Gamma^{{1\over2}}_{\beta\gamma,\alpha}(x,y,p)_\lambda + \Gamma^{{1\over2}}_{\gamma\alpha,\beta}(x,y,p)_\lambda\}\nonumber \\
+ {3\over 8} \epsilon_{i'j'k'}\epsilon_{ijk}\{x^2 - {4\over 3} y^2 + {1\over m^2} (p\cdot x)^2 - {4\over 3}{1\over m^2}
(p\cdot y)^2] \}\Gamma^{{1\over2}}_{\alpha\beta,\gamma}(x,y,p)_\lambda.
\end{eqnarray}
\item $L = 2$ $\underline{70}$-plet\\
{\bf $S = {1\over2}$ Octet}\\
\begin{eqnarray}
B^{JM}_{\alpha\beta\gamma,ijk}(x,y)^m_l = {1\over2} \epsilon_{i'j'k'}\sum_{\lambda_1,\lambda_2}
C^{JM}_{2\lambda{1\over2}c}
C^{2\lambda}_{1\lambda_1,1\lambda_2}\{x\cdot e^{\lambda_1}(p) y\cdot e^{\lambda_2}(p)\nonumber \\
(\Gamma^{{1\over2}}_{\alpha\gamma,\beta}(x,y,p)_c \epsilon_{ikm}\delta_{jl}
 + \Gamma^{{1\over2}}_{\gamma\beta,\alpha}(x,y,p)_c \epsilon_{jkm}\delta_{il}) \nonumber \\
+ {1\over4}[x\cdot e^{\lambda_1}(p) x\cdot e^{\lambda_2}(p)
- {4\over3} y\cdot e^{\lambda_1}(p) y\cdot e^{\lambda_2}(p)] \nonumber \\
(\Gamma^{{1\over2}}_{\gamma\beta,\alpha}(x,y,p)_c \epsilon_{kjm}\delta_{il}
+ \Gamma^{{1\over2}}_{\gamma\alpha,\beta}(x,y,p)_c \epsilon_{ijm}\delta_{kl}\}
\end{eqnarray}
{\bf $S = {3\over2}$ Octet}\\
\begin{eqnarray}
B^{JM}_{\alpha\beta\gamma,ijk}(x,y)^m_l =  {1\over2}\epsilon_{i'j'k'}\sum_{\lambda_1,\lambda_2}C^{JM}_{2\lambda{3\over2}c}
C^{2\lambda}_{1\lambda_1,1\lambda_2}\Gamma^{{3\over2}}_{\alpha\gamma\beta}(x,y,p)_c\nonumber \\
\{x\cdot e^{\lambda_1}(p) y\cdot e^{\lambda_2}(p)\epsilon_{ijm}\delta_{kl}
+ {1\over4}[x\cdot e^{\lambda_1}(p) x\cdot e^{\lambda_2}(p)\nonumber \\
- {4\over3} y\cdot e^{\lambda_1}(p) y\cdot e^{\lambda_2}(p)][\Gamma^{{1\over2}}_{\gamma\beta,\alpha}(x,y,p)_c + 
\Gamma^{{1\over2}}_{\gamma\alpha,\beta}(x,y,p)_c]\}.
\end{eqnarray}
{\bf $S = {1\over2}$ Decuplet}\\
\begin{eqnarray}
B^{JM, lmn}_{\alpha\beta\gamma,ijk}(x,y) = {1\over2}\epsilon_{i'j'k'} d^{lmn}_{ijk}
\sum_{\lambda_1,\lambda_2} C^{JM}_{2\lambda{1\over2}c}
C^{2\lambda}_{1\lambda_1,1\lambda_2} \nonumber \\
\{x\cdot e^{\lambda_1}(p) y\cdot e^{\lambda_2}(p)\Gamma^{{1\over2}}_{\gamma\beta,\alpha}(x,y,p)_c
+ {1\over4}(x\cdot e^{\lambda_1}(p) x\cdot e^{\lambda_2}(p)\nonumber \\
- {4\over3} y\cdot e^{\lambda_1}(p) y\cdot e^{\lambda_2}(p))(\Gamma^{{1\over2}}_{\gamma\beta,\alpha}(x,y,p)_c
+ \Gamma^{{1\over2}}_{\gamma\alpha,\beta}(x,y,p)_c)\}.
\end{eqnarray}
{\bf $S = {1\over2}$ Singlet}\\
\begin{eqnarray}
B^{JM}_{\alpha\beta\gamma}(x,y) = {1\over2}\epsilon_{i'j'k'}\epsilon_{ijk}\sum_{\lambda_1,\lambda_2} C^{JM}_{2\lambda{1\over2}c}
C^{2\lambda}_{1\lambda_1,1\lambda_2}x\cdot e^{\lambda_1}(p) y\cdot e^{\lambda_2}(p) \nonumber \\
\{\Gamma^{{1\over2}}_{\beta\gamma,\alpha}(x,y,p)_c
+ \Gamma^{{1\over2}}_{\alpha\gamma,\beta}(x,y,p)_c\}\nonumber \\
+ {3\over8}\epsilon_{i'j'k'}\epsilon_{ijk}\sum_{\lambda_1,\lambda_2} C^{JM}_{2\lambda{1\over2}c}
C^{2\lambda}_{1\lambda_1,1\lambda_2}\{x\cdot e^{\lambda_1}(p) x\cdot e^{\lambda_2}(p) \nonumber \\
-{4\over3} y\cdot e^{\lambda_1}(p) y\cdot e^{\lambda_2}(p)\}
\Gamma^{{1\over2}}_{\alpha\beta,\gamma}(x,y,p)_c.
\end{eqnarray}
\item $L = 1$ $\underline{20}$-plet\\
{\bf $S = {1\over2}$ Octet}\\
\begin{eqnarray}
B^{JM}_{\alpha\beta\gamma,ijk}(x,y)^m_l = \epsilon_{i'j'k'}\sum_{\lambda_1,\lambda_2}C^{JM}_{1\lambda{1\over2}c}
C^{1\lambda}_{1\lambda_1,1\lambda_2} x\cdot e^{\lambda_1}(p) y\cdot e^{\lambda_2}(p)\nonumber \\
\{\Gamma^{{1\over2}}_{\beta\gamma,\alpha}(x,y,p)_c \epsilon_{kim}\delta_{jl}
+ \Gamma^{{1\over2}}_{\gamma\alpha,\beta}(x,y,p)_c \epsilon_{kjm}\delta_{il}\}.
\end{eqnarray}
{\bf $S = {3\over2}$ Singlet}\\
\begin{eqnarray}
B^{JM}_{\alpha\beta\gamma}(x,y) = \epsilon_{i'j'k'}\epsilon_{ijk}\sum_{\lambda_1,\lambda_2} C^{JM}_{1\lambda{3\over2}c}
C^{1\lambda}_{1\lambda_1,1\lambda_2}x\cdot e^{\lambda_1}(p) y\cdot e^{\lambda_2}(p)
\Gamma^{{3\over2}}_{\alpha\beta\gamma}(x,y,p)_c .
\end{eqnarray}
\end{enumerate}
The $\Gamma^{{1\over2}}_{\alpha\beta,\gamma}(x,y,p)_c$ and the $\Gamma^{{3\over2}}_{\alpha\beta\gamma}(x,y,p)_c$
take the same expressions of Eq. (35) with different $f_{1,2}(x,y)$.

\pagebreak
\begin{flushleft}
{\bf Figure Captions}
\end{flushleft}
{\bf Fig. 1.} Ratio of electric and magnetic form factors of proton.
\\ {\bf Fig. 2.} Ratio of electric and magnetic form factors of proton.
\\{\bf Fig. 3.} Eclectic form factor of neutron.
\\ {\bf Fig. 4.} Electric form factor of neutron.
\\ {\bf Fig. 5} The ratio $\frac{\mu_p G^p_E}{G^p_M}$. 
%The data are from Ref. [1] and [C. B. Crawford , PRL 98 , 052301 (2007); 
%              E. Geis et al., PRL 101, 042501 (2008).]
\\ {\bf Fig. 6} Charge form factor of neutron. 
%the data are taken from Refs. [29,30,31,32,33,34,35,36,37],references therein. 
\\{\bf Fig. 7} Magnetic form factor of $p\rightarrow\Delta$.
\\{\bf Fig. 8(a)} Magnetic form factor of $p\rightarrow\Delta$.
\\{\bf Fig. 8(b)} Magnetic form factor of $p\rightarrow\Delta$ with a possible SU(6) symmetry breaking effect.
\\{\bf Fig. 9} Cross Section of virtual scalar photon.
\\{\bf Fig. 10} Axial-vector form factor of $p\rightarrow n$
\\{\bf Fig. 11} Cross Section of $\nu_\mu + n \rightarrow p + \mu^-$. E is the average neutrino energy.  
\\{\bf Fig. 12} Cross Section of $\nu_\mu + n \rightarrow p + \mu^-$. E is the average neutrino energy.
\\{\bf Fig. 13} Cross Section of $\bar{\nu}_\mu + p \rightarrow n + \mu^+$. E is the average antineutrino energy.  
\\{\bf Fig. 14} Cross Section of $\bar{\nu}_\mu + p \rightarrow \Lambda + \mu^+$.
\\{\bf Fig. 15} Cross Section of $\bar{\nu}_\mu + p \rightarrow \Sigma^0 + \mu^+$
\\{\bf Fig. 16} Cross Section of $\nu_\mu + p \rightarrow \Delta^{++} + \mu^-$.
\\{\bf Fig. 17} Differential cross Section of $\nu_\mu + n \rightarrow \Delta^{++} + \mu^-$, $\frac{d\sigma}{d q^2}$.
\\{\bf Fig. 18} Cross Section of $\nu_\mu + p \rightarrow \Delta^{++} + \mu^-$.
\\{\bf Fig. 19} Differential cross Section of $\nu_\mu + n \rightarrow \Delta^{++} + \mu^-$, $\frac{d\sigma}{d q^2}$.

%\begin{figure}
%\begin{center}
%\includegraphics[width=7in, height=7in]{fig5.ps}
%FIG. 5.
%\end{center}
%\end{figure}

%\begin{figure}
%\begin{center}
%\includegraphics[width=7in, height=7in]{fig6.ps}
%FIG. 6.
%\end{center}
%\end{figure}

%\begin{figure}
%\begin{center}
%\includegraphics[width=7in, height=7in]{fig7.ps}
%FIG. 7.
%\end{center}
%\end{figure}

\end{document}